%% file: main.tex
\makeatletter
\let\QFkernelDocument\document
\let\QFkernelEndDocument\enddocument
\let\QFkernelOutputPage\@outputpage
\makeatother
\documentclass{rQUF2e}
\makeatletter
\let\document\QFkernelDocument
\let\enddocument\QFkernelEndDocument
\let\@outputpage\QFkernelOutputPage
\makeatother

\usepackage{epstopdf}
\usepackage{etoolbox}
\usepackage[a4paper,total={39pc,676pt},centering]{geometry}
\usepackage{mathrsfs}
\usepackage{placeins}
\usepackage{xurl}
\usepackage[hidelinks]{hyperref}
\theoremstyle{plain}
\newtheorem{theorem}{Theorem}
\newtheorem{lemma}{Lemma}
\newtheorem{proposition}{Proposition}
\theoremstyle{remark}
\newtheorem{remark}{Remark}

\makeatletter
\patchcmd{\@maketitle}
  {{\receivedfont{\rm{(}}\@received{\rm{)}}\par}\vskip18pt}
  {}
  {}
  {\PackageError{roughfmm}{Could not remove the rQUF receipt date placeholder}{Check the rQUF class version.}}
\makeatother

\begin{document}

\pagestyle{headings}
\markboth{}{Rough SABR Forward Market Model}
\input{title.tex}

\input{RoughSABRFMM.tex}

\end{document}

%% file: title.tex
\title{Rough SABR Forward Market Model}

\author{
  Reo Adachi$^\ast$,
  Masaaki Fukasawa$^\dagger$\textsuperscript{a},
  Naoki Iida$^\ast$,
  Mitsumasa Ikeda$^\ast$,
  Yo Nakatsu$^\ast$,
  Ryota Tsurumi$^{\ast \dagger}$\textsuperscript{b},
  and Tomohisa Yamakami$^{\ast \ddagger}$\textsuperscript{c}
}

\affil{
  $^\ast$ Mizuho--DL Financial Technology Co., Ltd., Kojimachi-odori Building, 2-4-1 Kojimachi, Chiyoda-ku, Tokyo 102-0083, Japan. \\
  $^\dagger$ Graduate School of Engineering Science, The University of Osaka, 1-3 Machikaneyama, Toyonaka, Osaka 560--8531, Japan. \\
  $^\ddagger$ Graduate School of Economics, The University of Tokyo, 7-3-1 Hongo, Bunkyo-ku, Tokyo 113-0033, Japan. \\
  \textsuperscript{a} Email: \texttt{fukasawa@sigmath.es.osaka-u.ac.jp} \\
  \textsuperscript{b} The corresponding author. Email addresses:
  \texttt{ryota-tsurumi@fintec.co.jp} and \texttt{u693807i@ecs.osaka-u.ac.jp}. \\
  \textsuperscript{c} Email: \texttt{tomohisa-yamakami@fintec.co.jp}
}

\maketitle
\thispagestyle{empty}

\begin{abstract}
This paper advances interest rate modeling in the post-LIBOR era by introducing rough stochastic volatility into the Forward Market Model (FMM). We establish a rigorous asymptotic expansion of swaption implied volatility, connecting the FMM to a rough Bergomi-type framework for forward swap rates. This contribution bridges the gap between Heath-Jarrow-Morton (HJM)-consistent forward term rate models and forward swap rate models with stochastic volatility, offering a parsimonious yet precise framework for modeling swaption volatility surfaces. In calibrations to USD SOFR swaption data, the rough specification yields volatility-of-volatility estimates that vary less across option expiries than those obtained under the non-rough specification. When calibrated to short-tenor smiles and at-the-money quotes for longer tenors, the model also reproduces the main features of longer-tenor implied volatility smiles away from the money. The results suggest that allowing for a Hurst exponent below one half can improve the fit to observed swaption smiles. The proposed approach enhances the representation of persistent skew and term structure, addressing key challenges in modern fixed income markets.
\end{abstract}


\begin{classcode}
G13
\end{classcode}

%% file: RoughSABRFMM.tex
\section{Introduction}

Interest rate modeling serves as the cornerstone for pricing, hedging, and managing risk in fixed income markets. A diverse array of models has been developed to track the evolution of interest rates across maturities, striving to achieve a balance between theoretical rigor, empirical realism, and computational feasibility.

The Heath-Jarrow-Morton (HJM) framework provides a general arbitrage-free methodology for modeling the term structure of interest rates. By specifying the risk-neutral dynamics of forward rates, it accommodates a wide range of market behaviors. However, the HJM framework relies on infinitesimal forward rates, which are abstract quantities not directly observable or traded. To enable practical implementation and calibration, tractable submodels are required, focusing on the dynamics of market-observable quantities such as forward swap rates or rates linked to overnight indexed swaps (OIS).

Historically, the LIBOR Market Model (LMM), also referred to as the Brace-Gatarek-Musiela (BGM) model, dominated interest rate derivative modeling. The LMM's assumption of lognormal dynamics for forward LIBOR rates under their respective forward measures aligns with market practice, particularly in pricing caps, floors, and swaptions using Black-type formulas. Nonetheless, the model's reliance on deterministic volatility structures limits its ability to capture complex strike-dependent features of implied volatility surfaces, such as the observed volatility smile and skew in swaption markets. These limitations spurred the development of stochastic volatility extensions.

~\cite{Rebonato} enhanced the LMM by introducing stochastic volatility and providing connections to the SABR model for forward swap rates. The SABR model, characterized by its flexible parametric form, captures the intricacies of implied volatility skew and smile, making it a standard tool for swaption volatility modeling. Rebonato's work offers an approximate mapping between SABR dynamics and modified LMM volatility structures, bridging the gap between practical volatility surface modeling and arbitrage-free interest rate models.

In this work, we build upon this trajectory by integrating rough stochastic volatility into the Forward Market Model (FMM), an arbitrage-free framework tailored to the post-LIBOR era and developed by \cite{LM}. We rigorously derive an asymptotic expansion for swaption implied volatility, linking the model to a rough Bergomi-type framework for forward swap rate dynamics. This formulation establishes a direct connection between an HJM-consistent interest rate model and a rough volatility framework for swaption pricing, offering a parsimonious yet precise representation of swaption volatility surfaces, including their persistent skew and term structure.

Rough volatility models, in which volatility has lower H\"older regularity than Brownian motions, are known to explain a power-law type term structure of the volatility skew \citep[see, e.g.,][]{F}.
The roughness has a market microstructural foundation~\cite{EFR}, which potentially applies to bond markets and hence, forward term rates. Indeed, by a calibration experiment, we demonstrate that a rough volatility model exhibits a better performance than the lognormal SABR model, its non-rough counterpart.

\citet{HullWhite} use the sensitivities of the forward swap rate to its constituent forward rates to approximate swaption volatilities in the LIBOR market model. \citet[Proposition~3.2]{WuZhang} give an explicit expression for these sensitivities in a stochastic volatility extension of that model. We use the same sensitivity representation for the forward term rates of the FMM. Our contribution is to show how these sensitivities enter the short-time implied volatility expansion under rough volatility and to construct an approximate forward swap rate model that reproduces this expansion near the money.

In Section 2, we introduce the basic notation and assumptions, recall the exact swap-rate sensitivities, and give bounds used in the short-time analysis. Section 3 reviews the HJM framework and presents a rough SABR forward market model. In Section 4, we derive a short-term asymptotic expansion of the swaption implied volatility under this model and construct a forward swap rate model that produces the same expansion. All proofs are provided in Section 5, while Section 6 reports a calibration experiment assessing the effectiveness of the proposed models.

\section{Forward term rates and a forward swap rate}
\subsection{Notation}
Here we introduce notation and basic assumptions.
Let $M = \{M_t\}$ denote a money market account process
\begin{equation}\label{eq:mma}
	M_t = \exp\left\{ \int_0^t r_s \mathrm{d}s \right\},
\end{equation}
where $r = \{r_t\}$ is a piecewise continuous adapted process,
called a spot interest rate,
defined on a filtered probability space
$(\Omega,\mathscr{F},\mathsf{P},\{\mathscr{F}_t\}_{t\in [0,\infty)})$.
We assume $\mathscr{F}_0$ to consist of the events with probability $0$ or $1$.
Let $P_t(T)$ denote the price of the zero coupon bond with face value $1$ and maturity $T$ at time $t \leq T$. We extend $P_t(T)$ for $t \geq T$ by
\begin{equation}\label{ext}
	P_t(T) = \exp\left\{\int_T^t r_s \, \mathrm{d}s\right\} = \frac{M_t}{M_T}.
\end{equation}

In light of the fundamental theorem of asset pricing, we assume that
there exists an equivalent probability measure $\mathsf{Q}$,
called a risk neutral measure,
such that 
\begin{equation}\label{H8}
	\mathsf{E}_\mathsf{Q}\left[ \frac{P_t(u)}{M_t} \bigg| \mathscr{F}_s\right]
	= \frac{P_{s\wedge t}(u)}{M_{s\wedge t}}
\end{equation}
for all   $s,t,u \in [0,\infty)$,
or equivalently, for all   $s,t \in [0,\infty)$,
\begin{equation}\label{H8p}
	P_s(t) =  \mathsf{E}_\mathsf{Q}\left[\frac{M_s}{M_t}\bigg|\mathscr{F}_s\right]
	=  \mathsf{E}_\mathsf{Q}\left[\exp\left(-\int_s^t r_u \,\mathrm{d}u  \right)\bigg|\mathscr{F}_s\right].
\end{equation}
We also assume that 
$P(T) = \{P_t(T)\}$ is a continuous process for each $T\geq 0$.

For a tenor structure $0 = T_0 < T_1 < T_2 < \dots < T_N$, 
we define the forward term rate $R^j$ for the term $(T_{j-1},T_j]$ by
\begin{equation*}
	R^j_t = \frac{1}{\theta_j} \left( \frac{P_t(T_{j-1})}{P_t(T_j)} -1 \right), \ \ j=1,2,\dots, N,
\end{equation*}
where $\theta_j = T_j - T_{j-1}$.
Remark that $R^j_t = R^j_{T_j}$ for $t \geq T_j$ for each $j$.
For a given pair $(I,J)$ with $1 \leq I < J \leq N$,
the forward swap rate $S^{IJ} = \{S^{IJ}_t\}$
for the period $(T_I,T_J]$
is defined by
\begin{equation}\label{swaprate}
	S^{IJ}_t =  \frac{P_t(T_I)-P_t(T_J)}{A^{IJ}_t}, \ \ 
	A^{IJ}_t = \sum_{j=I + 1}^J \theta_j P_t(T_j).
\end{equation}
Since $P(T)$ is continuous and adapted,
so are $R^j$, $A^{IJ}$ and $S^{IJ}$.
Except in Section~6, the pair $(I,J)$ is fixed.
Therefore, we ease notation by setting $S^{} = S^{IJ}$ and $A = A^{IJ}$ in the sequel.

For
$T\geq 0$, the $T$-forward measure $\mathsf{Q}^T$ is defined by
\begin{equation*}
	\mathsf{Q}^T(F) = \frac{1}{P_0(T)}\mathsf{E}_\mathsf{Q}\left[ \frac{1}{M_T} ; F\right], \ \ F \in \mathscr{F}.
\end{equation*}
For a process $X$, $X/M$ is martingale under $\mathsf{Q}$ if and only if
$X/P(T)$ is a martingale under $\mathsf{Q}^T$.
In particular, $R^j$ is a martingale under $\mathsf{Q}^{T_j}$ for each $j$.
By the fundamental theorem of asset pricing, a no arbitrage price of an European payoff $f(R^j_T)$  with maturity $ T$ at time $t$ is given by 
\begin{equation*}
	M_t \mathsf{E}_\mathsf{Q}\left[\frac{f(R^j_T)}{M_T}\Bigg|\mathscr{F}_t\right] = P_t(T)\mathsf{E}_{\mathsf{Q}^T}[f(R^j_T)|\mathscr{F}_t].
\end{equation*}
In particular, the unique no arbitrage price at time $t$ for
the payoff $R^j_{T_j} - K$ at $T_j$ is
\begin{equation*}
	M_t\mathsf{E}_\mathsf{Q}\left[\frac{R^j_{T_j}-K}{M_T}\Bigg|\mathscr{F}_t\right]
	= P_t(T_j)(R^j_t-K)
\end{equation*}
for any constant $K$ and $t\geq 0$. The unique no arbitrage price at time $t$ of a backward-looking fixed-floating payer's swap with a fixed rate $K$, where the holder receives $\theta_j R^j_{T_j}$ and pay $\theta_j K$ at $T_j$, $j=I + 1,\dots,J$, is therefore
\begin{equation*}
	\sum_{j=I + 1}^J \theta_j P_t(T_j)(R^j_t-K) =
	P_t(T_I)-P_t(T_J) -  K \sum_{j=I + 1}^J \theta_j P_t(T_j)
	= (S^{}_t-K)A_t
\end{equation*}
which is zero if and only if $K = S^{}_t$ defined by \eqref{swaprate}.
This means that the forward swap rate $S^{}$ is the unique fair rate for this swap contract.  

The forward swap measure $\mathsf{Q}^\ast$ is defined by
\begin{equation*}
	\mathsf{Q}^{\ast}(F) = \frac{1}{A_0}\mathsf{E}_\mathsf{Q}\left[\frac{A_{T_N}}{M_{T_N}} ; F\right], \ \ F \in \mathscr{F},
\end{equation*}
where $A$ is defined by \eqref{swaprate}.
Note that $A/M$ is a martingale under $\mathsf{Q}$ and that
$A_t/M_t = A_{T_N}/M_{T_N}$ for $t \geq T_N$.
For a process $X$, $X/M$ is martingale under $\mathsf{Q}$ if and only if
$X/A$ is a martingale under $\mathsf{Q}^\ast$.
In particular, $S^{}$ is a martingale under $\mathsf{Q}^\ast$.
A no arbitrage price at time $t$ of the swaption with strike $K$ and maturity $T$
is given by
\begin{equation*}
	M_t \mathsf{E}_\mathsf{Q}\left[\frac{(S^{}_T-K)_+A_T}{M_T} \bigg| \mathscr{F}_t\right]
	=  A_t \mathsf{E}^\ast[(S^{}_T-K)_+ | \mathscr{F}_t],
\end{equation*}
where $\mathsf{E}^\ast$ denotes the expectation under $\mathsf{Q}^\ast$.

\subsection{Swap-rate sensitivities}
We first recall the exact sensitivities of the forward swap rate with respect to the forward term rates. The coefficients $\Pi^j$ below are the partial derivatives of $S$ with respect to $R^j$. An equivalent expression was given by \citet[Proposition~3.2]{WuZhang} in the LIBOR market model. We state the formula in our notation and collect the bounds used in the proof of Theorem~\ref{thm1}. The proof of the lemma is given in Section~\ref{sec:prooflem1}.
\begin{lemma}\label{lem1}
	Let $R^{j\ast}$ denote the local martingale part of the forward term rate $R^j$
	under $\mathsf{Q}^\ast$  for each $j=I + 1,\dots, J$.
	Then,
	\begin{equation}\label{sde}
		\mathrm{d}S^{} = \sum_{j=I + 1}^J 
		\Pi^j \mathrm{d}R^{j\ast},
	\end{equation}
	where
	\begin{equation*}
		\Pi^j = \frac{\theta_j P(T_j)}{AP(T_{j-1})}  \left(P(T_J)+ S^{}\sum_{k=j}^J \theta_k P(T_k) \right).
	\end{equation*}
	Moreover, if $R^j\geq 0$  for all $j=I + 1,\dots, J$, then
	$\Pi^j R^j \leq S^{}$   for all $j = I +1,\dots, J$. 
	Further, if $R^j>0$  for all $j=I + 1,\dots, J$, then
	\begin{equation}\label{bounded}
		\mathrm{d} \frac{\Pi^jR^j}{S^{}}
		= \sum_{i=I+1}^J X^{ji} \frac{\mathrm{d}R^i}{R^i} + 
		\sum_{i,k=I+1}^J Y^{jik}\mathrm{d} \langle \log R^i, \log R^k\rangle
	\end{equation}
	for some bounded processes $X^{ji}$ and $Y^{jik}$.
\end{lemma}

\begin{remark}\label{rem1}
	Writing
	\begin{equation*}
		\Pi^j =  \frac{\theta_j P(T_j)}{A} C^j,
	\end{equation*}
	we have a recursion formula for $C^j$:
	\begin{equation*}
		C^{I+1}  = 1, \ \ C^{j+1} = (1 + \theta_jR^j)C^j - \theta_jS^{}.
	\end{equation*}
	From this we observe that
	$C^j = 1$ for all $j$ if and only if $R^j = S^{}$ for all $j$.
\end{remark}

By Lemma~\ref{lem1}, the quadratic variation of $S$ admits the following representation, which motivates freezing the sensitivities at time zero:
\begin{equation*}
	\langle  S^{} \rangle_t = \sum_{i,j=I+1}^J \int_0^t \Pi^i_s \Pi^j_s \mathrm{d}
	\langle  R^i,  R^j \rangle_s \approx
	\sum_{i,j=I+1}^J  \Pi^i_0 \Pi^j_0 
	\langle  R^i,  R^j \rangle_t
\end{equation*}
Viewing the swap rate as a function of the forward term rates, its first-order linearization about the initial forward curve is given by
\begin{equation*}
	S^{}_t \approx S^{}_0 +  \sum_{j=I+1}^{J} \Pi^j_0 (R^j_t-R^j_0).
\end{equation*}
For comparison, one may instead freeze the annuity weights in the weighted-average representation:
\begin{equation*}
	S^{}_t = \sum_{j=I+1}^J \frac{\theta_jP_t(T_j)}{A_t} R^j_t \approx \sum_{j=I+1}^{J} \frac{\theta_jP_0(T_j)}{A_0} R^j_t
\end{equation*}
The two sets of weights coincide when the constituent forward rates are all equal at time zero, as follows from Remark~\ref{rem1}. For a non-flat initial forward curve, freezing the annuity weights need not give the same linearization. For a discussion of freezing approximations, see \citet{GP} and the references therein. In Section 4, we use the exact sensitivities evaluated at time zero to derive the short-time implied volatility expansion under rough volatility.

\section{A Forward Market Model with rough volatility}
\subsection{The HJM model}
We begin with recalling the Heath-Jarrow-Morton (HJM) framework.
The HJM model assumes 
\begin{equation}\label{def:forward}
	P_t(T) = \exp\left\{-\int_t^T f_t(s) \,\mathrm{d}s\right\}
\end{equation}
for all $t$ and $T$
with a family of continuous adapted processes $\{f_t(s)\}$ of the form
\begin{equation*}
	f_t(s) =f_0(s) + \int_0^t \mu_u(s) \, \mathrm{d}u + \int_0^t \sigma_u(s)^\top \mathrm{d}W_u
\end{equation*}
called the forward rate,
where $W$ is an $n$-dimensional Brownian motion under $\mathsf{Q}$ and $\{\mu_u(s)\}$ and $\{\sigma_u(s)\}$ 
are integrable processes in suitable senses.
By \eqref{ext}, we have $f_t(s) = r_s$ for $t \geq s$. 
This requires  $\mu_u(s) = 0$ and $\sigma_u(s)=0$ for $u >s$.

Further, to be consistent with the martingale property of $P(T)/M$ under $\mathsf{Q}$,
we have 
\begin{equation}\label{HJM2}
	\frac{P_t(T)}{M_t} = 
	P_0(T) \exp\left\{
	-\frac{1}{2}\int_0^t \left|\Sigma_u(T)\right|^2 \mathrm{d}u - \int_0^t \Sigma_u(T)^\top \mathrm{d}W_u
	\right\}
\end{equation}
for all $T \geq 0$, where
\begin{equation*}
	\Sigma_u(T) = \int_0^T \sigma_u(s)\,\mathrm{d}s = \int_u^T \sigma_u(s)\,\mathrm{d}s,
\end{equation*}
or equivalently,
\begin{equation}\label{HJM}
	\mu_u(s) = \sigma_u(s)^\top\int_0^s \sigma_u(v)\mathrm{d}v.
\end{equation}

From \eqref{HJM2} and \eqref{eq:drift1} in Appendix~A, we derive the dynamics of forward term rates 
\begin{equation}\label{ftr}
	\begin{split}
		\mathrm{d}R^j_t= & \
		\frac{1 + \theta_jR^j_t}{\theta_j} \left(\Sigma_t(T_j)- \Sigma_t(T_{j-1}) \right)^\top\mathrm{d}W_t  + \sum_{i=1}^j
		\frac{\theta_i}{1 + \theta_i R^i_t}
		\mathrm{d}\langle R^i,R^j \rangle_t
	\end{split}
\end{equation}
for $j=1,\dots, N$ under the HJM model.

Under the particular structure
\begin{equation}\label{cheyette}
	\sigma_u(s) = Z_u g(s) 1_{[0,s]}(u)
\end{equation}
with an $n\times m$ matrix-valued process $Z$ and an $\mathbb{R}^m$-valued deterministic continuous function $g$
for the HJM model~\eqref{HJM}, we have 
\begin{equation*}
	\Sigma_u(T) = Z_u (G(u\vee T) - G(u)), \ \ 
	G(u) = \int_0^u g(v)\, \mathrm{d}v
\end{equation*}
and
\begin{equation*}
	\begin{split}
		f_t(s) &= f_0(s) + \int_0^t \sigma_u(s)^\top \int_0^s \sigma_u(v)\, \mathrm{d}v \, \mathrm{d}u + \int_0^t \sigma_u(s)^\top \mathrm{d}W_u\\
		&= f_0(s) +  g(s)^\top \int_0^{t \wedge s}  Z_u^\top Z_u\int_u^s g(v)\, \mathrm{d}v \, \mathrm{d}u + g(s)^\top \int_0^{t\wedge s} Z_u^\top \mathrm{d}W_u\\
		&=  f_0(s) +  g(s)^\top \int_0^{t \wedge s}\mathrm{d}H_u (G(s)-G(u))
		+ g(s)^\top \int_0^{t\wedge s}  Z_u^\top\mathrm{d}W_u \\
		&=  f_0(s) +  g(s)^\top H_{t\wedge s}(G(s)-G(t\wedge s)) 
		+ g(s)^\top F_{t\wedge s},
	\end{split}
\end{equation*}
where
\begin{equation*}
	\ \ H_t = \int_0^t Z_u^\top Z_u\mathrm{d}u, \ \ 
	F_t =    \int_0^t
	H_u g(u) \mathrm{d}u+  \int_0^t
	Z_u^\top \mathrm{d}W_u.
\end{equation*}
In particular,
\begin{equation*}
	r_t = f_t(t) = f_0(t) + g(t)^\top F_t
\end{equation*}
and for $t\leq T$,
\begin{equation*}
	\begin{split}
		P_t(T) =  \frac{P_0(T)}{P_0(t)}\exp\left\{
		- \int_t^T
		g(s)^\top H_t (G(s)-G(t))\, \mathrm{d}s
		- 
		(G(T)-G(t))^\top F_t
		\right\}
	\end{split}
\end{equation*}
by \eqref{HJM2}.


	\subsection{The Forward Market Model}
	Here we describe a forward market model following \cite{LM}.
	A forward market model assumes \eqref{cheyette} with $m=n = N$ and
	\begin{equation*}
		g(s) = (g_1(s),\dots,g_N(s))^\top, \ \ g_j(s) = g_j(s)1_{(T_{j-1},T_j]}(s), \ \ j=1,\dots,N.
	\end{equation*}
	In this case,  only the $j$th element of $G(T_j \vee t)-G(T_{j-1}\vee t)$ is nonzero. Therefore,
	putting
	\begin{equation*}
		\gamma_j(t) = \int_{T_{j-1}\vee t}^{T_j \vee t} g_j(s)\, \mathrm{d}s,\ \ 
		\sigma^j_t =   \frac{1 + \theta_jR^j_t}{\theta_j} Z^j_t,
		\ \ j=1,\dots,N,
	\end{equation*}
	where $Z^j$ is the $j$th column vector of $Z$, by \eqref{ftr},
	we have
	\begin{equation*}
		\mathrm{d}R^j_t = \gamma_j(t)(\sigma^j_t)^\top
		\left[\mathrm{d}W_t + \sum_{i=1}^j  \frac{\theta_i}{1 + \theta_i R^i_t}
		\gamma_i(t)\sigma^i_t \mathrm{d}t
		\right], \ \ j=1,\dots, N.
	\end{equation*}
	Note that $\gamma_j(t)$ is constant until $t \leq T_{j-1}$ and $\gamma_j(t) = 0$ for $t \geq T_j$.
	In terms of
	\begin{equation*}
		\tilde{W}^j_t := \int_0^t \frac{1}{|\sigma^j_s|} (\sigma^j_s)^\top
		\mathrm{d}W_s
		=  \int_0^t \frac{1}{|Z^j_s|} (Z^j_s)^\top
		\mathrm{d}W_s
		, \ \ j=1,\dots, N,
	\end{equation*}
	we have
	\begin{equation*}
		\langle \tilde{W}^j \rangle_t = t, \ \ 
		\langle \tilde{W}^i, \tilde{W}^j \rangle_t = 
		\int_0^t \rho^{ij}_s \, \mathrm{d}s, \ \ \rho^{ij}_t := 
		\frac{1}{|\sigma^i_t||\sigma^j_t|} (\sigma^i_t)^\top \sigma^j_t
	\end{equation*}
	and
	\begin{equation}\label{fmm}
		\mathrm{d}R^j_t = \gamma_j(t)|\sigma^j_t|
		\left[\mathrm{d}\tilde{W}^j_t + \sum_{i=1}^j \frac{\theta_i}{1 + \theta_i R^i_t}
		\gamma_i(t)|\sigma^i_t|\rho^{ij}_t \mathrm{d}t
		\right], \ \ j=1,\dots, N.
	\end{equation}
	
	The simplest among the forward market models is achieved by
	\begin{equation*}
		g_j(s) = \frac{1}{\theta_j}1_{(T_{j-1},T_j]}(s),\ \
		Z^j_t = \frac{\theta_j R^j_t} {1 + \theta_j R^j_t}\varphi_j,\ \ j=1,\dots,N
	\end{equation*}
	with constant vectors $\varphi_j \in \mathbb{R}^N$.
	Indeed, $R^j$ follows a time-inhomogeneous Black-Scholes model under $\mathsf{Q}^{T_j}$
	with volatility $\gamma_j(t)|\varphi_j|$, where
	\begin{equation}\label{gammaj}
		\gamma_j(t) =
		\begin{cases}
			1 & t \leq T_{j-1} \\
			(T_j-t)/(T_j-T_{j-1}), & t \in (T_{j-1},T_j],\\
			0 & t > T_j. 
		\end{cases}
	\end{equation}

	\subsection{A rough SABR forward market model}
	To accommodate a volatility skew for each caplet market, 
	we propose a local-stochastic volatility model of the form
	\begin{equation}\label{rSABR}
		\begin{split}
			& \mathrm{d}R^j_t = \gamma_j(t)\eta_j(R^j_t)\sqrt{V^j_t}
			\mathrm{d}W^{j\dagger}_t, \\ 
			& \log V^j_t =  \log \xi_j(t)  - \frac{1}{2}\int_0^t \zeta_j(t-s)^2\mathrm{d}s  +
			\int_0^t \zeta_j(t-s)\mathrm{d}\bar{W}_s,
		\end{split}
	\end{equation}
	where 
	\begin{equation*}
		\begin{split}
			&W^{j\dagger} =  W^j + \sum_{i=1}^j \int_0^\cdot \frac{\theta_i}{1 + \theta_i R^i_t} 
			\gamma_i(t)\eta_i(R^i_t)\sqrt{V^i_t}\rho_{ij} \mathrm{d}t, \\
			& \bar{W} =  W^0 + \sum_{i=1}^N \int_0^\cdot \frac{\theta_i}{1 + \theta_i R^i_t} 
			\gamma_i(t)\eta_i(R^i_t)\sqrt{V^i_t}\rho_{i0} \mathrm{d}t,
		\end{split}
	\end{equation*}
	$(W^0,W^1,\dots,W^N)$ is a correlated Brownian motion under $\mathsf{Q}$
	with
	\begin{equation}\label{rho}
		\langle W^j \rangle_t = t, \ \ 
		\langle W^i, W^j \rangle_t = 
		\rho_{ij}t
	\end{equation}
	for a constant correlation matrix $[\rho_{ij}]$,
	$\gamma_j$ is defined by \eqref{gammaj}, and $\xi_j$,
	$\eta_j$ and $\zeta_j$ are deterministic positive continuous functions on their respective domains. For $\zeta_j$, the domain is $(0,\infty)$ and
	the integrability
	\begin{equation*}
		\int_0^t \zeta_j(s)^2 \,\mathrm{d}s < \infty
	\end{equation*}
	is assumed for all $t>0$.
	The well-posedness of the stochastic differential equation \eqref{rSABR} under a suitable condition is proved in Appendix~C.
	
	This model is consistent to the HJM framework. Indeed, it is derived from \eqref{cheyette} by taking $m=n=N$ and
	\begin{equation*}
		g_j(s) = \frac{1}{\theta_j}1_{(T_{j-1},T_j]}(s),\ \
		Z^j_t = \frac{\theta_j \eta_j(R^j_t)}{1 + \theta_j R^j_t} \sqrt{V^j_t} \varphi_j,\ \ j=1,\dots,N,
	\end{equation*}
	where $\varphi_j \in \mathbb{R}^N$ are such constant vectors that
	\begin{equation*}
		(\varphi_i)^\top \varphi_j = \rho_{ij}, \ \ i,j = 1,\dots,N.
	\end{equation*}
	For any $[\rho_{ij}]$, such $[\varphi_1,\dots,\varphi_N]$ is obtained by the Cholesky decomposition. 
	Here, $W^j$ corresponds to $\tilde{W}^j $ in \eqref{fmm} for each $j$.
	The volatility processes $V^j$  are driven by a common Brownian motion $W^0$ which is correlated to $W^j$ according to \eqref{rho}.

	In order to clarify some advantages of the model, let us first consider a special case of model:
	\begin{equation*}
		\log \xi_j(t) = 2\log \alpha_j + \frac{1}{4}\kappa_j^2t,\ \     \eta_j(r) = |r + \delta_j|^{\beta_j},\ \ \zeta_j(t) = \kappa_j
	\end{equation*}
	for some positive constants $\alpha_j$, $\beta_j$, 
	$\delta_j$, and $\kappa_j$.
	Then, the model reduces to
	\begin{equation*}
		\begin{split}
			& \mathrm{d}R^j_t = \gamma_j(t)|R^j_t +\delta_j|^{\beta_j}\sqrt{V^j_t}
			\mathrm{d}W^{j\dagger}_t, \\ 
			& \mathrm{d} \sqrt{V^j_t} =  (\kappa_j/2) \sqrt{V^j_t} \mathrm{d}\bar{W}_t,\ \ \sqrt{V^j_0} = \alpha_j,
		\end{split}
	\end{equation*}
	which is a time-inhomogeneous shifted SABR model, widely used in practice.
	
	While these Markovian SABR models successfully recover the volatility smile of each caplet and its dynamics, 
	the common volatility factor $W^0$ for all $j$ is not considered to work sufficiently for the joint calibration for swaptions.
	The existing approach is to introduce respective Brownian motions as the driving factors of the volatility processes, which amounts to introducing a large number of parameters describing the correlation among those Brownian motions.
	Practical challenges arise from
	not only the large number of the parameters to be calibrated but also the constraint on the parameters for describing a correlation matrix. 
	
	Our approach is to add flexibility by going beyond the finite-dimensional Markovian framework with parsimonious parameters.
	The deterministic functions $\xi_j$ are introduced to endow the model with an infinite dimensional Markov property which ensures time-consistent nature of the model; see \cite{FHT} for details.
	
	We motivate the drift structure of $\bar{W}$.
	By \eqref{eq:drift1} in Appendix~A,
	\begin{equation*}
		\begin{split}
			&\sum_{i=1}^j \int_0^\cdot \frac{\theta_i}{1 + \theta_i R^i_t} 
			\gamma_i(t)\eta_i(R^i_t)\sqrt{V^i_t}\rho_{ij} \mathrm{d}t = - \mathrm{d}\left\langle W^j, \log \frac{P(T_j)}{M} \right\rangle,
			\\ & \sum_{i=1}^j \int_0^\cdot \frac{\theta_i}{1 + \theta_i R^i_t} 
			\gamma_i(t)\eta_i(R^i_t)\sqrt{V^i_t}\rho_{i0} \mathrm{d}t = - \mathrm{d}\left\langle W^0, \log \frac{P(T_j)}{M} \right\rangle
		\end{split}
	\end{equation*}
	for each $j = 1,\dots, N$ and so, $(W^{j\dagger},\bar{W}^{j\dagger})$ is a correlated two dimensional Brownian motion under $\mathsf{Q}^{T_j}$ by the Girsanov-Maruyama theorem, where
	\begin{equation*}
		\begin{split}
			\bar{W}^{j\dagger} &=
			W^0 + \sum_{i=1}^j \int_0^\cdot \frac{\theta_i}{1 + \theta_i R^i_t} 
			\gamma_i(t)\eta_i(R^i_t)\sqrt{V^i_t}\rho_{i0} \mathrm{d}t
			\\&= \bar{W} - \sum_{i=j+1}^N\int_0^\cdot \frac{\theta_i}{1 + \theta_i R^i_t} 
			\gamma_i(t)\eta_i(R^i_t)\sqrt{V^i_t}\rho_{i0} \mathrm{d}t.
		\end{split}
	\end{equation*}
	This implies that the drift of $\bar{W}$ under $\mathsf{Q}^{T_j}$ is nonpositive if all $\rho_{i0}$ are nonpositive, which enable us to show the following result, 
	meaning that for each $j$, under $\mathsf{Q}^{T_j}$, $R^j$ follows an extension of a rough SABR model~\citep{FHT,FG,AAA}.
	\begin{proposition}\label{prop1}
		If $(R^1,\dots, R^N)$ solves \eqref{rSABR} with $R^i_0 > 0$,
		\begin{equation}\label{eta0}
			\sup_{r>0} \frac{\eta_i(r)}{r} 
			< \infty
		\end{equation}
		and
		\begin{equation}\label{rho0}
			\rho_{0i} \leq 0
		\end{equation}
		for all $i=1,\dots, N$, then
		\begin{equation}\label{Rj}
			\begin{split}
				R^j_t &= R^j_0 \exp\left(
				\int_0^t \frac{\eta(R^j_s)}{R^j_s} \gamma_j(s)\sqrt{V^j_s} \, \mathrm{d}W^{j\dagger}_s
				- \frac{1}{2} \int_0^t \left|\frac{\eta(R^j_s)}{R^j_s} \gamma_j(s)\right|^2V^j_s \, \mathrm{d}s
				\right),\\
				V^j_t &= \xi_j(t)\exp\left(
				\int_0^t \zeta_j(t-s)\,\mathrm{d}\bar{W}_s
				-\frac{1}{2}\int_0^t \zeta_j(t-s)^2 \,\mathrm{d}s
				\right)
			\end{split}
		\end{equation}
		and $R^j$ is a martingale under $\mathsf{Q}^{T_j}$ for each $j=1,\dots, N$.
	\end{proposition}
	
	The proof is given in Section~\ref{sec:proofprop1}.

	\section{A rough Bergomi forward swap model}
	\subsection{An asymptotic expansion of the implied volatility}
	The swaption implied volatility $\sigma(k,t)$ at time $0$ is defined as the unique volatility parameter that equates the Black-Scholes put price with the put option price written on the forward swap rate $S^{}$:
	\begin{equation*}
		p_\mathrm{BS}\left(k,\sigma(k,t)\sqrt{t}\right) = \mathsf{E}^\ast\left[\left(e^k - \frac{S^{}_t}{S^{}_0}\right)_+ \bigg| \mathscr{F}_0 \right],
	\end{equation*}
	where 
	\begin{equation*}
		p_\mathrm{BS}(k,v) = 
		e^k\Phi(-d_-(k,v)) - \Phi(-d_+(k,v)), \ \ d_\pm(k,v) = - \frac{k}{v} \pm \frac{v}
		2.
	\end{equation*}
	To derive an asymptotic formula for $\sigma(k,t)$, we focus on a simplified version of \eqref{rSABR}.
	Namely, we assume  for all $i$,
	\begin{equation}\label{zeta}
		\zeta_i(t) = \kappa t^{H-1/2}
	\end{equation}
	with $\kappa > 0$ and $H \in (0,1/2)$ and that $\eta_i$ are nonnegative $C^2$ functions with 
	\begin{equation}\label{eta}
		\sup_{r>0} \frac{\eta_i(r)}{r}  + 
		\sup_{r>0} |\eta_i^\prime(r)| +  \sup_{r>0} r |\eta_i^{\prime\prime}(r)|
		< \infty.
	\end{equation}
	
	Put
	\begin{equation}\label{vt}
		v(t) =  \sum_{i,j=I +1}^J \rho_{ij}  \pi_i\pi_j \sqrt{\xi_i(t)\xi_j(t)}, \ \ 
		\pi_j = \frac{\Pi^j_0 \eta_j(R^j_0)}{S^{}_0}.
	\end{equation}
	
	\begin{theorem}\label{thm1} 
		If $v(0)>0$ and $(R^1,\dots, R^N)$ solves \eqref{rSABR} with $R^i_0 > 0$, \eqref{rho0}, \eqref{zeta}, and \eqref{eta}, then
		\begin{equation*}
			\sigma(k,t)
			= \sqrt{\bar{v}(t)}
			\left( 1  
			+ \psi k t^{H-1/2} \right) + o(t^H)
		\end{equation*}
		uniformly in $k \in \{k\in \mathbb{R}; |k| \leq a\sqrt{t}\} $ as $t \to 0$ for any $a>0$, where
		\begin{equation*}
			\begin{split}
				&\bar{v}(t) =    \int_0^1 v(ts)\, \mathrm{d}s\,, \\
				&\psi = \frac{\kappa}{(2H+1)(H+3/2)v(0)} \sum_{j=I + 1}^J \rho_{0j}  \pi_j \sqrt{\xi_j(0)}.
			\end{split}
		\end{equation*}
	\end{theorem}
	The proof is given in Section~\ref{sec:proof1}

	\subsection{A forward swap model with consistent asymptotics}
	Here we propose an approximate model for the forward swap rate $S^{}$
	under \eqref{rSABR} with \eqref{zeta}, \eqref{eta} and \eqref{rho0}. For $v(0)>0$, define $S^{\ast}$ by
	\begin{equation}\label{rBergomi}
		\begin{split}
			&\frac{\mathrm{d}S^{\ast}_t}{S^{\ast}_t}= \sqrt{V_t} \,\mathrm{d}W^{\ast}_t, \\ 
			&    V_t = v(t)\exp\left(
			\int_0^t \zeta(t-s)\, \mathrm{d}W^{0\ast}_s - \frac{1}{2}\int_0^t \zeta(t-s)^2\, \mathrm{d}s
			\right)
		\end{split}
	\end{equation}
	with $S^{\ast}_0 = S^{}_0$,
	where $\zeta(t) = \kappa t^{H-1/2}$ and
	$(W^{0\ast},W^{\ast})$
	is a correlated Brownian motion under $\mathsf{Q}^\ast$ such that
	\begin{equation}\label{rhomap}
		\langle W^{0\ast}, W^\ast \rangle_t = \rho t, \ \ \rho = 
		\frac{1}{\sqrt{v(0)}} \sum_{j=I + 1}^J \rho_{0j}  \pi_j \sqrt{\xi_j(0)}.
	\end{equation}
	Such a Brownian motion exists; let $W^{0\ast}$ be the local martingale part of $W^0$ under $\mathsf{Q}^\ast$ and
	\begin{equation*}
		W^\ast = \frac{1}{\sqrt{v(0)}} \sum_{j=I + 1}^J \pi_j\sqrt{\xi_j(0)} W^{j\ast},
	\end{equation*}
	where 
	$W^{j\ast}$ is the local martingale part of $W^{j\dagger}$ under $\mathsf{Q}^\ast$.
	This means that $|\rho|\leq 1 $.
	
	This is an approximative model in the sense that the implied volatility under $S^{\ast}$ has the same short-time asymptotics as 
	$\sigma(k,t)$. To be precise, let $\sigma^\ast(k,t)$ denote the implied volatility, that is, the solution to
	\begin{equation*}
		p_\mathrm{BS}\left(k,\sigma^\ast(k,t)\sqrt{t}\right) = \mathsf{E}^\ast\left[\left(e^k - \frac{S^{\ast}_t}{S^{\ast}_0}\right)_+ \bigg| \mathscr{F}_0 \right]
	\end{equation*}
	for $k\in \mathbb{R}$ and $t \geq 0$.
	Noting that $S^{\ast}$ follows a rough Bergomi model under $\mathsf{Q}^\ast$,
	we deduce the next result by combining Theorem~\ref{thm1} above, 
	Theorem~2.1 and Remark~2.5 of \cite{F}.
	
	\begin{theorem}\label{thm2}
		If $v(0)>0$ and $(R^1,\dots, R^N)$ solves \eqref{rSABR} with $R^i_0 > 0$, \eqref{rho0}, \eqref{zeta}, and \eqref{eta}, then
		\begin{equation*}
			\sigma^\ast(k,t)
			=  \sigma(k,t) + o(t^H)
		\end{equation*}
		uniformly in $k \in \{k\in \mathbb{R}; |k| \leq a\sqrt{t}\} $ as $t \to 0$ for any $a>0$.
	\end{theorem}
	
	\begin{remark}
		The forward swap rate model \eqref{rBergomi} depends on $[\rho_{ij}]_{1 \leq i,j \leq N}$ 
		only via $v(t)$ defined by \eqref{vt}.
	\end{remark}
	
	\begin{remark}
		~\cite{Rebonato}
		proposed a similar but different mapping between the correlation parameters of their SABR-LMM model and an approximate forward swap rate model.
		While the derivation was heuristic in \cite{Rebonato}, our mapping \eqref{rhomap} is justified by Theorem~\ref{thm2}.
	\end{remark}
	
	\section{Proofs}
	
	\subsection[Proof of Lemma]{Proof of Lemma~\ref{lem1}}\label{sec:prooflem1}
	We write $F$ to represent a process of finite variation, which may change line by line.
	Since $S^{}$ is a martingale under $\mathsf{Q}^\ast$, it suffices to show
	\begin{equation}\label{toshow}
		\mathrm{d}S^{} = \sum_{j=I + 1}^J \Pi^j \, \mathrm{d}R^j + \mathrm{d}F.
	\end{equation}
	We put $P^j = P(T_j)$ for brevity. 
	Since
	\begin{equation*}
		\frac{P^I}{A} =  \frac{P^J}{A} + S^{},
	\end{equation*}
	using It\^o's formula, we obtain
	\begin{equation*}
		\begin{split}
			\mathrm{d}S^{} & = \frac{\mathrm{d}P^I}{A} - \frac{\mathrm{d}P^J}{A} - S^{} \frac{\mathrm{d}A}{A} + \mathrm{d}F \\
			&= \frac{P^J}{A}\left( \frac{\mathrm{d}P^I}{P^I} - \frac{\mathrm{d}P^J}{P^J} \right) +  S^{} \left( \frac{\mathrm{d}P^I}{P^I} - \frac{\mathrm{d}A}{A} \right)+ \mathrm{d}F \\
			&= \frac{P^J}{A}\left( \frac{\mathrm{d}P^I}{P^I} - \frac{\mathrm{d}P^J}{P^J} \right) +  \frac{S^{}}{A} \sum_{k=I+1}^J \theta_kP^k \left( \frac{\mathrm{d}P^I}{P^I} - \frac{\mathrm{d}P^k}{P^k} \right)+ \mathrm{d}F \\
			&= \frac{P^J}{A}\sum_{j=I+1}^J \left( \frac{\mathrm{d}P^{j-1}}{P^{j-1}} - \frac{\mathrm{d}P^j}{P^j} \right) +  \frac{S^{}}{A} \sum_{k=I+1}^J \theta_kP^k \sum_{j=I+1}^k \left( \frac{\mathrm{d}P^{j-1}}{P^{j-1}} - \frac{\mathrm{d}P^j}{P^j} \right)+ \mathrm{d}F \\
			&=
			\sum_{j=I+1}^J \left(\frac{P^J}{A} + \frac{S^{}}{A} \sum_{k=j}^J\theta_kP^k\right) \left(\frac{\mathrm{d}P^{j-1}}{P^{j-1}} - \frac{\mathrm{d}P^j}{P^j} \right) + \mathrm{d}F.
		\end{split}
	\end{equation*}
	Then, \eqref{toshow} follows from another application of It\^o's formula:
	\begin{equation*}
		\mathrm{d}R^j = \frac{P^{j-1}}{\theta_jP^j} \left(\frac{\mathrm{d}P^{j-1}}{P^{j-1}} - \frac{\mathrm{d}P^j}{P^j} \right)  + \mathrm{d}F.
	\end{equation*}
	
	Since
	\begin{equation}\label{decompose}
		\begin{split}
			\frac{R^j \Pi^j}{S^{}} 
			&= \frac{1}{S^{}A}\frac{\theta_jR^j}{1 + \theta_j R^j}
			\left(P^J +  
			S^{}\sum_{k=j}^J\theta_kP^k\right)
			\\ & = \frac{\theta_jR^j}{1 + \theta_j R^j}
			\left(\frac{P^J}{P^I-P^J} +  \frac{1}{A}\sum_{k=j}^J\theta_kP^k\right)
			\\ & = \frac{\theta_jR^j}{1 + \theta_j R^j}
			\left(\frac{1}{\prod_{i=I+1}^J(1 + \theta_i R^i) -1} +  \frac{1}{A}\sum_{k=j}^J\theta_kP^k\right),
		\end{split}
	\end{equation}
	if all $R^j$ are nonnegative,
	then we have
	\begin{equation*}
		\frac{R^j \Pi^j}{S^{}}  \leq 
		\frac{\theta_jR^j}{1 + \theta_j R^j}
		\left(\frac{1}{1 + \theta_j R^j-1} + 1\right) = 1.
	\end{equation*}
	Now, suppose $R^j > 0$ for all $j$. Note that
	\begin{equation*}
		\frac{A}{P^J} = \theta_J +  \sum_{k=I+1}^{J-1} \theta_k \prod_{i=k+1}^{J} \frac{P^{i-1}}{P^i}  
		=  \theta_J +  \sum_{j=I+1}^{J-1} \theta_k \prod_{i=k+1}^{J} (1 + \theta_iR^i)
	\end{equation*}
	so that
	\begin{equation*}
		\frac{1}{A}\sum_{k=j}^J\theta_kP^k
		=  \frac{P^J}{A} \sum_{k=j}^J\theta_k\frac{P^k}{P^J} =
		\frac{\theta_J +  \sum_{k=j}^{J-1} \theta_k \prod_{i=k+1}^{J} (1 + \theta_iR^i)}
		{\theta_J +  \sum_{k=I+1}^{J-1} \theta_k \prod_{i=k+1}^{J} (1 + \theta_iR^i)}.
	\end{equation*}
	From this, we observe that 
	\begin{equation}\label{a1}
		\mathrm{d} \log \left(
		\frac{1}{A}\sum_{k=j}^J\theta_kP^k \right)
		= \sum_{i=I+1}^J \hat{X}^{ji} \frac{\mathrm{d}R^i}{R^i} + 
		\sum_{i,k=I+1}^J \hat{Y}^{jik}\mathrm{d} \langle \log R^i, \log R^k\rangle
	\end{equation}
	for bounded processes $\hat{X}^{ji}$ and $\hat{Y}^{jik}$.
	Also,
	\begin{equation}\label{a2}
		\mathrm{d}\log \left(\prod_{i=I+1}^J(1 + \theta_i R^i) -1
		\right)
		=
		\sum_{i=I+1}^J \tilde{X}^{i} \frac{\mathrm{d}R^i}{R^i} + 
		\sum_{i,k=I+1}^J \tilde{Y}^{ik}\mathrm{d} \langle \log R^i, \log R^k\rangle
	\end{equation}
	for bounded processes $\tilde{X}^{i}$ and $\tilde{Y}^{ik}$ since
	\begin{equation}\label{b1}
		\frac{1}{\prod_{i=I+1}^J(1 + \theta_i R^i) -1} 
		\leq \frac{1}{\theta_j R^j\prod_{i=I+1, i \neq j}^J(1 + \theta_i R^i) }
	\end{equation}
	for any $j = I+1, \dots, J$.
	By \eqref{decompose},
	\begin{equation*}
		\mathrm{d} \frac{R^j \Pi^j}{S^{}} 
		=  \mathrm{d}G^j + \mathrm{d}H^j,
	\end{equation*}
	where 
	\begin{equation*}
		G^j = \frac{\theta_jR^j}{1 + \theta_j R^j}
		\frac{1}{\prod_{i=I+1}^J(1 + \theta_i R^i) -1}, \ \ 
		H^j = \frac{\theta_jR^j}{1 + \theta_j R^j}
		\frac{1}{A}\sum_{k=j}^J\theta_kP^k.
	\end{equation*}
	By \eqref{b1}, we have
	\begin{equation*}
		0 < G^j
		\leq \frac{1}{\prod_{i=I+1}^J(1 + \theta_i R^i)} \leq 1,
	\end{equation*}
	and by \eqref{a2}, we observe that
	\begin{equation*}
		\begin{split}
			\mathrm{d} G^j &= G^j\,  \mathrm{d}\log G^j + \frac{1}{2} G^j\, \mathrm{d}  \langle \log G^j \rangle
			\\& =    \sum_{i=I+1}^J \check{X}^{ji} \frac{\mathrm{d}R^i}{R^i} + 
			\sum_{i,k=I+1}^J \check{Y}^{jik}\mathrm{d} \langle \log R^i, \log R^k\rangle
		\end{split}
	\end{equation*}
	for bounded processes $\check{X}^{ji}$ and $\check{Y}^{jik}$.
	Further by \eqref{a1},
	\begin{equation*}
		\begin{split}
			\mathrm{d} H^j &= H^j\,  \mathrm{d}\log H^j + \frac{1}{2} H^j\, \mathrm{d}  \langle \log H^j \rangle \\
			&  =
			\sum_{i=I+1}^J \bar{X}^{ji} \frac{\mathrm{d}R^i}{R^i} + 
			\sum_{i,k=I+1}^J \bar{Y}^{jik}\mathrm{d} \langle \log R^i, \log R^k\rangle
		\end{split}
	\end{equation*}
	for bounded processes $\bar{X}^{ji}$ and $\bar{Y}^{jik}$.
	
	\subsection[Proof of Proposition]{Proof of Proposition~\ref{prop1}}\label{sec:proofprop1}
	The representation \eqref{Rj} under \eqref{eta0} follows that the unique solution to the equation
	\begin{equation*}
		\mathrm{d}R = R \mathrm{d}X
	\end{equation*}
	is given by
	\begin{equation*}
		R = R_0 \exp\left(X - \frac{1}{2} \langle X \rangle \right)
	\end{equation*}
	for any continuous semimartingale $X$. Apply this to
	\begin{equation*}
		X_t = \int_0^{t \wedge \tau_n} \frac{\eta(R^j_s)}{R^j_s} \gamma_j(s)\sqrt{V^j_s} \, \mathrm{d}W^{j\dagger}_s,
	\end{equation*}
	where
	\begin{equation*}
		\tau_n  = \inf\left\{t \geq 0; R^j_t = \frac{1}{n}\right\}
	\end{equation*}
	for each $n=1,2\dots,$ and then argue that $\lim \tau_n = \infty$.
	To prove that $R^j$ is a martingale, we extend the argument of \cite{Gassiat}.
	We already know that $R^j$ is a nonnegative local martingale by \eqref{Rj},
	it suffices to show $\mathsf{E}_{\mathsf{Q}^{T_j}}[R^j_T] = R^j_0$ for any $T \geq 0$. Let
	\begin{equation*}
		\hat{\tau}_n = \inf\{t \geq 0; Y_t = n\}, \ \ Y_t = \int_0^t \zeta_j(t-s)\,\mathrm{d}\bar{W}_s
	\end{equation*}
	and notice that $\{R^j_{t \wedge \hat{\tau}_n}\}$ is a positive martingale under $\mathsf{Q}^{T_j}$ for each $n$ by \eqref{Rj}. Define an equivalent probability measure $\hat{\mathsf{Q}}_n$ by
	\begin{equation*}
		\frac{\mathrm{d}\hat{\mathsf{Q}}_n}{\mathrm{d}\mathsf{Q}^{T_j}} 
		= \frac{R^j_{\hat{\tau_n}}}{R^j_0}.
	\end{equation*}
	Then,
	\begin{equation*}
		\begin{split}
			R^j_0  =     \mathsf{E}_{\mathsf{Q}^{T_j}}[R^j_{T \wedge \hat{\tau}_n}] & = 
			\mathsf{E}_{\mathsf{Q}^{T_j}}[R^j_{\hat{\tau}_n} 1_{\{\hat{\tau}_n \leq T\}}]
			+  \mathsf{E}_{\mathsf{Q}^{T_j}}[R^j_T 1_{\{\hat{\tau}_n >  T\}}] \\
			&= R^j_0 \hat{\mathsf{Q}}_n(\hat{\tau}_n \leq T) + \mathsf{E}_{\mathsf{Q}^{T_j}}[R^j_T 1_{\{\hat{\tau}_n > T\}}].
		\end{split}
	\end{equation*}
	By the monotone convergence theorem, we have
	\begin{equation*}
		\mathsf{E}_{\mathsf{Q}^{T_j}}[R^j_T 1_{\{\hat{\tau}_n > T\}}] \to  \mathsf{E}_{\mathsf{Q}^{T_j}}[R^j_T]
	\end{equation*}
	as $n\to \infty$. Under $\hat{\mathsf{Q}}_n$, by the Girsanov-Maruyama theorem,
	\begin{equation*}
		\begin{split}
			\hat{W}^n:=\bar{W}^{j\dagger} - \langle \bar{W}^{j\dagger}, \log R^j \rangle_{\cdot \wedge \hat{\tau}_n} =
			\bar{W} & - \sum_{i=j+1}^N\int_0^\cdot \frac{\theta_i}{1 + \theta_i R^i_t} 
			\gamma_i(t)\eta_i(R^i_t)\sqrt{V^i_t}\rho_{i0} \mathrm{d}t \\ & - 
			\int_0^{\cdot \wedge \hat{\tau}_n}\frac{\eta(R^j_s)}{R^j_s} \gamma_j(s)\sqrt{V^j_s} \rho_{j0}\, \mathrm{d}s
		\end{split}
	\end{equation*}
	is a Brownian motion. Under \eqref{eta}, we have
	\begin{equation*}
		Y_t = \int_0^t \zeta_j(t-s)\,\mathrm{d}\bar{W}_s \leq 
		\int_0^t \zeta_j(t-s)\,\mathrm{d}\hat{W}^n_s =:\hat{Y}^n_t.
	\end{equation*}
	The law of $\hat{Y}^n$ under $\hat{\mathsf{Q}}_n$ does not depend on $n$ and
	\begin{equation*}
		\hat{\mathsf{Q}}_n(\hat{\tau}_n \leq T) 
		\leq   \hat{\mathsf{Q}}_n\left(\sup_{t \leq T}\hat{Y}^n_t \geq n\right) \to 0
	\end{equation*}
	as $n\to \infty$, which concludes the proof.
	
	\subsection[Proof of Theorem]{Proof of Theorem~\ref{thm1}}\label{sec:proof1}
	Put
	\begin{equation*}
		v^\ast(t) = \mathsf{E}^\ast[V^\ast_t | \mathscr{F}_0], \ \   V^\ast_t = \sum_{i,j=I+1}^J  \hat{\Pi}^i_t \hat{\Pi}^j_t \sqrt{V^i_t} \sqrt{V^j_t}\rho_{ij}, \ \
		\hat{\Pi}^j_t = \frac{\Pi^j_t\eta_j(R^j_t)}{S^{}_t}.
	\end{equation*}
	By Lemma~\ref{lem1}, we have
	\begin{equation*}
		\mathrm{d} \langle \log S^{} \rangle_t = V^\ast_t \,\mathrm{d}t.
	\end{equation*}
	Therefore by Theorem~2.1 of \cite{F}, it suffices to show  (i):
	\begin{equation*}
		v(t) = v^\ast(t) + o(t^H),
	\end{equation*}
	(ii):
	\begin{equation*}
		\frac{1}{t^H} \left(\frac{V^\ast_t}{v^\ast(t)} - 1 \right) 
	\end{equation*}
	is uniformly (in $t$) integrable,  and (iii):
	\begin{equation*}
		\left( \frac{1}{\sqrt{t}}\left(\frac{S^{}_t}{S^{}_0} -1 \right), \frac{1}{t^H }\left(\frac{V^\ast_t}{v^\ast(t)} - 1 \right) \right)
	\end{equation*}
	converges in law to the normal distribution with mean $0$ and covariance $[\Sigma_{ij}]$,
	\begin{equation*}
		\Sigma_{12} 
		= \frac{2\kappa}{2H+1} \sum_{j=I + 1}^J \rho_{0j}  \pi_j \sqrt{\xi_j(0)}
	\end{equation*}
	as $t\to 0$. 
	As a preliminary, first note that by \eqref{Rj}
	all $R^j$ are positive and so,
	by Lemma~\ref{lem1}, we have that $\Pi^j R^j \leq S^{}$ for all $j$ and \eqref{bounded}. Moreover,
	this also implies that for any $p \geq 1$ and $T>0$,
	there exists a constant $C_p>0$ such that
	\begin{equation*}
		\mathsf{E}^\ast\left[ \langle \log R^j \rangle_t^p\right] \leq C_p t^p.
	\end{equation*}
	for all $j$ and $t \in [0,T]$, which in turn implies 
	that  for any $p \geq 1$ and $T>0$, there exists a constant $C_p>0$ such that
	\begin{equation}\label{L2conv}
		\mathsf{E}^\ast\left[ \left|\hat{\Pi}^j_t - \pi_j\right|^{2p}\right] \leq C_p t^p
	\end{equation}
	for all $j$ and $t \in [0,T]$ by \eqref{bounded} and \eqref{eta}.
	Second,
	by \eqref{eq:drift2}, we have
	\begin{equation*}
		\begin{split}
			\bar{W}^\ast &:= W^0 +  \int_0^\cdot \sum_{i=1}^J \left(\sum_{k=(I+1)\vee i}^J \frac{\theta_k P(T_k)}{A}  \right) \frac{\theta_i}{1 + \theta_i R^i} \mathrm{d} 
			\langle W^0, R^i \rangle 
		\end{split}
	\end{equation*}
	is a Brownian motion under $\mathsf{Q}^\ast$.
	Note that $\bar{W} = \bar{W}^\ast + D$ and
	\begin{equation*}
		V^j_t = \xi_j(t)\exp\left(\kappa \int_0^t(t-s)^{H-1/2}\mathrm{d}D_s \right) V^0_t,
	\end{equation*}
	where
	\begin{equation*}
		D = \sum_{i=J+1}^N \int_0^\cdot \frac{\theta_i}{1 + \theta_i R^i} \mathrm{d}
		\langle W^0, R^i \rangle + \sum_{i=I+2}^J \int_0^\cdot
		\left(\sum_{k=I+1}^{i-1} \frac{\theta_k P(T_k)}{A}\right)
		\frac{\theta_i}{1 + \theta_i R^i} \mathrm{d}\langle W^0, R^i \rangle
	\end{equation*}
	and
	\begin{equation*}
		V^0_t = \exp\left(\kappa\int_0^t (t-s)^{H-1/2} \mathrm{d}\bar{W}^\ast_s - \frac{\kappa^2 t^{2H}}{4H}\right).
	\end{equation*}
	
	Let us show (i). Using $\mathsf{E}^\ast[V^0_t] = 1$, we have
	\begin{equation*}
		\begin{split}
			v^\ast(t)-v(t)  =& \sum_{i,j=I+1}^J\mathsf{E}^\ast\left[
			(\hat{\Pi}^i_t\hat{\Pi}^j_t - \pi_i \pi_j)
			\sqrt{V^i_t} \sqrt{V^j_t}
			\right] \rho_{ij} \\
			& +  v(t)
			\mathsf{E}^\ast\left[ \left(
			\exp\left(\kappa \int_0^t(t-s)^{H-1/2}\mathrm{d}D_s \right) - 1\right)V^0_t
			\right].
		\end{split}
	\end{equation*}
	Under \eqref{rho0},
	\begin{equation}\label{neg}
		\kappa \int_0^t(t-s)^{H-1/2}\mathrm{d}D_s  \leq 0
	\end{equation}
	and so
	\begin{equation*}
		\begin{split}
			0 &\leq  \mathsf{E}^\ast\left[V^0_t\left( 1- 
			\exp\left(\kappa \int_0^t(t-s)^{H-1/2}\mathrm{d}D_s \right) \right)
			\right] \\ & \leq -
			\mathsf{E}^\ast\left[ V^0_t
			\kappa \int_0^t(t-s)^{H-1/2}\mathrm{d}D_s 
			\right] = O(t^{H+1/2}).
		\end{split}
	\end{equation*}
	By \eqref{L2conv} and \eqref{neg}, we also have
	\begin{equation*}
		\mathsf{E}^\ast\left[
		(\hat{\Pi}^i_t\hat{\Pi}^j_t - \pi_i \pi_j)
		\sqrt{V^i_t} \sqrt{V^j_t}
		\right]  = O(t^{1/2}),
	\end{equation*}
	which concludes (i).
	
	Next we show (ii).
	By the above computation for (i), we observe that
	\begin{equation*}
		\mathsf{E}^\ast[|V^\ast_t - v(t) V^0_t|^2] = O(t).
	\end{equation*}
	Therefore, it suffices to show
	that $ t^{-H} (V^0_t -1)$ is uniformly integrable.
	By a straightforward computation,
	\begin{equation*}
		\mathsf{E}^\ast[|t^{-H} (V^0_t -1)|^2] =
		t^{-2H}(\mathsf{E}^\ast[|V^0_t|^2] -1) = 
		\frac{1}{t^{2H}}\left(\exp\left( \frac{\kappa^2 t^{2H}}{2H}\right) -1 \right) = O(1),
	\end{equation*}
	which implies the uniform integrability.
	
	Finally, we show (iii). By Lemma~\ref{lem1} and \eqref{L2conv}, we have
	\begin{equation*}
		\frac{1}{\sqrt{t}}\left(\frac{S^{}_t}{S^{}_0} -1 \right)
		= \sum_{j=I+1}^J \pi_j \sqrt{\xi_j(t)} \frac{W^{j*}_t}{\sqrt{t}}+ o_p(1),
	\end{equation*}
	where $W^{j\ast}$ is the local martingale part of $W^{j\dagger}$ under $\mathsf{Q}^\ast$. 
	Also,
	\begin{equation*}
		\frac{1}{t^H }\left(\frac{V^\ast_t}{v^\ast(t)} - 1 \right) = \frac{1}{t^H }\left(V^0_t- 1 \right)
		+ o_p(1) = \frac{\kappa}{t^H }\int_0^t (t-s)^{H-1/2}\mathrm{d}\bar{W}^\ast_s + o_p(1)
	\end{equation*}
	as seen above. These leading terms are Gaussian with covariance
	\begin{equation*}
		\mathsf{E}^\ast\left[
		\left(\sum_{j=I+1}^J \pi_j \sqrt{\xi_j(t)} \frac{W^{j*}_t}{\sqrt{t}}\right)
		\frac{\kappa}{t^H }\int_0^t (t-s)^{H-1/2}\mathrm{d}\bar{W}^\ast_s
		\right]= \frac{2\kappa}{2H+1}\sum_{j=I+1}^J \pi_j \sqrt{\xi_j(t)}\rho_{j0},
	\end{equation*}
	which concludes (iii).
	
	\section{Numerical Analysis}
	\subsection{Market Data and Separate Tenor Calibration}\label{sec:separateCalibration}
	Here we examine the effectiveness of the proposed model by a 
	calibration experiment to market data.
	We use USD secured overnight financing rate (SOFR) swap rate and SOFR swaption smile data 
	on 9 December 2024 extracted from Bloomberg\footnote{Supported by the Center for Advanced Research in Finance (CARF)}.
	
	The smile data consist of swaption implied volatilities (IVs) for strikes in the set
	$$
	\mathcal{K} = K_{\mathrm{ATM}} + \{-2\%, -1\%, -0.5\%,- 0.25\%, 0\%, 0.25\%, 0.5\%, 1\%, 2\%\}, $$
	where $K_{\mathrm{ATM}}$ is the ATM strike, and for
	some maturities and tenors  designated by S in Table~\ref{table:swaption_dataset}.
	For example, we have a  3Y $\times$ 5Y smile, that is
	for a swaption with maturity $t = T^I = 3$ years and
	$T^J = T^I + 5$ years.
	\begin{table}[]
		\tablefont
		\centering
		\caption{Swaption Dataset (row:maturity, column:underlying tenor)}
		\label{table:swaption_dataset}
		\scalebox{1.0}{
			\begin{tabular}{|c|c|c|c|c|c|c|c|c|c|c|}
				\hline
				& 1Y & 2Y & 3Y & 4Y & 5Y & 6Y & 7Y & 8Y & 9Y & 10Y \\
				\hline
				1Y & S & S & & & S & & & & & S \\
				\hline
				2Y &   &   & & &   & & & & & \multicolumn{1}{c}{} \\
				\cline{1-10}
				3Y & S & S & & & S & & & &   \multicolumn{2}{c}{} \\
				\cline{1-9}
				4Y &   &   & & &   & & &     \multicolumn{3}{c}{} \\
				\cline{1-8}
				5Y & S & S & & & S & &       \multicolumn{4}{c}{} \\
				\cline{1-7}
				6Y &   &   & & &   &         \multicolumn{5}{c}{} \\
				\cline{1-6}
				7Y &   &   & & &             \multicolumn{6}{c}{} \\
				\cline{1-5}
				8Y &   &   & &               \multicolumn{7}{c}{} \\
				\cline{1-4}
				9Y &   &   &                 \multicolumn{8}{c}{} \\
				\cline{1-3}
				10Y& S &                     \multicolumn{9}{c}{} \\
				\cline{1-2}
			\end{tabular}
		}
	\end{table}
	For other maturities and tenors, say, 1Y $\times$ 3Y, we have ATM volatility only.
	The convention of the obtained swaption volatility is that the underlying swap has annual payment intervals in both legs.
	So, the year-fraction $\theta_j = T_j - T_{j-1}$ is 1 year for all $j$.
	
	First, we focus on smiles with 1 year tenor, that is
	1Y $\times$ 1Y, 
	3Y $\times$ 1Y, 
	5Y $\times$ 1Y,  and
	10Y $\times$ 1Y.
	The underlying forward swap rates are related each other, but in order to examine whether our rough volatility model provides a reasonable term structure,
	we pretend as if they are independent and calibrate the rough Bergomi forward swap model \eqref{rBergomi} to obtain volatility of volatility parameter $\kappa$, while $H$ fixed, separately for each smile.
	Assuming $\mathsf{E}^\ast[\sqrt{V_{t}^{\ast}}]=\sqrt{V_{0}^{\ast}}$, we set initial variance curve as $v(t)=v(0)\exp(\frac{\kappa^2t^{2H}}{8H})$.
	Under this condition, when $H=0.5$, the model coincides with the log normal SABR model.
	
	The result is given in Table~\ref{tb:separateCalibration}.
	The model IVs under different values of $H$ are calculated by the Monte-Carlo method with 1,000,000 paths, using the hybrid scheme \citep{BLP} with the Black-Scholes model paths as control variates.
	Since the model with $H=0.5$ is a SABR model, we also present results calibrated using the Hagan formula at the first row.
	
	Discrepancies in the parameter $\kappa$ are observed when comparing the results between the Hagan formula and Monte Carlo simulation ($H=0.5$).
	These discrepancies become more pronounced as the maturity increases.
	The Hagan formula performs  well when $\kappa^{2}T$ is small enough.
	In this swaption data, the estimated values of $\kappa$ are mostly greater than 1 and deviate significantly from the Monte Carlo results, suggesting that the assumptions underlying the formula may not be satisfied.
	
	We observe from Table~\ref{tb:separateCalibration} that the calibrated values of $\kappa$ fall into smaller ranges when $H < 0.5$ than when $H=0.5$.
	This result suggests the possibility of a parsimonious modeling of the forward swap rate processes with a single value of $\kappa$ by choosing $H<0.5$.
	
	\begin{table}[]
		\tablefont
		\centering
		\caption{Model Calibration Results for Different Hurst Exponents}
		\label{tb:separateCalibration}
		\begin{tabular}{l|cccc}
			\hline
			& \textbf{1Y} & \textbf{3Y} & \textbf{5Y} & \textbf{10Y} \\
			\hline
			Hagan & 1.743 & 1.095 & 1.001 & 0.842 \\
			\hline
			H=0.5  & 1.952 & 1.328 & 1.415 & 3.525 \\
			H=0.45  & 1.821 & 1.299 & 1.394 & 1.938 \\
			H=0.4  & 1.693 & 1.266 & 1.372 & 1.739 \\
			H=0.35 & 1.566 & 1.230 & 1.347 & 1.635 \\
			H=0.3 & 1.442 & 1.188 & 1.318 & 1.568 \\
			H=0.25 & 1.320 & 1.142 & 1.282 & 1.516 \\
			H=0.2 & 1.201 & 1.091 & 1.241 & 1.473 \\
			H=0.15 & 1.083 & 1.033 & 1.190 & 1.417 \\
			H=0.1 & 0.960 & 0.961 & 1.116 & 1.048 \\
			H=0.05 & 0.782 & 0.835 & 0.944 & 0.762 \\
			\hline
		\end{tabular}
	\end{table}
	
	\subsection{Calibration of the Proposed Model}\label{sec:calibration}
	We calibrate the proposed model \eqref{rSABR} with two steps.
	As the first-step calibration, we estimate the parameters $\kappa, \alpha_{i}, \rho_{0i}, i=1,\cdots,N$ using swaption IVs with underlying of 1Y tenor.
	Since $\theta_j = 1$, we have $J = I+1$ for each of such smiles.
	Therefore, $S^{IJ} = R^J$ and the swaption has the same value as the caplet (1Y tenor swaption) with maturity $T_I$,
	Based on this observation, we have calibrated the proposed model \eqref{rSABR} with \eqref{zeta} and $\eta_j(r) = r$.
	We have parameterized the function $\xi_j(t)$ 
	by $\alpha_j > 0$ as
	\begin{equation*}
		\xi_j(t)
		= \alpha_j^2 \exp\left\{
		\frac{\kappa^2 t^{2H}}{8H} - \sum_{i=j+1}^N \frac{\theta_i\eta_i(R^i_0)}{1 + \theta_i R^i_0}
		\alpha_i \rho_{0i} \kappa \int_0^t (t-s)^{H-1/2} \gamma_i(s)\mathrm{d}s
		\right\},
	\end{equation*}
	where $T_N =$ 11 years, so that $\mathsf{E}^{T_j}[\sqrt{V^j_t}]\approx \alpha_j$.
	Note that
	\begin{equation*}
		\int_0^t (t-s)^{H-1/2} \gamma_i(s)\mathrm{d}s = 
		\frac{t^{H+1/2}}{H+1/2} - \frac{(t-T_{i-1})_+^{H+3/2} - (t-T_i)_+^{H+3/2}}{(H+1/2)(H+3/2)(T_i-T_{i-1})}.
	\end{equation*}
	
	Furthermore, we define the model smile discrepancy in the first-step calibration as the sum of squared differences between the market IVs in smile tenors and those generated by the mapped SMM model \eqref{rBergomi}.
	The parameters $\kappa, \rho_{0i}$ (for $i=2,4,6,11$) are estimated by minimizing the model smile discrepancy.
	Given the parameters $\kappa, \rho_{0i}$ for $i=2,4,6,11$, the remaining $\rho_{0i}$ are determined by linear interpolation and flat extrapolation.
	We determine $\alpha_{i}$ such that the ATM swaption price computed under the mapped rough Bergomi forward swap model \eqref{rBergomi} matches the market IV.
	Swaption prices under the mapped model are computed using the Monte Carlo simulation.
	Instead of the Monte Carlo, we have also examined the use of the rough SABR formula proposed in \cite{FG}; the results are reported in Appendix~B.

	The first-step calibration is performed with variations in the Hurst exponent $H$.
	Table \ref{tb:firstStepCalibration} reports the estimated parameters, under each condition, along with the corresponding model smile discrepancy expressed as the root mean squared error (RMSE).
	The RMSE-minimizing Hurst exponent is $H=0.2$. These values deviate substantially from $H=0.5$.
	This suggests that the assumption of a low Hurst exponent effectively captures the term structure of $\kappa$.
	The same value, $H=0.2$, minimizes the first-step calibration RMSE when we increase the number of time steps from 200 to 1000 (Appendix~D).
	On two additional market dates, the RMSE is again minimized at a Hurst exponent below $0.5$, although the minimizing value differs across dates (Appendix~E).
	
	\begin{table}[]
		\tablefont
		\centering
		\caption{First-Step Calibration Results}
		\label{tb:firstStepCalibration}
		\begin{tabular}{lc|cccc|cccc|c}
			\hline
			Hurst Exponent & $\kappa$ & $\alpha_{2}$ & $\alpha_{4}$ & $\alpha_{6}$ & $\alpha_{11}$ & $\rho_{0,2}$ & $\rho_{0,4}$ & $\rho_{06}$ & $\rho_{0,11}$ & RMSE \\
			\hline
			H = 0.5  & 1.72065 & 0.315 & 0.315 & 0.343 & 0.401 & -63.9\% & -39.8\% & -46.7\% & -47.0\% & 0.728\% \\
			H = 0.45 & 1.62013 & 0.314 & 0.311 & 0.330 & 0.368 & -63.3\% & -40.1\% & -46.9\% & -47.0\% & 0.678\% \\
			H = 0.4  & 1.52652 & 0.312 & 0.305 & 0.318 & 0.339 & -62.6\% & -40.4\% & -47.3\% & -47.4\% & 0.625\% \\
			H = 0.35 & 1.43774 & 0.309 & 0.300 & 0.307 & 0.314 & -61.9\% & -40.8\% & -48.1\% & -48.4\% & 0.574\% \\
			H = 0.3  & 1.35083 & 0.304 & 0.293 & 0.296 & 0.292 & -61.4\% & -41.5\% & -49.2\% & -49.9\% & 0.532\% \\
			H = 0.25 & 1.26378 & 0.298 & 0.284 & 0.283 & 0.271 & -61.2\% & -42.4\% & -50.8\% & -52.2\% & 0.507\% \\
			H = 0.2  & 1.17658 & 0.288 & 0.271 & 0.268 & 0.251 & -61.5\% & -43.8\% & -53.2\% & -55.3\% & 0.505\% \\
			H = 0.15 & 1.08537 & 0.271 & 0.253 & 0.248 & 0.229 & -63.3\% & -46.5\% & -57.1\% & -60.6\% & 0.530\% \\
			H = 0.1  & 0.98173 & 0.241 & 0.223 & 0.218 & 0.199 & -69.7\% & -52.6\% & -65.6\% & -71.3\% & 0.585\% \\
			H = 0.05 & 0.81767 & 0.186 & 0.173 & 0.169 & 0.154 & -97.1\% & -74.2\% & -94.2\% & -99.8\% & 0.707\% \\
			\hline
		\end{tabular}
	\end{table}

	As the second-step of the calibration, we estimate the correlation matrix among underlying assets $R^{j}$.
	In this paper, we estimate the correlation matrix of$(W_{0}, W_{1}, \cdots, W_{N})$, namely $(N+1)\times(N+1)$-matrix with indices $i,j=0,\cdots,N$.
	We construct the correlation matrix by a parametrized lower triangular matrix $B$, that is,
	\begin{equation*}
		b_{i,j} =
		\begin{cases}
			\cos\omega_{i,j}\prod_{k=0}^{j-1}\sin\omega_{i,k}, & 0 < j < i\\
			\prod_{k=0}^{j-1}\sin\omega_{i,k}, & j=i.
		\end{cases},
	\end{equation*}
	with $\prod_{k=0}^{-1} \cdot = 1$ for consistency. Then, $\Sigma = BB^{\mathsf{T}}$ satisfies the properties of a correlation matrix \citep{DBM}.
	To embed the correlation estimated in the first-step calibration, we set $\rho_{i,0} = \cos\omega_{i,0}$ for $i = 1, \cdots, N$.
	Since the correlation related to $W_1$ cannot be estimated from the available swaptions, we impose the condition $\mathrm{d}\langle W_{1},W_{2}\rangle=\mathrm{d}t$
	This is achieved by setting $\cos\omega_{2,1} = 1$ and $\cos\omega_{i,2} = 0$ for $i \geq 3$.
	
	To calibrate the correlation matrix, we use ATM swaptions with an underlying tenor of 2 years or longer.
	Similar to the first-step calibration, the FMM \eqref{rSABR} is mapped to rough Bergomi forward swap model \eqref{rBergomi}, and swaption prices are then computed using either Monte Carlo simulation or the rough SABR formula (in Appendix~B).
	Since the mapping only involves correlations of interest rate related to $R^i$ included in the underlying swap,
	incremental estimation is possible by using co-terminal swaptions.
	Specifically, we treat the $i$-th row of $b_{i,j}$, as the search domain and calibrate them by minimizing the sum of squared differences between the model IV
	and the market IV for the $i$-Y co-terminal swaptions -- i.e., the set of swaptions for which the sum of the maturity and the underlying tenor equals $i$ years.
	Once the estimation is complete, we increment the row index and repeat the process.
	We estimate the correlation matrices with H = 0.5 and H = 0.2.
	
	The obtained $\rho_{ij} (i,j=1,\cdots,N)$ are shown in Table \ref{table:corrMCH5}, \ref{table:corrMCH2}.
	There is a tendency for the correlation between underlying assets with distant maturities to be lower.
	Additionally, when the Hurst exponent decreases, the correlation matrix tends to be estimated higher.

	\begin{table}[htbp]
		\tablefont
		\centering
		\caption{Correlation Matrix Among Interest Rate Factors. (H=0.5)}
		\label{table:corrMCH5}
		\begin{tabular}{cccccccccc}
			\hline
			1.000 & 0.8981 & 0.7390 & 0.7547 & 0.6308 & 0.4311 & 0.3862 & 0.3669 & 0.1519 & 0.4204 \\
			0.8981 & 1.000 & 0.9060 & 0.7303 & 0.6194 & 0.4596 & 0.3431 & 0.3240 & 0.0400 & 0.3230 \\
			0.7390 & 0.9060 & 1.000 & 0.8733 & 0.7541 & 0.6294 & 0.4746 & 0.4584 & 0.06683 & 0.3049 \\
			0.7547 & 0.7303 & 0.8733 & 1.000 & 0.8592 & 0.7046 & 0.6047 & 0.5897 & 0.1996 & 0.3979 \\
			0.6308 & 0.6194 & 0.7541 & 0.8592 & 1.000 & 0.8671 & 0.7051 & 0.7034 & 0.3427 & 0.3830 \\
			0.4311 & 0.4596 & 0.6294 & 0.7046 & 0.8671 & 1.000 & 0.8729 & 0.8814 & 0.6186 & 0.5210 \\
			0.3862 & 0.3431 & 0.4746 & 0.6047 & 0.7051 & 0.8729 & 1.000 & 0.9993 & 0.8787 & 0.8546 \\
			0.3669 & 0.3240 & 0.4584 & 0.5897 & 0.7034 & 0.8814 & 0.9993 & 1.000 & 0.8833 & 0.8414 \\
			0.1519 & 0.0400 & 0.06683 & 0.1996 & 0.3427 & 0.6186 & 0.8787 & 0.8833 & 1.000 & 0.8904 \\
			0.4204 & 0.3230 & 0.3049 & 0.3979 & 0.3830 & 0.5210 & 0.8546 & 0.8414 & 0.8904 & 1.000 \\
			\hline
		\end{tabular}
	\end{table}
	
	\begin{table}[htbp]
		\tablefont
		\centering
		\caption{Correlation Matrix Among Interest Rate Factors. (H=0.2)}
		\label{table:corrMCH2}
		\begin{tabular}{cccccccccc}
			\hline
			1.000 & 0.9593 & 0.8681 & 0.8775 & 0.8604 & 0.7131 & 0.6896 & 0.7382 & 0.5948 & 0.6950 \\
			0.9593 & 1.000 & 0.9728 & 0.9453 & 0.9217 & 0.8318 & 0.7761 & 0.8298 & 0.6773 & 0.7761 \\
			0.8681 & 0.9728 & 1.000 & 0.9464 & 0.9196 & 0.8861 & 0.8084 & 0.8621 & 0.7144 & 0.8042 \\
			0.8775 & 0.9453 & 0.9464 & 1.000 & 0.9477 & 0.9024 & 0.8942 & 0.9261 & 0.8168 & 0.8925 \\
			0.8604 & 0.9217 & 0.9196 & 0.9477 & 1.000 & 0.9507 & 0.8967 & 0.9384 & 0.8390 & 0.8905 \\
			0.7131 & 0.8318 & 0.8861 & 0.9024 & 0.9507 & 1.000 & 0.9534 & 0.9779 & 0.9240 & 0.9464 \\
			0.6896 & 0.7761 & 0.8084 & 0.8942 & 0.8967 & 0.9534 & 1.000 & 0.9932 & 0.9879 & 0.9995 \\
			0.7382 & 0.8298 & 0.8621 & 0.9261 & 0.9384 & 0.9779 & 0.9932 & 1.000 & 0.9683 & 0.9909 \\
			0.5948 & 0.6773 & 0.7144 & 0.8168 & 0.8390 & 0.9240 & 0.9879 & 0.9683 & 1.000 & 0.9879 \\
			0.6950 & 0.7761 & 0.8042 & 0.8925 & 0.8905 & 0.9464 & 0.9995 & 0.9909 & 0.9879 & 1.000 \\
			\hline
		\end{tabular}
	\end{table}
	
	\subsection{Validation of Market Fitting for Rough SABR Forward Market Model}\label{section:ValidationOfMarketFitting}
	We have calibrated the parameters of the proposed model~\eqref{rSABR} in the preceding subsection.
	Now we examine whether our approximation to the forward swap rate by \eqref{rBergomi}
	works or not, and whether the calibrated model is useful to predict out-of-sample volatility smiles.
	
	For both parameter sets obtained in the previous section,
	we calculate the swaption IVs for the maturities and underlying where market smile data is available, and compare them with market prices.
	We plot the following IVs:
	\begin{itemize}
		\item Swaption IVs computed using Monte Carlo simulation under the FMM model ``FMM'',
		\item Swaption IVs under our approximative swap rate model \eqref{rBergomi} using the same calibrated parameters ``Mapped SMM'',
		\item Market IVs ``Market''.
	\end{itemize}
	Error bars corresponding to the $\pm2$ standard deviations are shown for both ``FMM'' and ``Mapped SMM''.
	
	\begin{figure}[!tbp]
		\centering
		\begin{tabular}{cc}
			\includegraphics[width=0.47\textwidth]{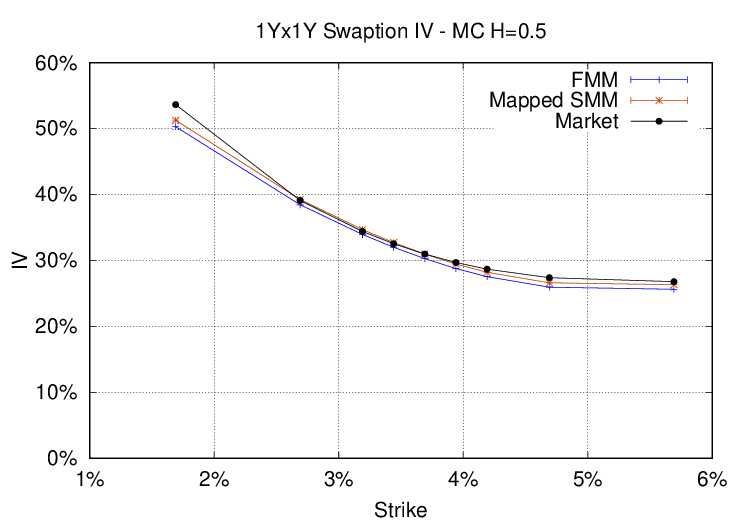} &
			\includegraphics[width=0.47\textwidth]{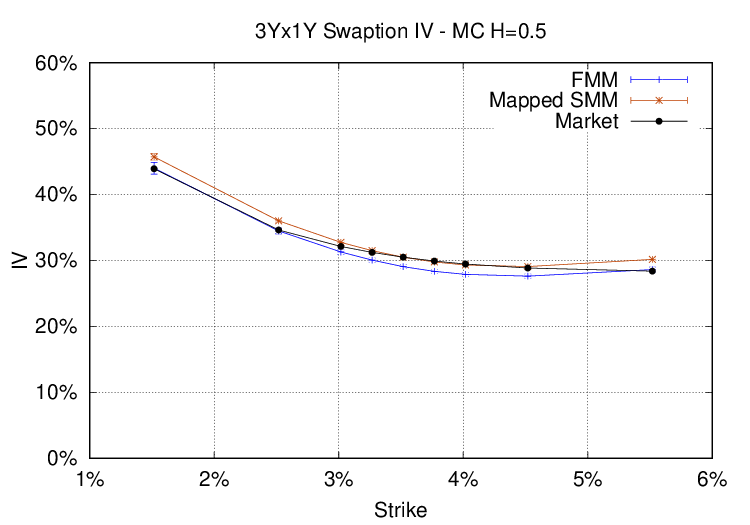} \\
			\includegraphics[width=0.47\textwidth]{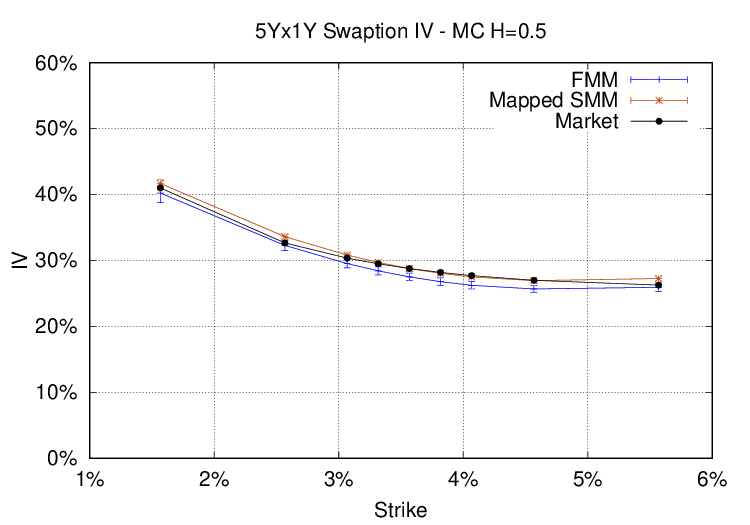} &
			\includegraphics[width=0.47\textwidth]{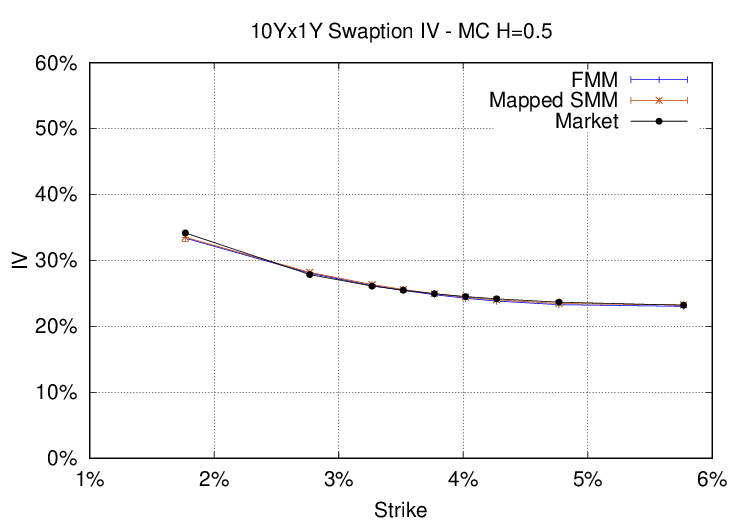}\\
		\end{tabular}
		\caption{Model and Market IVs of 1Y tenor at $H=0.5$}
		\label{figure:MCH50A}
	\end{figure}

	\begin{figure}[!tbp]
		\centering
		\begin{tabular}{cc}
			\includegraphics[width=0.47\textwidth]{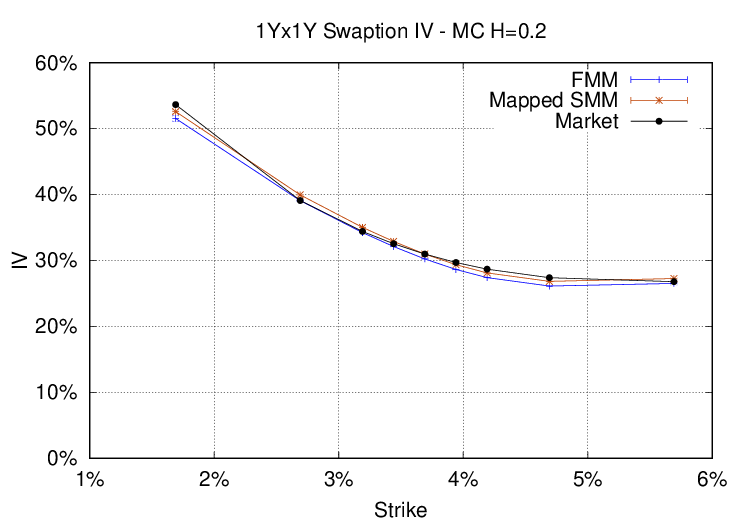} &
			\includegraphics[width=0.47\textwidth]{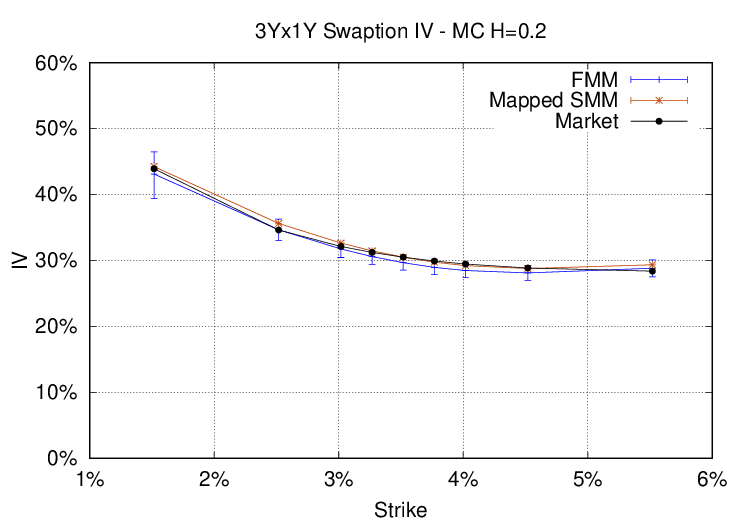} \\
			\includegraphics[width=0.47\textwidth]{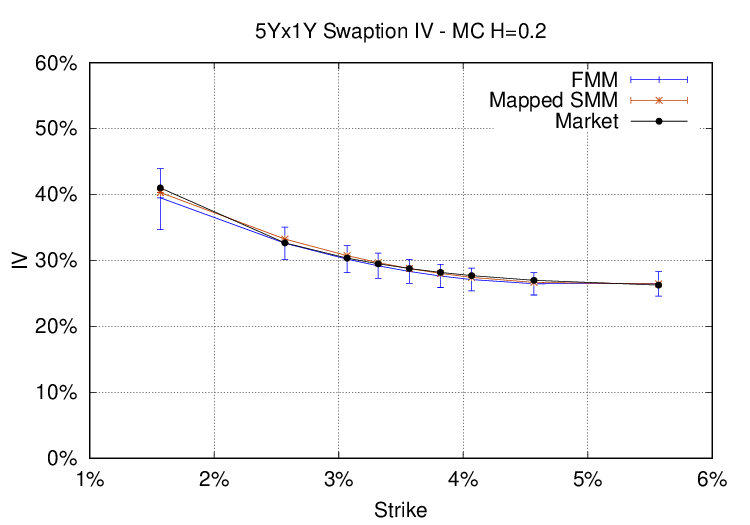} &
			\includegraphics[width=0.47\textwidth]{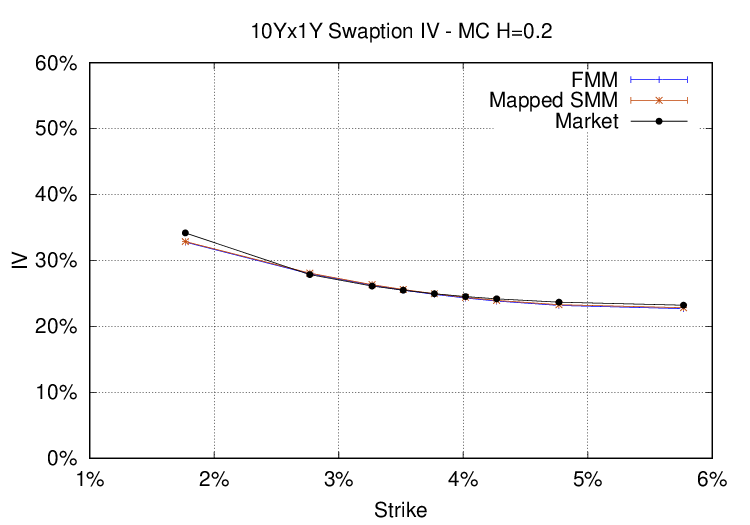}\\
		\end{tabular}
		\caption{Model and Market IVs of 1Y tenor at $H=0.2$}
		\label{figure:MCH20A}
	\end{figure}
	
	Figure \ref{figure:MCH50A}, \ref{figure:MCH20A}
	show the quality of in-sample fit, as the calibration was done using swaptions with 1Y tenor.
	The fit of FMM to Market is almost perfect for 10Y $\times$ 1Y,
	while their volatility levels slightly deviate for other maturities.
	This is explained by our model on the drift;
	we have introduced negative drift terms of volatility in \eqref{rSABR} for $j < N$ to ensure the martingale property of $R^j$ under $\mathsf{Q}^{T_j}$.
	For $T_N = 11$Y, there is no drift.
	The drift terms would induce some errors to our calibration of $\alpha_j$ for $j < N$.
	The approximative swap rate model (Mapped SMM), in spite of its much simpler structure,
	performs remarkably well to reproduce Market values.
	Not only the skew is accurate but also the level deviations in FMM are corrected.
	This would be explained by the fact that there is no drift of volatility in \eqref{rBergomi}.

\begin{figure}[!tbp]
		\centering
		\begin{tabular}{cc}
			\includegraphics[width=0.47\textwidth]{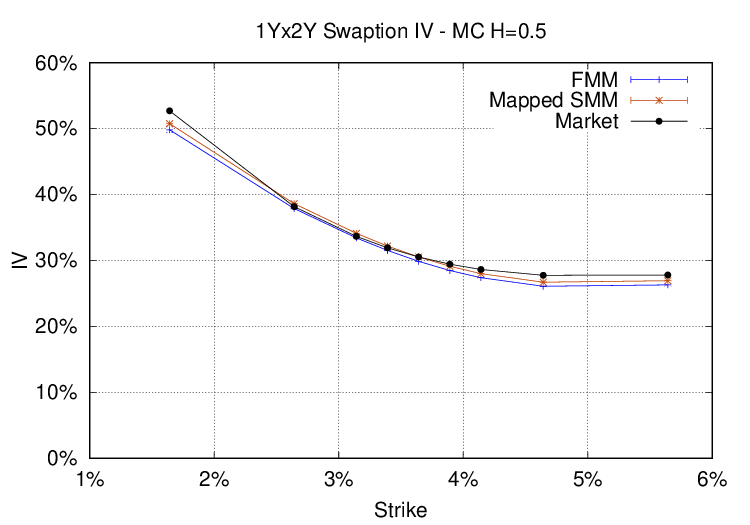} &
			\includegraphics[width=0.47\textwidth]{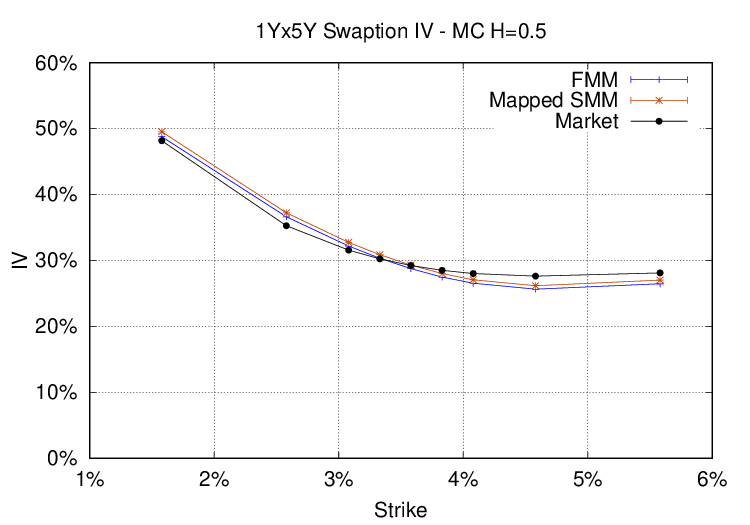} \\
			\includegraphics[width=0.47\textwidth]{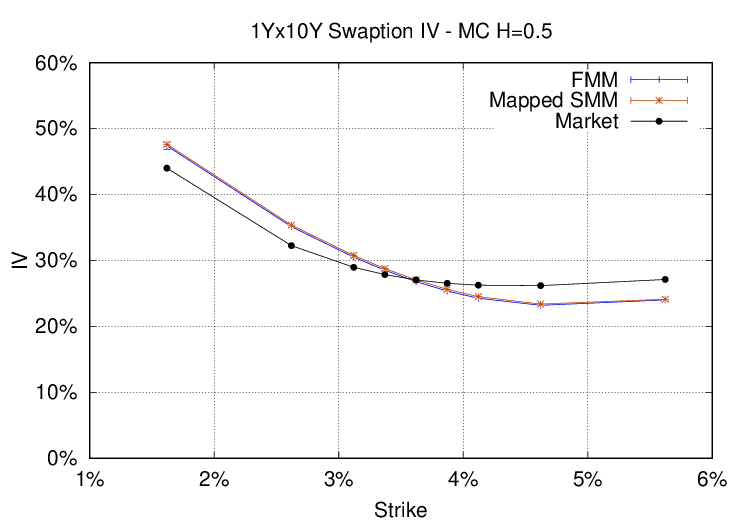} &
			\includegraphics[width=0.47\textwidth]{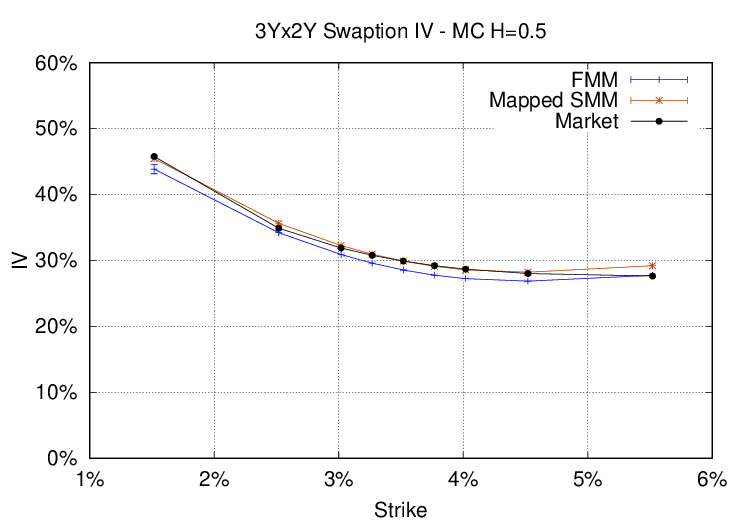} \\
			\includegraphics[width=0.47\textwidth]{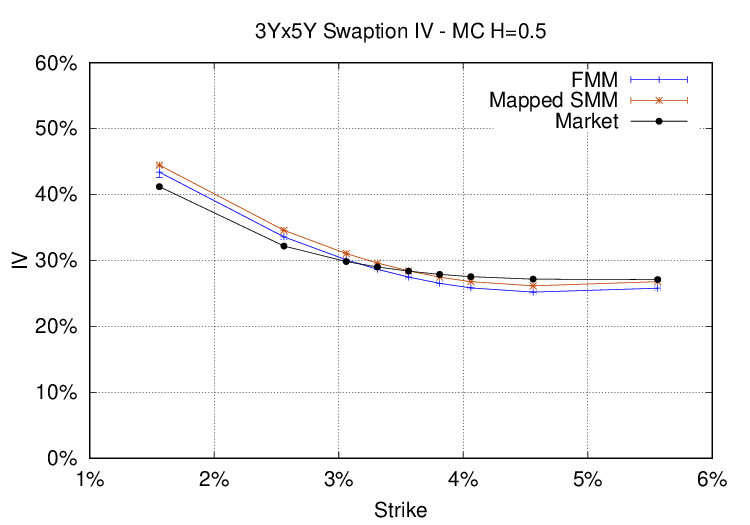} &
			\includegraphics[width=0.47\textwidth]{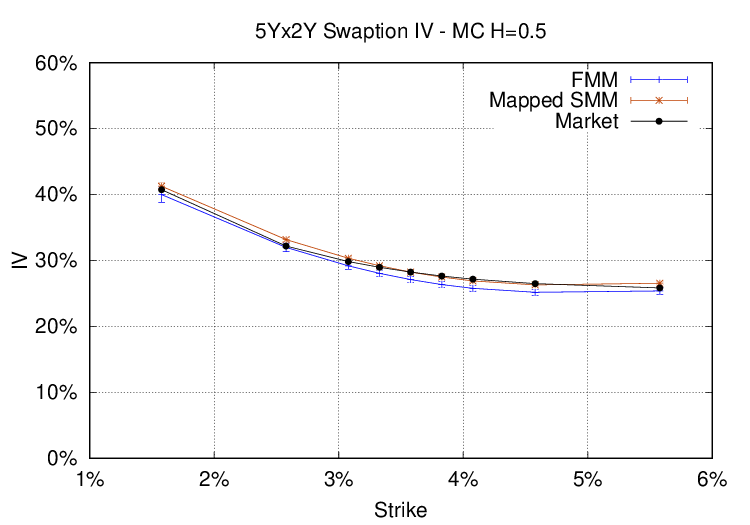} \\
			\includegraphics[width=0.47\textwidth]{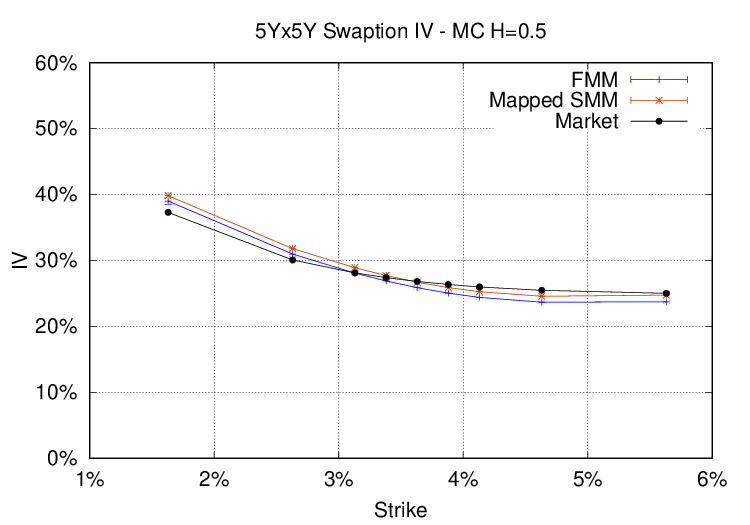} &
			\\
		\end{tabular}
		\caption{Model and Market IVs for tenors of 2Y or longer at $H=0.5$}
		\label{figure:MCH50B}
	\end{figure}

\begin{figure}[!tbp]
		\centering
		\begin{tabular}{cc}
			\includegraphics[width=0.47\textwidth]{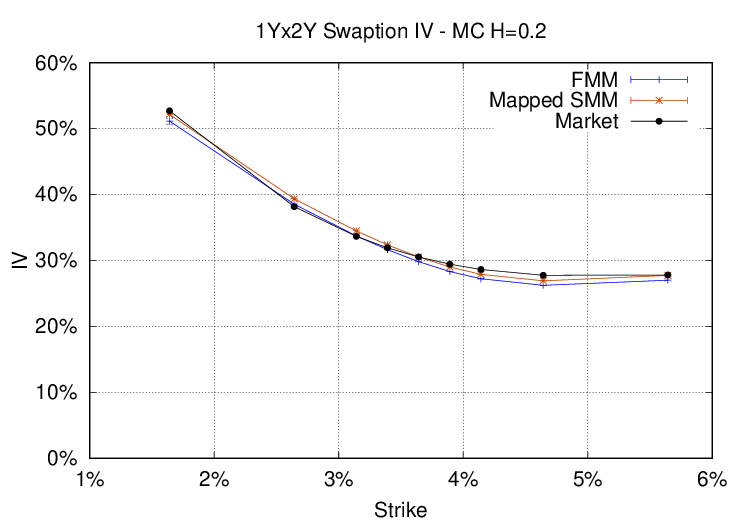} &
			\includegraphics[width=0.47\textwidth]{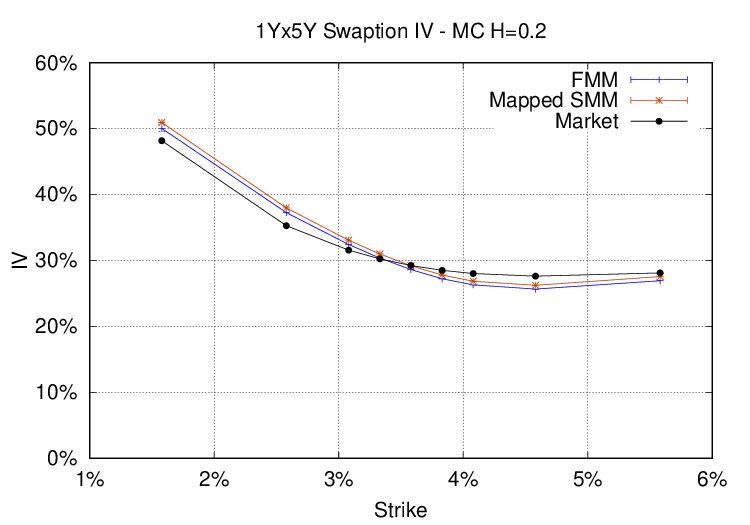} \\
			\includegraphics[width=0.47\textwidth]{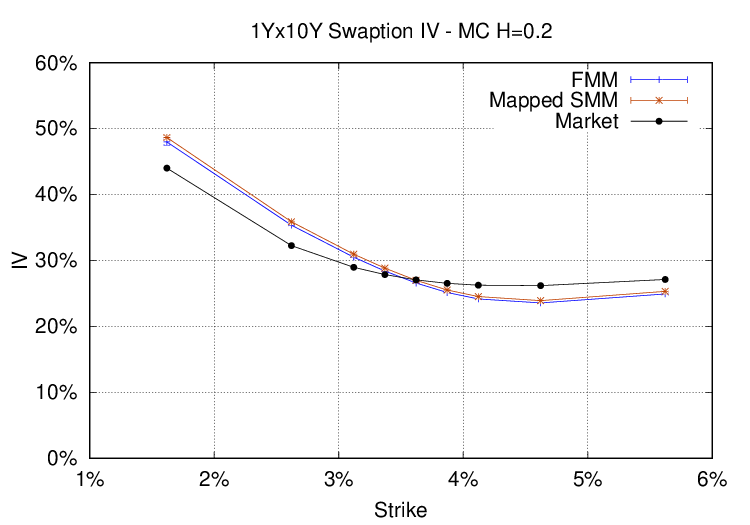} &
			\includegraphics[width=0.47\textwidth]{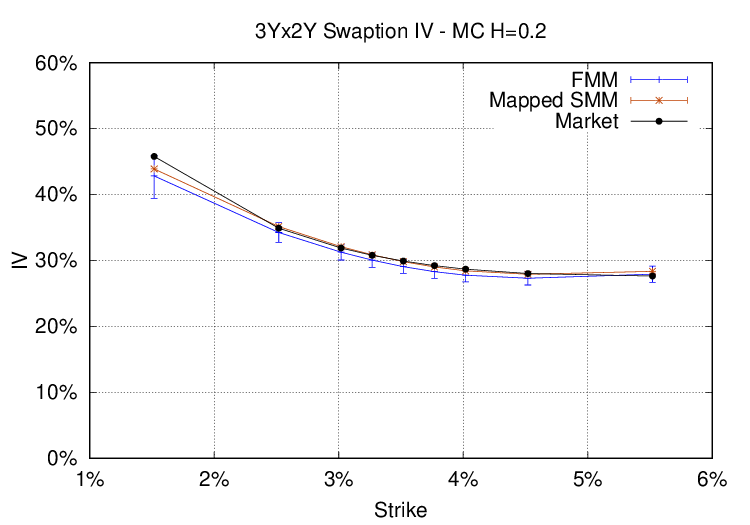} \\
			\includegraphics[width=0.47\textwidth]{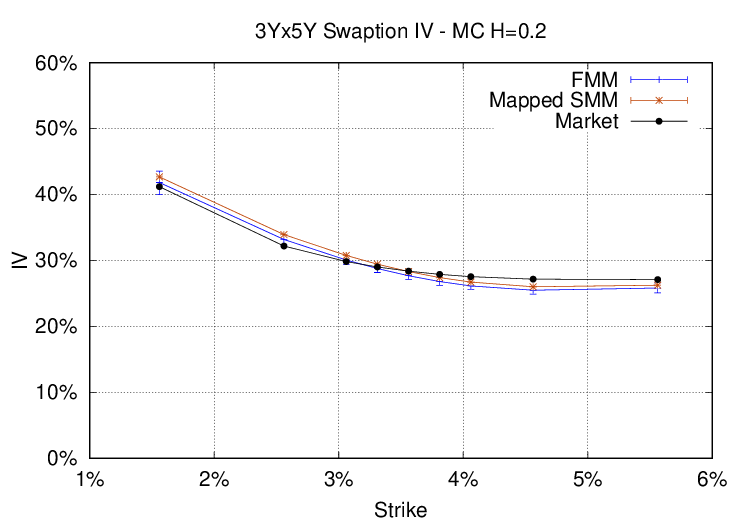} &
			\includegraphics[width=0.47\textwidth]{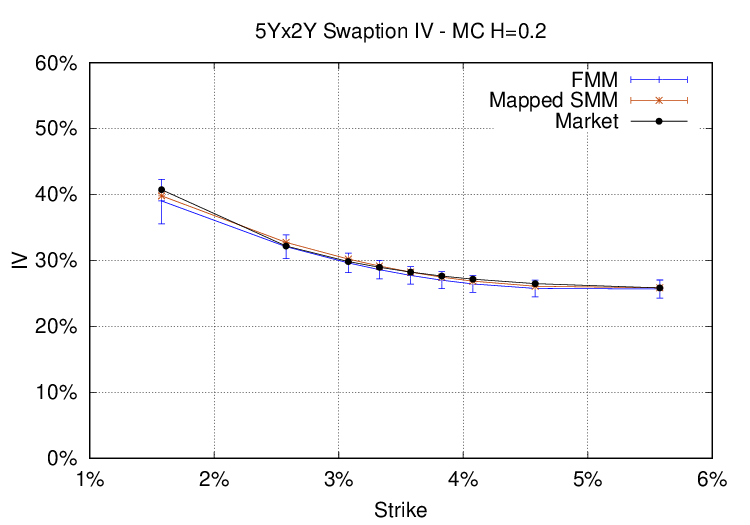} \\
			\includegraphics[width=0.47\textwidth]{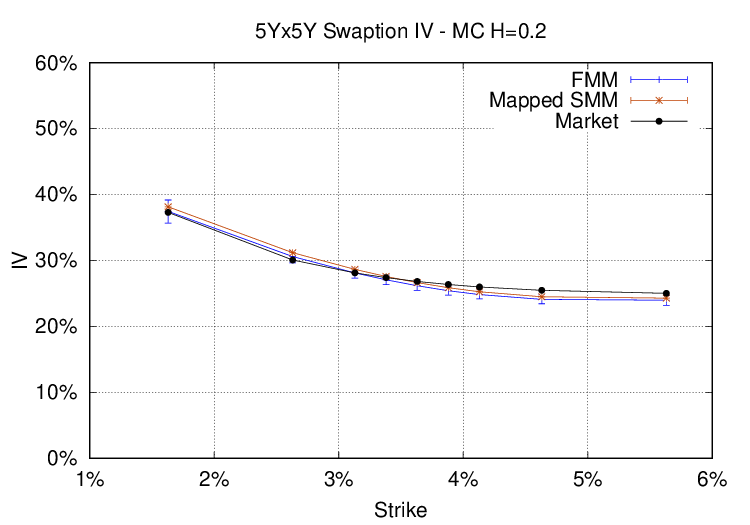} &
			\\
		\end{tabular}
		\caption{Model and Market IVs for tenors of 2Y or longer at $H=0.2$}
		\label{figure:MCH20B}
	\end{figure}

	In Figure \ref{figure:MCH50B}, \ref{figure:MCH20B}, we present volatility smiles for tenors of 2Y or longer.
	These show the quality of out-of-sample fit, as the calibration was done without using these smile data other than the ATM volatility values.
	
	First, we notice that FMM and Mapped SMM are close, meaning that our approximative model \eqref{rBergomi} works well, even though the case $H=0.5$ is not included in our theoretical justification by Theorem~\ref{thm2}.
	
	Second, we observe that these model IVs successfully predict Market values, suggesting that the proposed model \eqref{rSABR} that describes the joint distribution of the forward term rate processes is useful to relate the caplet skew (that is, the swaption skew with 1 year tenor) to the swaption skew with different tenor.

	However, for the 1Y $\times$ 5Y and 1Y $\times$ 10Y swaptions, the implied volatilities of both the FMM and the mapped SMM decrease more steeply with strike around the money than the market IVs.
	This discrepancy is particularly pronounced for the 1Y $\times$ 10Y swaption and is also visible on the two additional market dates (Figures~\ref{figure:20240607MCH50B}--\ref{figure:20240607MCH20B} and~\ref{figure:20250610MCH50B}--\ref{figure:20250610MCH10B} in Appendix~E).
	For these longer swap tenors, only ATM quotes enter the calibration.
	The skew is therefore determined indirectly by the common volatility factor and its correlations with the forward rates underlying each swap.
	One possible explanation is that the resulting dependence between the swap rate and its volatility is too strongly negative.
	
	Table~\ref{table:RMSEComparison} presents the RMSE values in
	the swaption volatility fit to compare the performances under $H=0.5$ and $H=0.2$.  We observe that the  values under $H=0.2$ are superior to those under $H=0.5$ most of the case, suggesting rough volatility.
	We observe a similar pattern on the two additional market dates: for both the FMM and the mapped SMM, the value of $H$ selected in the first-step calibration gives a lower smile RMSE than $H=0.5$ for most of the maturity--tenor pairs (Appendix~E).
	
	\begin{table}[]
		\tablefont
		\centering
		\caption{RMSE Comparison between FMM and Mapped SMM under Different Hurst Exponents}
		\label{table:RMSEComparison}
		\begin{tabular}{|c|cc|cc|}
			\hline
			& \multicolumn{2}{c|}{FMM} & \multicolumn{2}{c|}{Mapped SMM} \\
			& H = 0.5 & H = 0.2 & H = 0.5 & H = 0.2 \\
			\hline
			1Yx1Y & 1.42\% & 1.03\% & 0.87\% & 0.61\% \\
			3Yx1Y & 1.09\% & 0.71\% & 1.00\% & 0.52\% \\
			5Yx1Y & 1.06\% & 0.63\% & 0.55\% & 0.37\% \\
			10Yx1Y & 0.34\% & 0.54\% & 0.28\% & 0.49\% \\\hline
			1Yx2Y & 1.35\% & 1.02\% & 0.85\% & 0.67\% \\
			1Yx5Y & 1.18\% & 1.43\% & 1.15\% & 1.56\% \\
			1Yx10Y & 2.27\% & 2.26\% & 2.29\% & 2.40\% \\
			3Yx2Y & 1.24\% & 1.20\% & 0.61\% & 0.70\% \\
			3Yx5Y & 1.42\% & 1.03\% & 1.50\% & 1.01\% \\
			5Yx2Y & 0.98\% & 0.73\% & 0.49\% & 0.42\% \\
			5Yx5Y & 1.25\% & 0.82\% & 1.15\% & 0.71\% \\
			\hline
		\end{tabular}
	\end{table}
	
	In summary, this calibration experiment suggests that the proposed model \eqref{rSABR} for forward term rates and the approximation \eqref{rBergomi} to forward swap rates effectively reproduce the swaption market volatilities.

	One possible extension is to replace the power-law kernel in \eqref{zeta} with a gamma kernel, $\zeta(u)=\kappa u^{H-1/2}e^{-\lambda u}$, with $\lambda>0$, as considered by \citet{FukasawaHirano}.
	The exponential factor preserves the short-time roughness while damping the effect of past volatility shocks.
	If the persistence of these shocks contributes to the steep model skew, this modification may improve the fit.
	Whether it does so after recalibration remains to be tested.
	
	\FloatBarrier
	\section{Conclusion}
	In this study, we propose the rough SABR Forward Market Model, which incorporates the rough nature of interest rate volatility, with the aim of effectively capturing the term structure of swaption skew.
	We develop an approximate skew formula based on the proposed model and construct an approximate forward swap rate model that preserves the original model's asymptotic behavior. This leads to the derivation of parameter interrelationships.
	
	Our numerical analysis and calibration to USD SOFR swaption market data demonstrate the model's practical validity and robustness.
	Notably, even with limited data, the model successfully reproduces implied volatility skews across a wide range of maturities and tenors.
	The results indicate that lower Hurst exponent yield better alignment with market observations.
	Moreover, the introduced approximation techniques are shown to be accurate and simplify the parameter estimation process.
	
	Overall, this research bridges advanced probabilistic modeling and practical derivative pricing, offering valuable tools and insights for both academics and market practitioners.

	\section*{Declaration of generative AI use}
	During the revision of this manuscript, GitHub Copilot in Visual Studio Code was used to assist with mathematical discussion, proof review, literature searches, and manuscript and figure preparation. The authors take full responsibility for the content and references of the manuscript.

	\bibliographystyle{rQUF}
	\bibliography{refs}
	\markboth{}{Rough SABR Forward Market Model}

	\section*{Appendix A}
	%

	Here we describe the drift terms of the forward term rates.
	Note first that
	\begin{equation*}
		\begin{split}
			& \mathsf{E}_\mathsf{Q}\left[ \frac{\mathrm{d}\mathsf{Q}^{T}}{\mathrm{d}\mathsf{Q}} \bigg| \mathscr{F}_t \right] = \frac{1}{P_0(T)} \frac{P_t(T)}{M_t}, \\
			& \mathsf{E}_\mathsf{Q}\left[ \frac{\mathrm{d}\mathsf{Q}^\ast}{\mathrm{d}\mathsf{Q}} \bigg| \mathscr{F}_t \right] = \frac{1}{A_0} \frac{A_t}{M_t}
		\end{split}
	\end{equation*}
	for all $t \geq 0$.
	By the Girsanov-Maruyama theorem,
	a continuous adapted process $X$ is a local martingale under $\mathsf{Q}^T$ if and only if 
	\begin{equation*}
		X + \left\langle X, \log \frac{P(T)}{M} \right\rangle
	\end{equation*}
	is a local martingale under $\mathsf{Q}$.
	In particular,
	\begin{equation*}
		R^j + \left\langle R^j, \log \frac{P(T_j)}{M} \right\rangle
	\end{equation*}
	is a continuous local martingale under $\mathsf{Q}$ because
	$R^j$ is so under $\mathsf{Q}^{T_j}$.
	Since
	\begin{equation}\label{eq:decom}
		\frac{P(T_j)}{P(T_0)} = \prod_{i=1}^j  \frac{P(T_i)}{P(T_{i-1})} = \prod_{i=1}^j \frac{1}{1 + \theta_iR^i}
	\end{equation}
	and $P(T_0) = P(0) = M$,
	we have
	\begin{equation}\label{eq:drift1}
		- \mathrm{d} \left\langle X, \log \frac{P(T_j)}{M} \right\rangle 
		= \sum_{i=1}^j \frac{\theta_i}{1 + \theta_i R^i}\mathrm{d} \langle R^i,X\rangle
	\end{equation}
	for a continuous semimartingale $X$.
	
	Similarly, by the Girsanov-Maruyama theorem,
	a continuous adapted process $X$ is a local martingale under $\mathsf{Q}$ if and only if 
	\begin{equation*}
		X - \left\langle X, \log \frac{A}{M} \right\rangle
	\end{equation*}
	is a local martingale under $\mathsf{Q}^\ast$.
	Note that
	\begin{equation*}
		\begin{split}
			\mathrm{d}\langle X, \log A \rangle 
			&= \frac{1}{A} \mathrm{d} 
			\langle X, A \rangle  = \sum_{k=I+1}^J \frac{\theta_k P(T_k)}{A} \mathrm{d} 
			\langle X, \log P(T_k) \rangle
			\\ & =  -\sum_{k=I+1}^J \frac{\theta_k P(T_k)}{A} \sum_{i=1}^k \frac{\theta_i}{1 + \theta_i R^i}\mathrm{d} 
			\langle R^i, X\rangle \\
			&= -\sum_{i=1}^J \left(\sum_{k=(I+1)\vee i}^J \frac{\theta_k P(T_k)}{A} \right)  \frac{\theta_i}{1 + \theta_i R^i}\mathrm{d} 
			\langle R^i, X\rangle 
		\end{split}
	\end{equation*}
	by \eqref{eq:decom}. This implies that
	\begin{equation}\label{eq:drift2}
		- \left\langle X, \log \frac{A}{M} \right\rangle
		= \sum_{i=1}^{I+1}  \frac{\theta_i}{1 + \theta_i R^i}\mathrm{d} 
		\langle R^i, X\rangle
		+ \sum_{i=I+2}^J \frac{\sum_{k=i}^J\theta_k P(T_k)}{A}  \frac{\theta_i}{1 + \theta_i R^i}\mathrm{d} 
		\langle R^i, X\rangle.
	\end{equation}
	

	\section*{Appendix B}
	
	In the numerical experiments in Section~6, option prices under the rough Bergomi model are computed using the Monte Carlo method.
	As an alternative to Monte Carlo pricing, the rough SABR formula proposed by \cite{FG} provides a short-term approximation
	for plain vanilla option prices within the rough SABR framework.
	This section presents the results of using the rough SABR formula in place of the Monte Carlo method.
	We observe that the rough SABR formula does not provide sufficiently accurate approximations for our purpose.
	
	Table \ref{tb:separateCalibrationF} presents the results of separate tenor calibration using the rough SABR formula,
	corresponding to Table \ref{tb:separateCalibration} in Section \ref{sec:separateCalibration}.
	The results obtained with the rough SABR formula at $H=0.5$ are close to those from the Hagan formula.
	In contrast, the results deviate significantly from Monte Carlo results in Table \ref{tb:separateCalibration},
	especially for longer maturities.
	Nevertheless, both approaches exhibit a similar tendency in which the range of $\kappa$ narrows as the Hurst exponent decreases.
	
	\begin{table}[]
		\tablefont
		\centering
		\caption{Model Calibration Results for Different Hurst Exponents by Rough SABR Formula}
		\label{tb:separateCalibrationF}
		\begin{tabular}{l|cccc}
			\hline
			& \textbf{1Y} & \textbf{3Y} & \textbf{5Y} & \textbf{10Y} \\
			\hline
			Hagan & 1.743 & 1.095 & 1.001 & 0.842 \\
			\hline
			H=0.5  & 1.730 & 1.102 & 0.994 & 0.851 \\
			H=0.45  & 1.605 & 1.081 & 1.001 & 0.887 \\
			H=0.4  & 1.484 & 1.057 & 1.004 & 0.921 \\
			H=0.35 & 1.365 & 1.029 & 1.002 & 0.953 \\
			H=0.3 & 1.250 & 0.997 & 0.997 & 0.982 \\
			H=0.25 & 1.139 & 0.962 & 0.986 & 1.006 \\
			H=0.2 & 1.031 & 0.922 & 0.969 & 1.025 \\
			H=0.15 & 0.927 & 0.878 & 0.947 & 0.945 \\
			H=0.1 & 0.827 & 0.830 & 0.919 & 1.005 \\
			H=0.05 & 0.731 & 0.779 & 0.883 & 0.740 \\
			\hline
		\end{tabular}
	\end{table}

	Table \ref{tb:firstStepCalibrationF} shows the results of the first-step calibration in Section \ref{sec:calibration},
	where option prices are computed using the rough SABR formula instead of the Monte Carlo method.
	Compared to Table \ref{tb:firstStepCalibration}, the results exhibit a tendency for the level of $\alpha_{i}$ to decrease as the Hurst exponent becomes smaller.
	
	\begin{table}[]
		\tablefont
		\centering
		\caption{Comparison of Monte Carlo and Rough SABR Formula Calibration Results}
		\label{tb:firstStepCalibrationF}
		\begin{tabular}{lc|cccc|cccc|c}
			\hline
			Hurst Exponent & $\kappa$ & $\alpha_{2}$ & $\alpha_{4}$ & $\alpha_{6}$ & $\alpha_{11}$ & $\rho_{1Y}$ & $\rho_{3Y}$ & $\rho_{5Y}$ & $\rho_{10Y}$ & RMSE \\
			\hline
			H = 0.5  & 1.18389 & 0.282 & 0.227 & 0.173 & 0.082 & -72.9\% & -37.1\% & -36.4\% & -26.7\% & 1.339\% \\
			H = 0.45 & 1.24617 & 0.274 & 0.217 & 0.165 & 0.082 & -69.5\% & -36.1\% & -36.5\% & -28.1\% & 1.159\% \\
			H = 0.4  & 1.30590 & 0.264 & 0.205 & 0.156 & 0.081 & -66.4\% & -35.3\% & -36.9\% & -29.8\% & 0.977\% \\
			H = 0.35 & 1.36412 & 0.251 & 0.191 & 0.146 & 0.079 & -63.6\% & -34.9\% & -37.5\% & -31.8\% & 0.795\% \\
			H = 0.3  & 1.42316 & 0.234 & 0.174 & 0.133 & 0.075 & -61.1\% & -34.7\% & -38.5\% & -34.2\% & 0.620\% \\
			H = 0.25 & 1.48775 & 0.210 & 0.152 & 0.116 & 0.067 & -58.9\% & -34.8\% & -39.8\% & -36.9\% & 0.463\% \\
			H = 0.2  & 1.56721 & 0.174 & 0.122 & 0.093 & 0.055 & -56.8\% & -35.3\% & -41.5\% & -40.2\% & 0.344\% \\
			H = 0.15 & 1.68225 & 0.120 & 0.080 & 0.060 & 0.035 & -55.0\% & -36.0\% & -43.7\% & -44.2\% & 0.305\% \\
			H = 0.1  & 1.88812 & 0.045 & 0.027 & 0.019 & 0.011 & -53.3\% & -37.1\% & -46.3\% & -48.9\% & 0.359\% \\
			H = 0.05 & 2.41007  & 3.37e-4 & 1.48e-4 & 9.25e-5 & 4.45e-5 & -51.9\% & -38.7\% & -49.6\% & -54.5\%& 0.464\% \\
			\hline
		\end{tabular}
	\end{table}
	
	Tables \ref{table:corrRSH5} and \ref{table:corrRSH15} present the results of the second-step correlation calibration for $H=0.5$ and $H=0.15$, respectively.
	Compared to Tables \ref{table:corrMCH5} and \ref{table:corrMCH2}, the estimated correlations tend to be higher when using the rough SABR formula.
	
	\begin{table}[htbp]
		\tablefont
		\centering
		\caption{Correlation Matrix Among Interest Rate Factors. (rough SABR formula, H=0.5)}
		\label{table:corrRSH5}
		\begin{tabular}{cccccccccc}
			\hline
			1.000 & 0.9726 & 0.9060 & 0.9045 & 0.9030 & 0.8938 & 0.8844 & 0.8746 & 0.8645 & 0.8541 \\
			0.9726 & 1.000 & 0.9796 & 0.9789 & 0.9782 & 0.9736 & 0.9687 & 0.9633 & 0.9576 & 0.9516 \\
			0.9060 & 0.9796 & 1.000 & 0.9999 & 0.9999 & 0.9996 & 0.9988 & 0.9976 & 0.9960 & 0.9939 \\
			0.9045 & 0.9789 & 0.9999 & 1.000 & 0.9999 & 0.9997 & 0.9990 & 0.9978 & 0.9963 & 0.9943 \\
			0.9030 & 0.9782 & 0.9999 & 0.9999 & 1.000 & 0.9998 & 0.9991 & 0.9981 & 0.9966 & 0.9947 \\
			0.8938 & 0.9736 & 0.9996 & 0.9997 & 0.9998 & 1.000 & 0.9998 & 0.9991 & 0.9981 & 0.9966 \\
			0.8844 & 0.9687 & 0.9988 & 0.9990 & 0.9991 & 0.9998 & 1.000 & 0.9998 & 0.9992 & 0.9981 \\
			0.8746 & 0.9633 & 0.9976 & 0.9978 & 0.9981 & 0.9991 & 0.9998 & 1.000 & 0.9998 & 0.9992 \\
			0.8645 & 0.9576 & 0.9960 & 0.9963 & 0.9966 & 0.9981 & 0.9992 & 0.9998 & 1.000 & 0.9998 \\
			0.8541 & 0.9516 & 0.9939 & 0.9943 & 0.9947 & 0.9966 & 0.9981 & 0.9992 & 0.9998 & 1.000 \\
			\hline
		\end{tabular}
	\end{table}
	
	\begin{table}[htbp]
		\tablefont
		\centering
		\caption{Correlation Matrix Among Interest Rate Factors. (rough SABR formula, H=0.15)}
		\label{table:corrRSH15}
		\begin{tabular}{cccccccccc}
			\hline
			1.000 & 0.9939 & 0.9772 & 0.9851 & 0.9915 & 0.9916 & 0.9918 & 0.9919 & 0.9921 & 0.9922 \\
			0.9939 & 1.000 & 0.9946 & 0.9980 & 0.9998 & 0.9998 & 0.9998 & 0.9999 & 0.9999 & 0.9999 \\
			0.9772 & 0.9946 & 1.000 & 0.9991 & 0.9965 & 0.9964 & 0.9963 & 0.9962 & 0.9961 & 0.9960 \\
			0.9851 & 0.9980 & 0.9991 & 1.000 & 0.9991 & 0.9991 & 0.9990 & 0.9990 & 0.9989 & 0.9989 \\
			0.9915 & 0.9998 & 0.9965 & 0.9991 & 1.000 & 1.000 & 1.000 & 0.9999 & 0.9999 & 0.9999 \\
			0.9916 & 0.9998 & 0.9964 & 0.9991 & 1.000 & 1.000 & 1.000 & 1.000 & 0.9999 & 0.9999 \\
			0.9918 & 0.9998 & 0.9963 & 0.9990 & 1.000 & 1.000 & 1.000 & 1.000 & 1.000 & 0.9999 \\
			0.9919 & 0.9999 & 0.9962 & 0.9990 & 0.9999 & 1.000 & 1.000 & 1.000 & 1.000 & 1.000 \\
			0.9921 & 0.9999 & 0.9961 & 0.9989 & 0.9999 & 0.9999 & 1.000 & 1.000 & 1.000 & 1.000 \\
			0.9922 & 0.9999 & 0.9960 & 0.9989 & 0.9999 & 0.9999 & 0.9999 & 1.000 & 1.000 & 1.000 \\
			\hline
		\end{tabular}
	\end{table}
	
	Finally, we compare the results of the FMM and the mapped SMM.
	For parameter sets obtained using rough SABR formula above, we calculate the swaption IVs for the maturities and underlying where market smile data is available, and
	compare them with market prices. We plot the following IVs:
	
	\begin{itemize}
		\item Swaption IVs computed using Monte Carlo simulation under the FMM model ``FMM'',
		\item Swaption IVs under our approximative swap rate model \eqref{rBergomi} using the Monte Carlo method ``Mapped SMM(MC)'',
		\item Swaption IVs under our approximative swap rate model \eqref{rBergomi} using the rough SABR formula ``Mapped SMM(formula)'',
		\item Market IVs ``Market''.
	\end{itemize}
	
	Error bars corresponding to the $\pm2$ standard deviations are shown for both ``FMM'' and ``Mapped SMM (MC)''.
	
\begin{figure}[!tbp]
		\centering
		\begin{tabular}{cc}
			\includegraphics[width=0.4\textwidth]{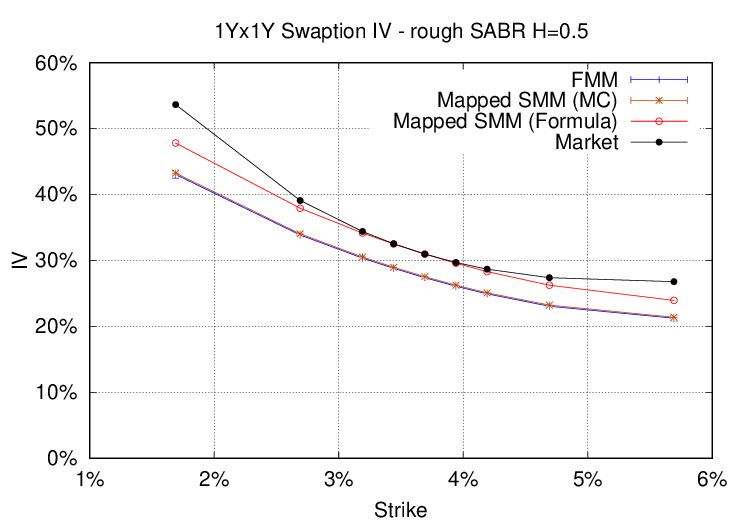} &
			\includegraphics[width=0.4\textwidth]{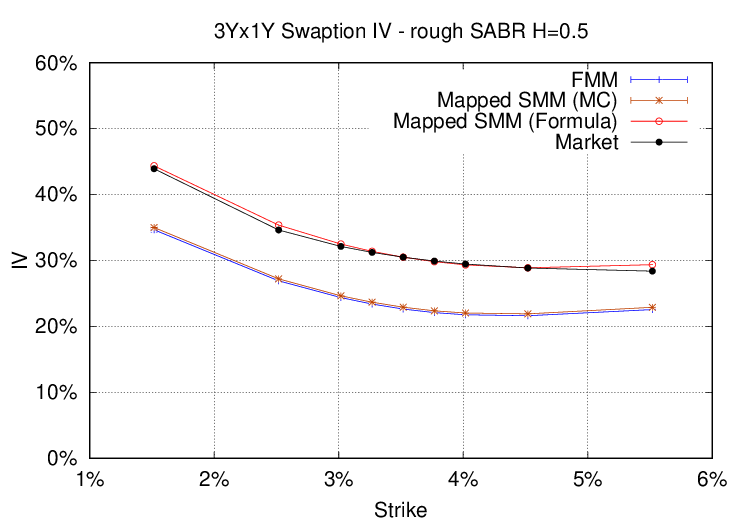} \\
			\includegraphics[width=0.4\textwidth]{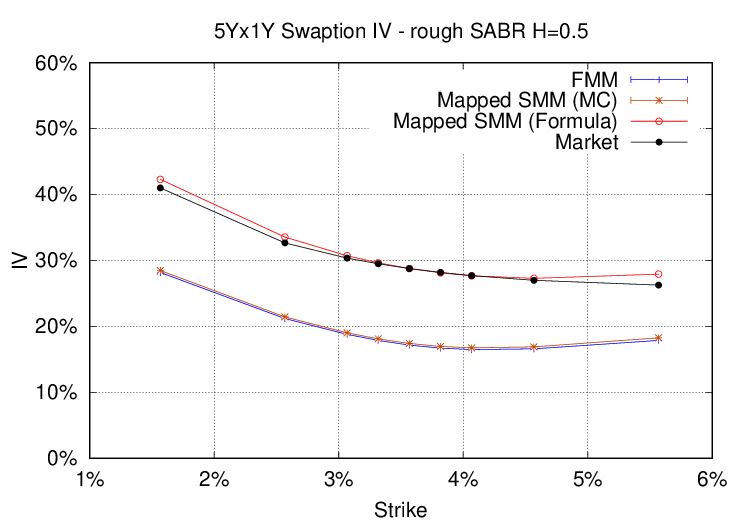} &
			\includegraphics[width=0.4\textwidth]{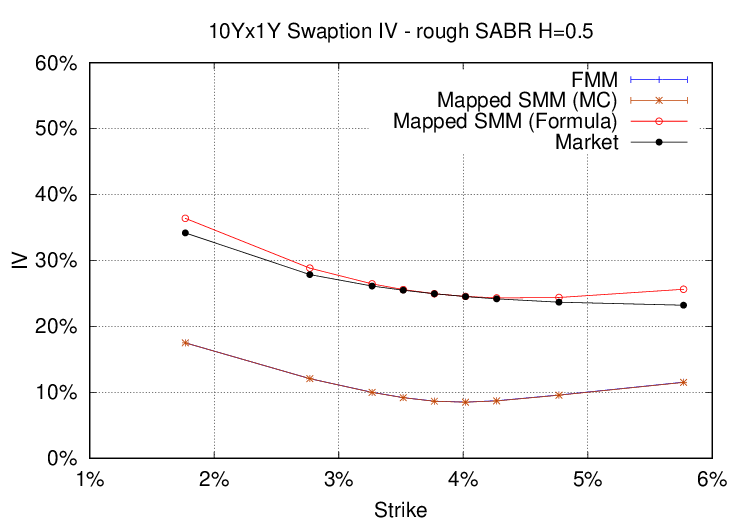}\\
		\end{tabular}
		\caption{Model and Market IVs of 1Y tenor calibrated with the rough SABR formula at $H=0.5$}
		\label{figure:RSH50A}
	\end{figure}

\begin{figure}[!tbp]
		\centering
		\begin{tabular}{cc}
			\includegraphics[width=0.4\textwidth]{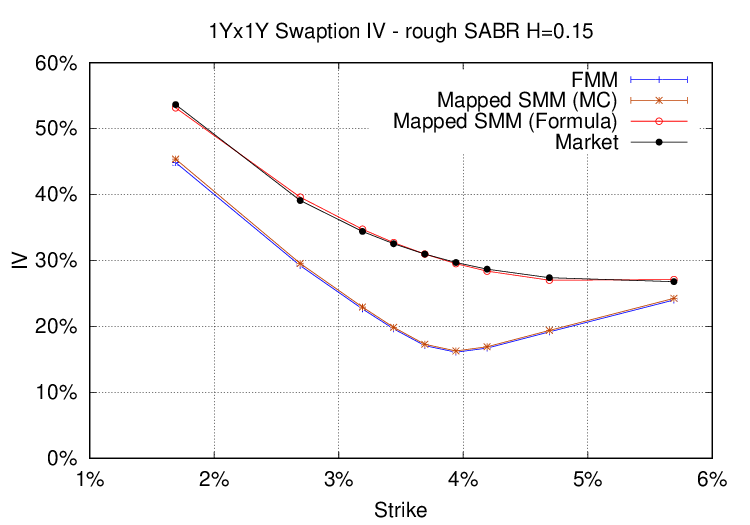} &
			\includegraphics[width=0.4\textwidth]{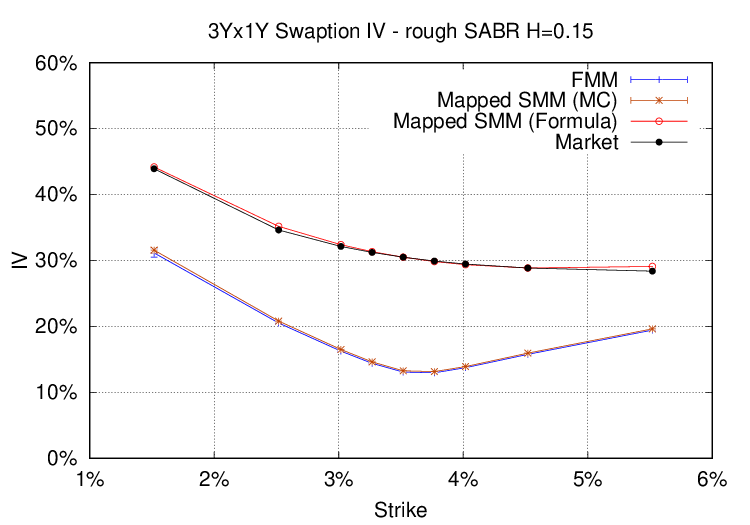} \\
			\includegraphics[width=0.4\textwidth]{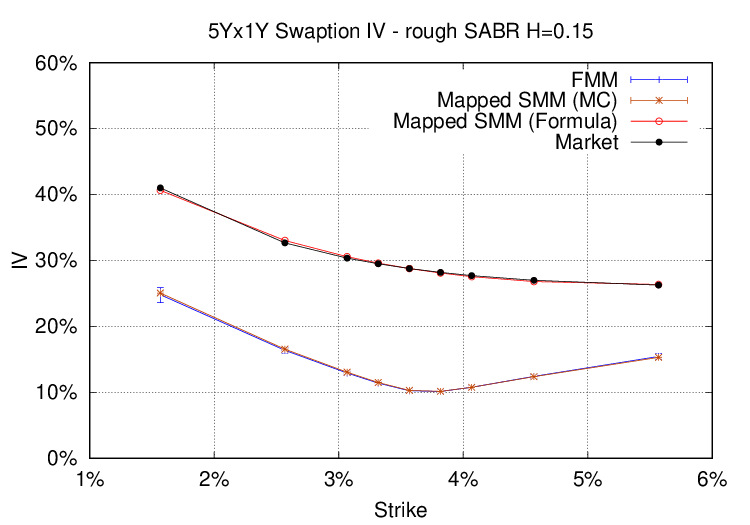} &
			\includegraphics[width=0.4\textwidth]{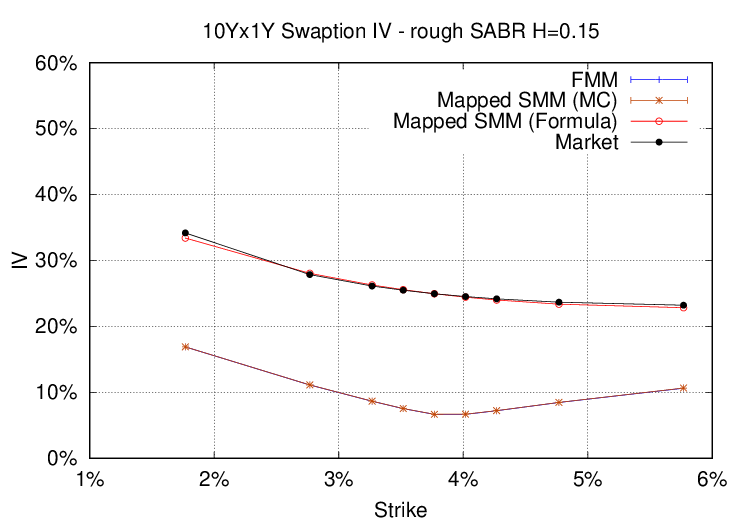}\\
		\end{tabular}
		\caption{Model and Market IVs of 1Y tenor calibrated with the rough SABR formula at $H=0.15$}
		\label{figure:RSH15A}
	\end{figure}

	Figure \ref{figure:RSH50A}, \ref{figure:RSH15A}
	show the quality of in-sample fit, as the calibration was done using swaptions with 1Y tenor.
	The fit of Mapped SMM (formula) to Market is good for all 1Y tenor data,
	however, FMM and Mapped SMM (MC) are largely deviate for other maturities.
	
\begin{figure}[!tbp]
		\centering
		\begin{tabular}{cc}
			\includegraphics[width=0.4\textwidth]{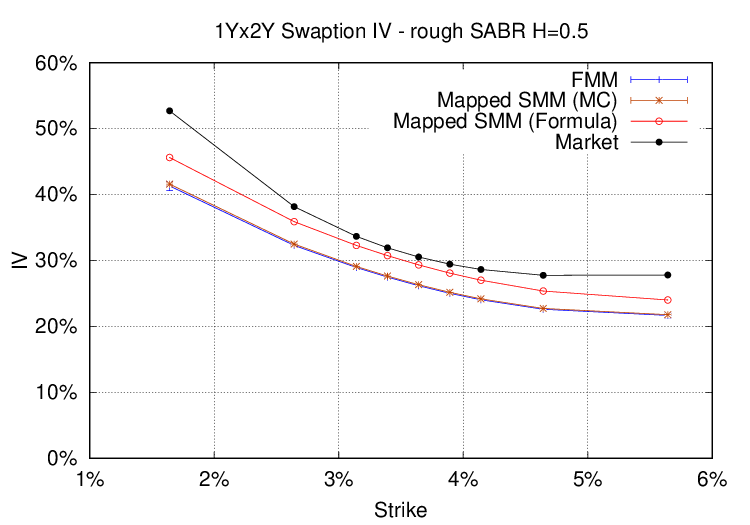} &
			\includegraphics[width=0.4\textwidth]{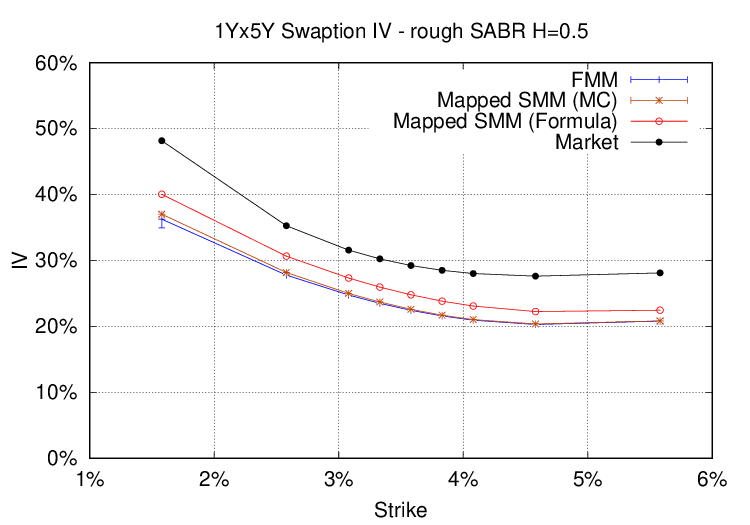} \\
			\includegraphics[width=0.4\textwidth]{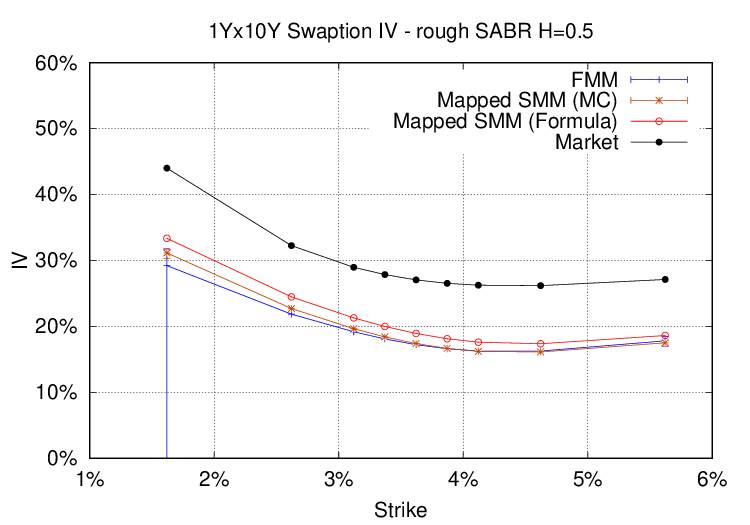} &
			\includegraphics[width=0.4\textwidth]{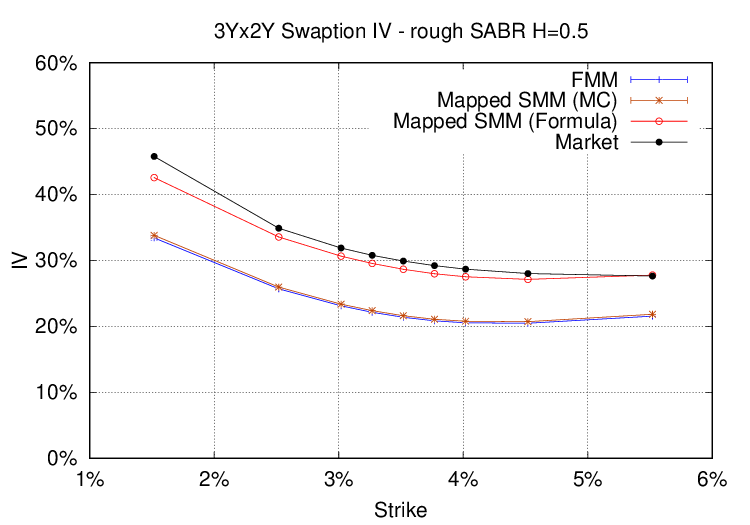} \\
			\includegraphics[width=0.4\textwidth]{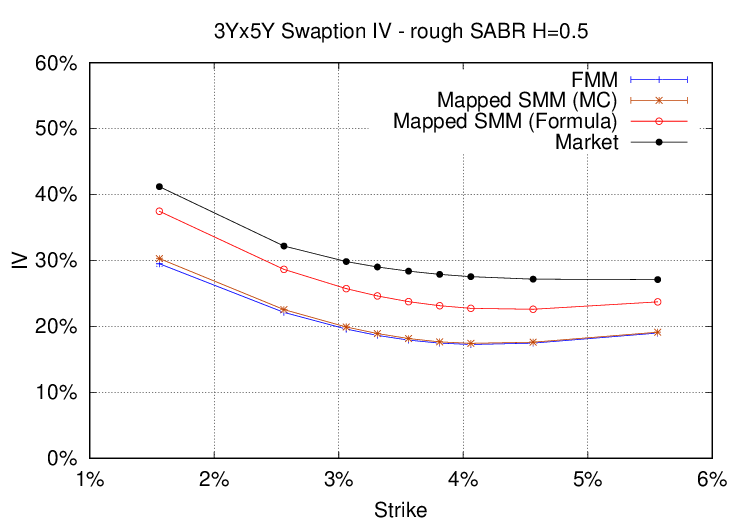} &
			\includegraphics[width=0.4\textwidth]{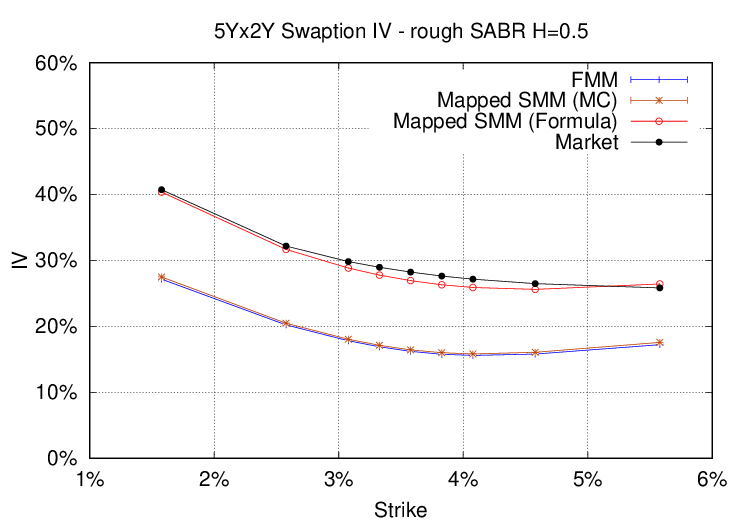} \\
			\includegraphics[width=0.4\textwidth]{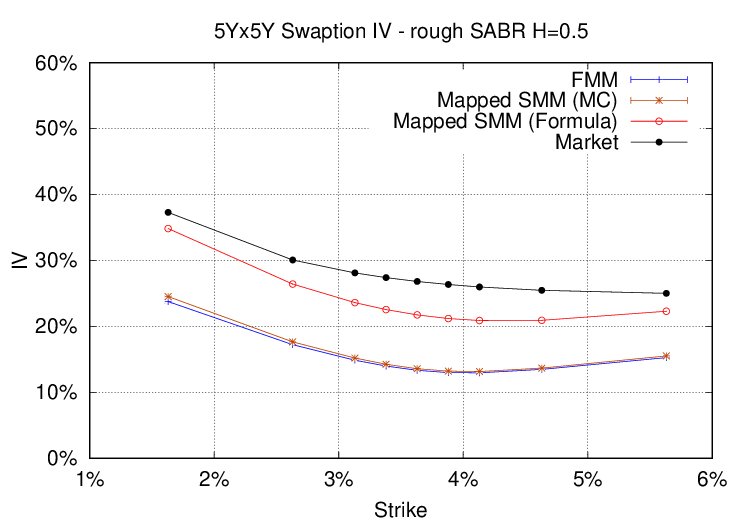} &
			\\
		\end{tabular}
		\caption{Model and Market IVs of non 1Y tenor calibrated with the rough SABR formula at $H=0.5$}
		\label{figure:RSH50B}
	\end{figure}

\begin{figure}[!tbp]
		\centering
		\begin{tabular}{cc}
			\includegraphics[width=0.4\textwidth]{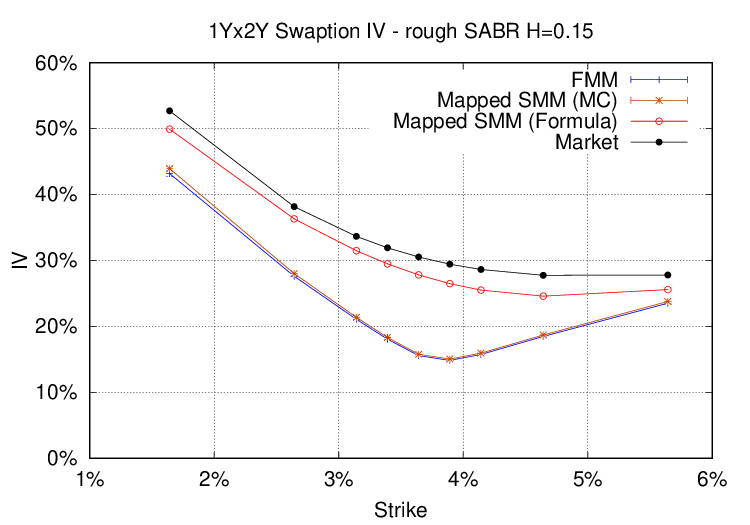} &
			\includegraphics[width=0.4\textwidth]{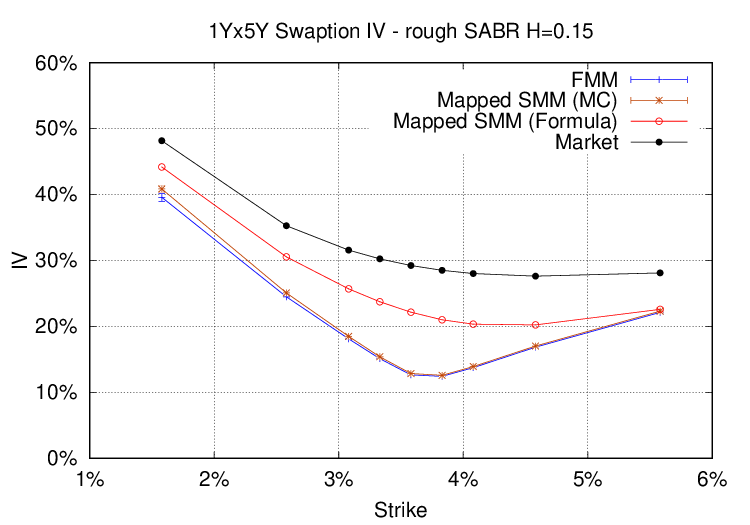} \\
			\includegraphics[width=0.4\textwidth]{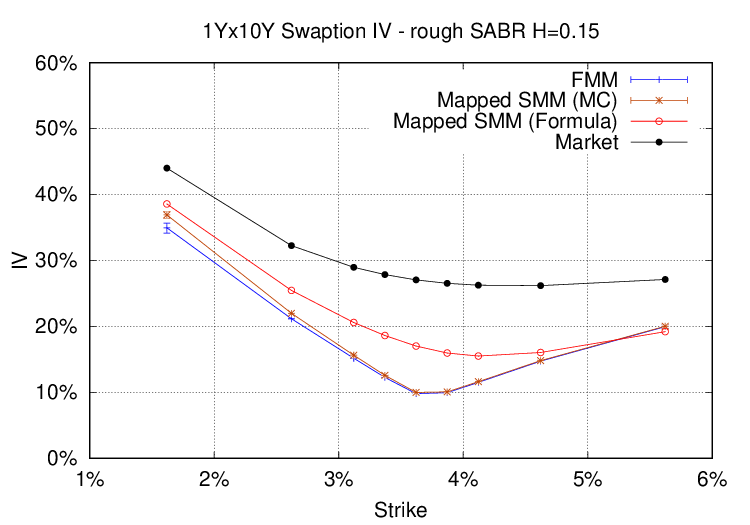} &
			\includegraphics[width=0.4\textwidth]{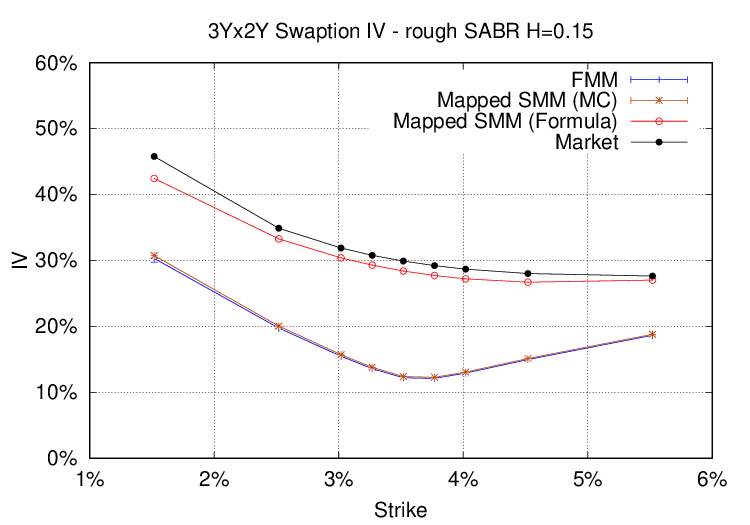} \\
			\includegraphics[width=0.4\textwidth]{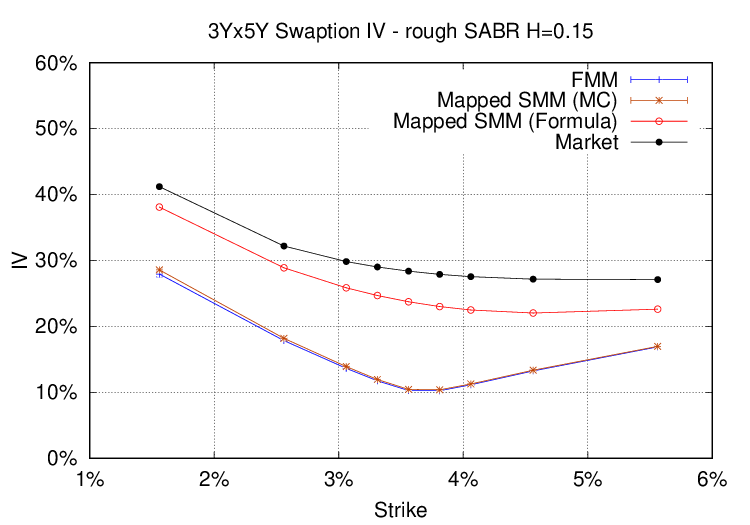} &
			\includegraphics[width=0.4\textwidth]{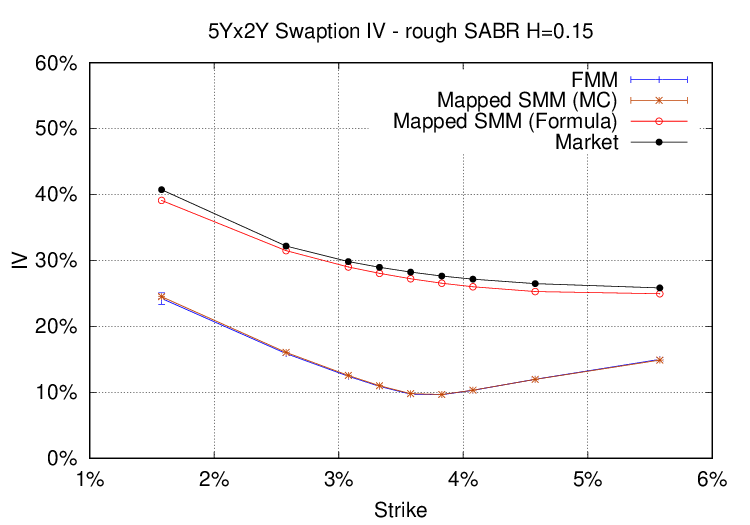} \\
			\includegraphics[width=0.4\textwidth]{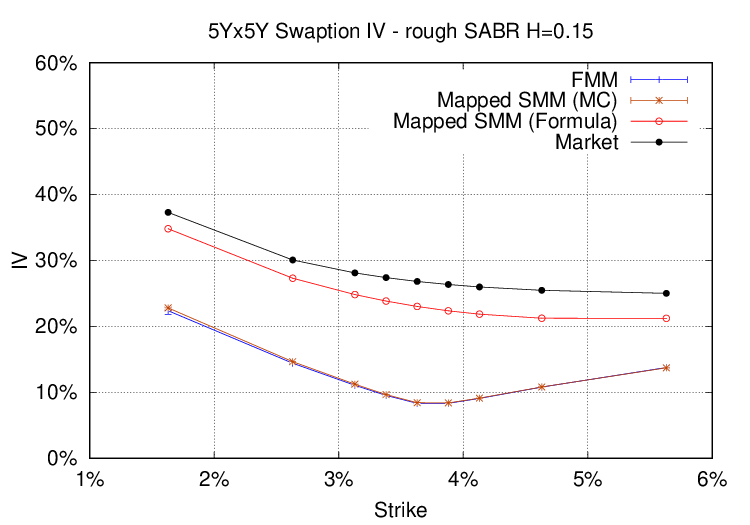} &
			\\
		\end{tabular}
		\caption{Model and Market IVs of non 1Y tenor calibrated with the rough SABR formula at $H=0.15$}
		\label{figure:RSH15B}
	\end{figure}

	In Figure \ref{figure:RSH50B}, \ref{figure:RSH15B}, we present volatility smiles for tenors of 2Y or longer.
	These results indicate that the Mapped SMM (formula) tends to deviate from market observations as the tenor of the underlying increases.
	Moreover, significant discrepancies were observed when compared with both the FMM and the Mapped SMM (MC).

	\FloatBarrier
	\section*{Appendix C}
	Here we discuss the well-posedness of the rough SABR forward market model~\eqref{rSABR}.
	We start by fixing a probability space
	$(\Omega,\mathscr{F},\mathsf{Q}^{T_N})$ and
	noting the well-posedness of the corresponding stochastic differential equation under $\mathsf{Q}^{T_N}$:
	\begin{equation}\label{rSABR2}
		\begin{split}
			& \mathrm{d}R^j_t = \gamma_j(t)\eta_j(R^j_t)\sqrt{V^j_t}
			\mathrm{d}W^{j\dagger}_t, \\ 
			& \log V^j_t =  \log \xi_j(t)  - \frac{1}{2}\int_0^t \zeta_j(t-s)^2\mathrm{d}s  +
			\int_0^t \zeta_j(t-s)\mathrm{d}\bar{W}^0_s,
		\end{split}
	\end{equation}
	where 
	\begin{equation*}
		W^{j\dagger} =  \bar{W}^j - \sum_{i=j+1}^N \int_0^\cdot \frac{\theta_i}{1 + \theta_i R^i_t} 
		\gamma_i(t)\eta_i(R^i_t)\sqrt{V^i_t}\rho_{ij} \mathrm{d}t
	\end{equation*}
	and 
	$(\bar{W}^0,\bar{W}^1,\dots,\bar{W}^N)$ is a correlated Brownian motion under $\mathsf{Q}^{T_N}$
	with
	\begin{equation*}
		\langle \bar{W}^j \rangle_t = t, \ \ 
		\langle \bar{W}^i, \bar{W}^j \rangle_t = 
		\rho_{ij}t.
	\end{equation*}
	\begin{lemma}\label{lem2}
		If $\eta_j$ is Lipschitz continuous with \eqref{eta0} 
		and $R^j_0 > 0$
		for each $j=1,\dots, N$,  then 
		there exists a unique strong solution
		$(R^1,\dots,R^N)$
		to \eqref{rSABR2}.  
	\end{lemma}
	\begin{proof}
		Noting that $V^j$ is explicit, the result follows from the standard argument.
	\end{proof}
	\begin{lemma}
		Let $(R^1,\dots,R^N)$ be a solution
		to \eqref{rSABR2}.  Then, under \eqref{eta0} and \eqref{rho0},
		\begin{equation*}
			\prod_{j=1}^N (1 + \theta_j R^j)
		\end{equation*}
		is a martingale under $\mathsf{Q}^{T_N}$.
	\end{lemma}
	\begin{proof}
		Let
		\begin{equation*}
			U = \sum_{j=1}^N U^j, \ \ 
			U_t^j = \int_0^t 
			\frac{\theta_j }{1 + \theta_jR^j_s}\gamma_j(s)\eta_j(R^j_s)\sqrt{V^j_s}
			\mathrm{d}\bar{W}^j_s.
		\end{equation*}
		Since
		\begin{equation*}
			\begin{split}
				\mathrm{d} \log
				\prod_{j=1}^N (1 + \theta_j R^j) 
				& = \sum_{j=1}^N \mathrm{d} \log (1 + \theta_jR^j)\\
				&= \sum_{j=1}^N \left\{
				\frac{\theta_j \mathrm{d}R^j}{1 + \theta_jR^j} 
				- \frac{1}{2} \frac{\theta_j^2 \mathrm{d}\langle R^j \rangle}{(1 + \theta_jR^j)^2} \right\}\\
				&= \sum_{j=1}^N 
				\mathrm{d}U^j 
				- \frac{1}{2}\sum_{i,j=1}^N 
				\mathrm{d}\langle U^i,U^j \rangle
				\\
				&= \mathrm{d}U - \frac{1}{2}\mathrm{d}\langle U \rangle,
			\end{split}
		\end{equation*}
		the product is a nonnegative local martingale. 
		It suffices then to show
		\begin{equation*}
			\mathsf{E}_{\mathsf{Q}^{T_N}}\left[ \prod_{j=1}^N \frac{1 + \theta_j R^j_{T_N}}{1 + \theta_j R^j_0} \right] = 1.
		\end{equation*}
		Let
		\begin{equation*}
			\tau_n = \inf\left\{t \geq 0; \max_{j=1,\dots,N} V^j_t = n\right\}.
		\end{equation*}
		for integers $n$ and define a sequence of probability measures $\mathsf{Q}^n$ by
		\begin{equation*}
			\frac{\mathrm{d}\mathsf{Q}^n}{\mathrm{d}\mathsf{Q}^{T_N}}
			= \exp\left\{U_{\tau_n} - \frac{1}{2}\langle U \rangle_{\tau_n}\right\}.
		\end{equation*}
		Here we have used \eqref{eta0}.
		Then, we have
		\begin{equation}\label{Gas2}
			\begin{split}
				1 &= \mathsf{E}_{\mathsf{Q}^{T_N}}\left[ 
				\exp\left\{U_{\tau_n \wedge T_N} - \frac{1}{2}\langle U \rangle_{\tau_n \wedge T_N}\right\}
				\right]
				\\
				&= \mathsf{E}_{\mathsf{Q}^{T_N}}
				\left[ 
				\exp\left\{U_{T_N} - \frac{1}{2}\langle U \rangle_{ T_N}\right\}
				; \tau_n > T_N
				\right] + 
				\mathsf{Q}^n(\tau_n \leq T_N).
			\end{split}
		\end{equation}
		By the Girsanov-Maruyama theorem,
		\begin{equation*}
			\hat{W}^n = \bar{W}^0 - \langle \bar{W}^0,U_{\cdot \wedge \tau_n}\rangle =
			\bar{W}^0 - 
			\sum_{j=1}^N \int_0^{\cdot \wedge \tau_n} \frac{\theta_j}{1 + \theta_j R^j_t} 
			\gamma_j(t)\eta_j(R^j_t)\sqrt{V^j_t}\rho_{0j} \mathrm{d}t
		\end{equation*}
		is a Brownian motion under $\mathsf{Q}^n$.
		Under \eqref{rho0},
		\begin{equation*}
			\begin{split}
				\log V^j_t &=  \log \xi_j(t)  - \frac{1}{2}\int_0^t \zeta_j(t-s)^2\mathrm{d}s  +
				\int_0^t \zeta_j(t-s)\mathrm{d}\bar{W}^0_s\\
				& \leq \log \xi_j(t)  - \frac{1}{2}\int_0^t \zeta_j(t-s)^2\mathrm{d}s  +
				\int_0^t \zeta_j(t-s)\mathrm{d}\hat{W}^n_s =: \log \hat{V}^{j,n}_t
			\end{split}
		\end{equation*}
		and so, 
		\begin{equation*}
			\mathsf{Q}^n(\tau_n \leq T_N) = \mathsf{Q}^n\left(\sup_{0 \leq t \leq T_N} \max_{j=1,\dots,N} V^j_t \geq n \right)
			\leq 
			\mathsf{Q}^n\left(\sup_{0 \leq t \leq T_N} \max_{j=1,\dots,N} \hat{V}^{j,n}_t \geq n \right).
		\end{equation*}
		The law of $\hat{V}^{j,n}$ under $\mathsf{Q}^n$ does not depend on $n$.
		Therefore, the right hand side converges to $0$ as $n\to \infty$.
		Together with the monotone convergence theorem, this implies that 
		the right hand side of \eqref{Gas2} converges to
		\begin{equation*}
			\mathsf{E}_{\mathsf{Q}^{T_N}}\left[ 
			\exp\left\{U_{T_N} - \frac{1}{2}\langle U \rangle_{ T_N}\right\}
			\right]
			= \mathsf{E}_{\mathsf{Q}^{T_N}}\left[ \prod_{j=1}^N \frac{1 + \theta_j R^j_{T_N}}{1 + \theta_j R^j_0} \right],
		\end{equation*}
		which concludes the proof.
	\end{proof}
	\begin{lemma}
		Let $(R^1,\dots,R^N)$ be a solution
		to \eqref{rSABR2} under \eqref{eta0} and \eqref{rho0}. 
		Define a probability measure $\mathsf{Q}$ by
		\begin{equation*}
			\frac{\mathrm{d}\mathsf{Q}}{\mathrm{d}\mathsf{Q}^{T_N}} = 
			\prod_{j=1}^N \frac{1 + \theta_j R^j_{T_N}}{1 + \theta_j R^j_0}.
		\end{equation*}
		Then, 
		\begin{equation*}
			\begin{split}
				W^j  
				:= \bar{W}^j -  \sum_{i=1}^N \int_0^\cdot \frac{\theta_i}{1 + \theta_i R^i_t} 
				\gamma_i(t)\eta_i(R^i_t)\sqrt{V^i_t}\rho_{ij} \mathrm{d}t,\ \ 
				j=0,1,\dots,N
			\end{split}
		\end{equation*}
		is a Brownian motion under $\mathsf{Q}$ with \eqref{rho},
		and 
		$(R^1,\dots,R^N)$ is a solution to \eqref{rSABR} with
		$\bar{W} = \bar{W}^0$.
	\end{lemma}
	\begin{proof}
		By the previous lemma, $\mathsf{Q}$ is a probability measure.
		The result follows from the standard application of the Girsanov-Maruyama theorem.
	\end{proof}
	\begin{theorem}
		If $\eta_j$ is Lipschitz continuous with \eqref{eta0} 
		and $R^j_0 > 0$ for each $j=1,\dots, N$,  then,
		there exists a weak solution to \eqref{rSABR} under \eqref{rho0}
		which is unique in law.
	\end{theorem}
	\begin{proof}
		The existence of a weak solution is from the preceding lemmas.
		For any solution, 
		\begin{equation*}
			\prod_{j=1}^N \frac{1}{1 + \theta_j R^i}
		\end{equation*} 
		is a bounded local martingale and hence a martingale. 
		We can therefore define a probability measure $\mathsf{Q}^{T_N}$ by
		\begin{equation*}
			\frac{\mathrm{d}\mathsf{Q}^{T_N}}{\mathrm{d}\mathsf{Q}} = \prod_{j=1}^N \frac{1 + \theta_j R^j_0}{1 + \theta_j R^j_{T_N}},
		\end{equation*}
		under which $(R^1,\dots,R^N)$ solves \eqref{rSABR2} with $\bar{W}^0 = \bar{W}$.
		Consequently, the law of $(R^1,\dots,R^N)$
		is determined by that under $\mathsf{Q}^{T_N}$ that is unique by Lemma~\ref{lem2}.
	\end{proof}

\section*{Appendix D}
To assess the adequacy of the time discretization used in the numerical experiments for the rough volatility model, we recalculated the results under the same conditions as in Table \ref{tb:firstStepCalibration}, increasing the number of time steps from 200 to 1000.
Comparing Tables~\ref{tb:firstStepCalibration} and~\ref{tb:firstStepCalibration20241209div1000}, we find that $H=0.2$ gives the smallest RMSE on both time grids.
At this value of $H$, the RMSE is $0.505\%$ on both grids to the precision reported.
The difference is more pronounced at $H=0.05$, where the RMSE decreases from $0.707\%$ to $0.673\%$.
Thus, the calibration at the smallest Hurst exponent considered is more sensitive to the time discretization, but refining the grid does not change the choice of $H$ in the main calibration experiment.
	\begin{table}[]
		\tablefont
		\centering
		\caption{First-Step Calibration Results 2024/12/9 div=1000}
		\label{tb:firstStepCalibration20241209div1000}
		\begin{tabular}{lc|cccc|cccc|c}
			\hline
			Hurst Exponent & $\kappa$ & $\alpha_{2}$ & $\alpha_{4}$ & $\alpha_{6}$ & $\alpha_{11}$ & $\rho_{0,2}$ & $\rho_{0,4}$ & $\rho_{0,6}$ & $\rho_{0,11}$ & RMSE \\
			\hline
			H=0.5 & 1.71818 & 0.315 & 0.316 & 0.344 & 0.404 & -63.8\% & -38.7\% & -45.0\% & -45.6\% & 0.729\%\\
			H=0.45 & 1.61779 & 0.314 & 0.311 & 0.331 & 0.370 & -63.1\% & -39.0\% & -45.6\% & -45.5\% & 0.680\% \\
			H=0.4 & 1.52474 & 0.312 & 0.306 & 0.319 & 0.341 & -62.4\% & -39.4\% & -46.4\% & -46.0\% & 0.628\%\\
			H=0.35 & 1.43641 & 0.309 & 0.300 & 0.307 & 0.315 & -61.7\% & -39.9\% & -47.3\% & -47.0\% & 0.579\%\\
			H=0.3 & 1.35068 & 0.305 & 0.293 & 0.296 & 0.292 & -61.1\% & -40.5\% & -48.4\% & -48.5\% & 0.536\%\\
			H=0.25 & 1.26563 & 0.298 & 0.284 & 0.283 & 0.271 & -60.7\% & -41.3\% & -49.8\% & -50.8\% & 0.508\%\\
			H=0.2 & 1.17889 & 0.288 & 0.272 & 0.268 & 0.251 & -60.9\% & -42.6\% & -51.9\% & -53.9\% & 0.505\%\\
			H=0.15 & 1.08942 & 0.271 & 0.253 & 0.248 & 0.227 & -62.3\% & -44.9\% & -55.4\% & -58.7\% & 0.530\%\\
			H=0.1 & 0.99336 & 0.239 & 0.221 & 0.215 & 0.195 & -67.4\% & -50.4\% & -63.2\% & -68.6\% & 0.580\%\\
			H=0.05 & 0.86356 & 0.171 & 0.158 & 0.154 & 0.139 & -94.1\% & -72.5\% & -92.6\% & -100.0\% & 0.673\%\\
			\hline
		\end{tabular}
	\end{table}

	\FloatBarrier
	\section*{Appendix E}
	To examine whether the calibration results depend on the market date, we repeat the two-step calibration of Section~\ref{sec:calibration} using USD SOFR swaption data from 7 June 2024 and 10 June 2025.
	Among the Hurst exponents considered, the first-step RMSE is minimized at $H=0.25$ on 7 June 2024 and at $H=0.1$ on 10 June 2025 (Tables~\ref{tb:firstStepCalibration20240607} and~\ref{tb:firstStepCalibration20250610}).
	The corresponding RMSEs are $0.547\%$ and $0.786\%$, compared with $0.665\%$ and $1.308\%$, respectively, at $H=0.5$.
	For the June 2025 data, increasing the number of time steps to 1000 leaves the minimizing value of $H$ unchanged (Table~\ref{tb:firstStepCalibration20250610-1000}).
	
	\begin{table}[]
		\tablefont
		\centering
		\caption{First-Step Calibration Results 2024/6/7}
		\label{tb:firstStepCalibration20240607}
		\begin{tabular}{lc|cccc|cccc|c}
			\hline
			Hurst Exponent & $\kappa$ & $\alpha_{2}$ & $\alpha_{4}$ & $\alpha_{6}$ & $\alpha_{11}$ & $\rho_{0,2}$ & $\rho_{0,4}$ & $\rho_{0,6}$ & $\rho_{0,11}$ & RMSE \\
			\hline
			H=0.5 & 2.11217 & 0.327 & 0.387 & 0.412 & 0.492 & -69.6\% & -52.1\% & -51.2\% & -51.5\% & 0.665\% \\
			H=0.45 & 1.96443 & 0.325 & 0.375 & 0.387 & 0.440 & -69.2\% & -52.2\% & -51.4\% & -51.2\% & 0.643\% \\
			H=0.4 & 1.82986 & 0.322 & 0.364 & 0.366 & 0.394 & -68.8\% & -52.6\% & -51.9\% & -51.4\% & 0.616\% \\
			H=0.35 & 1.70528 & 0.319 & 0.353 & 0.346 & 0.354 & -68.6\% & -53.2\% & -52.7\% & -52.2\% & 0.587\% \\
			H=0.3 & 1.58894 & 0.313 & 0.340 & 0.326 & 0.320 & -68.4\% & -54.0\% & -53.9\% & -53.6\% & 0.562\% \\
			H=0.25 & 1.47720 & 0.304 & 0.325 & 0.307 & 0.288 & -68.7\% & -55.3\% & -55.6\% & -56.0\% & 0.547\% \\
			H=0.2 & 1.36718 & 0.291 & 0.306 & 0.284 & 0.259 & -69.8\% & -57.5\% & -58.4\% & -59.6\% & 0.550\% \\
			H=0.15 & 1.25333 & 0.270 & 0.279 & 0.256 & 0.228 & -73.2\% & -61.7\% & -63.4\% & -65.7\% & 0.579\% \\
			H=0.1 & 1.12027 & 0.234 & 0.239 & 0.218 & 0.192 & -83.5\% & -71.9\% & -75.0\% & -79.0\% & 0.643\% \\
			H=0.05 & 0.86287 & 0.179 & 0.188 & 0.170 & 0.148 & -100.0\% & -100.0\% & -100.0\% & -100.0\% & 1.072\% \\
			\hline
		\end{tabular}
	\end{table}

	\begin{table}[htbp]
		\tablefont
		\centering
		\caption{Correlation Matrix Among Interest Rate Factors. (H=0.5)2024/6/7}
		\label{table:corrMCH520240607}
		\begin{tabular}{cccccccccc}
			\hline
			1.0000 & 0.8576 & 0.6392 & 0.4557 & 0.3055 & 0.2303 & 0.1265 & 0.0020 & 0.0406 & -0.0030 \\
			0.8576 & 1.0000 & 0.8271 & 0.6372 & 0.5041 & 0.2567 & 0.2025 & 0.1707 & 0.0549 & 0.0177 \\
			0.6392 & 0.8271 & 1.0000 & 0.7988 & 0.5858 & 0.4887 & 0.2948 & 0.1971 & 0.0747 & 0.0693 \\
			0.4557 & 0.6372 & 0.7988 & 1.0000 & 0.8178 & 0.6269 & 0.5196 & 0.3548 & 0.2752 & 0.1574 \\
			0.3055 & 0.5041 & 0.5858 & 0.8178 & 1.0000 & 0.8076 & 0.6396 & 0.5297 & 0.3755 & 0.2794 \\
			0.2303 & 0.2567 & 0.4887 & 0.6269 & 0.8076 & 1.0000 & 0.8223 & 0.6532 & 0.5506 & 0.4364 \\
			0.1265 & 0.2025 & 0.2948 & 0.5196 & 0.6396 & 0.8223 & 1.0000 & 0.8293 & 0.6622 & 0.5752 \\
			0.0020 & 0.1707 & 0.1971 & 0.3548 & 0.5297 & 0.6532 & 0.8293 & 1.0000 & 0.8533 & 0.6772 \\
			0.0406 & 0.0549 & 0.0747 & 0.2752 & 0.3755 & 0.5506 & 0.6622 & 0.8533 & 1.0000 & 0.8618 \\
			-0.0030 & 0.0177 & 0.0693 & 0.1574 & 0.2794 & 0.4364 & 0.5752 & 0.6772 & 0.8618 & 1.0000 \\
			\hline
		\end{tabular}
	\end{table}
	
	\begin{table}[htbp]
		\tablefont
		\centering
		\caption{Correlation Matrix Among Interest Rate Factors. (H=0.25)2024/6/7}
		\label{table:corrMCH2520240607}
		\begin{tabular}{cccccccccc}
			\hline
			1.0000 & 0.9512 & 0.8401 & 0.7411 & 0.6569 & 0.5685 & 0.5900 & 0.5424 & 0.5386 & 0.6232 \\
			0.9512 & 1.0000 & 0.9246 & 0.8521 & 0.8119 & 0.6716 & 0.6247 & 0.6113 & 0.5706 & 0.5912 \\
			0.8401 & 0.9246 & 1.0000 & 0.8977 & 0.7994 & 0.7735 & 0.6527 & 0.6180 & 0.5674 & 0.5427 \\
			0.7411 & 0.8521 & 0.8977 & 1.0000 & 0.9134 & 0.8335 & 0.8121 & 0.7014 & 0.6235 & 0.5789 \\
			0.6569 & 0.8119 & 0.7994 & 0.9134 & 1.0000 & 0.8941 & 0.8245 & 0.7931 & 0.7155 & 0.6462 \\
			0.5685 & 0.6716 & 0.7735 & 0.8335 & 0.8941 & 1.0000 & 0.9017 & 0.8234 & 0.7827 & 0.7333 \\
			0.5900 & 0.6247 & 0.6527 & 0.8121 & 0.8245 & 0.9017 & 1.0000 & 0.9000 & 0.8202 & 0.7980 \\
			0.5424 & 0.6113 & 0.6180 & 0.7014 & 0.7931 & 0.8234 & 0.9000 & 1.0000 & 0.9173 & 0.8132 \\
			0.5386 & 0.5706 & 0.5674 & 0.6235 & 0.7155 & 0.7827 & 0.8202 & 0.9173 & 1.0000 & 0.9212 \\
			0.6232 & 0.5912 & 0.5427 & 0.5789 & 0.6462 & 0.7333 & 0.7980 & 0.8132 & 0.9212 & 1.0000 \\
			\hline
		\end{tabular}
	\end{table}

	\begin{figure}[!tbp]
		\centering
		\begin{tabular}{cc}
			\includegraphics[width=0.47\textwidth]{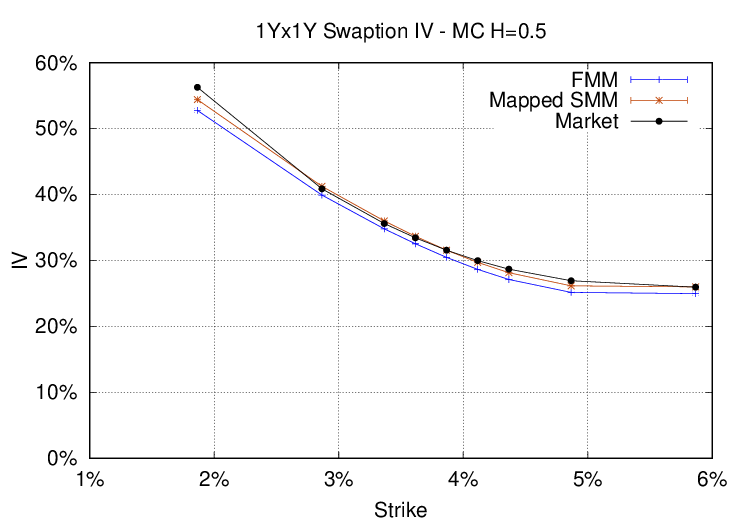} &
			\includegraphics[width=0.47\textwidth]{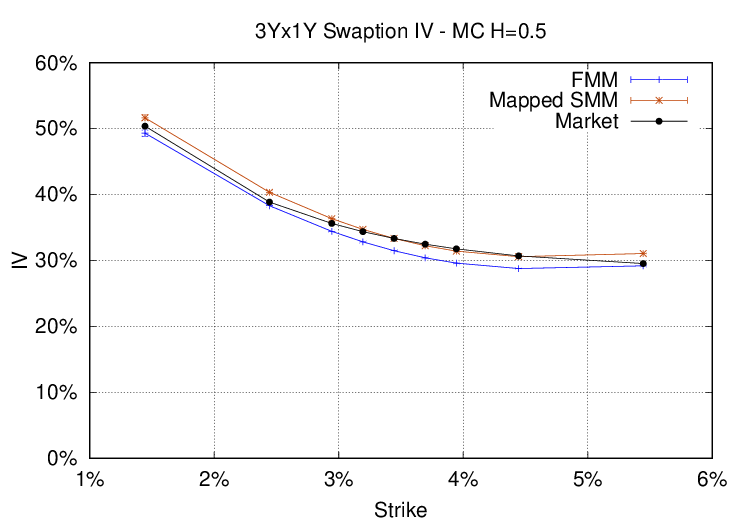} \\
			\includegraphics[width=0.47\textwidth]{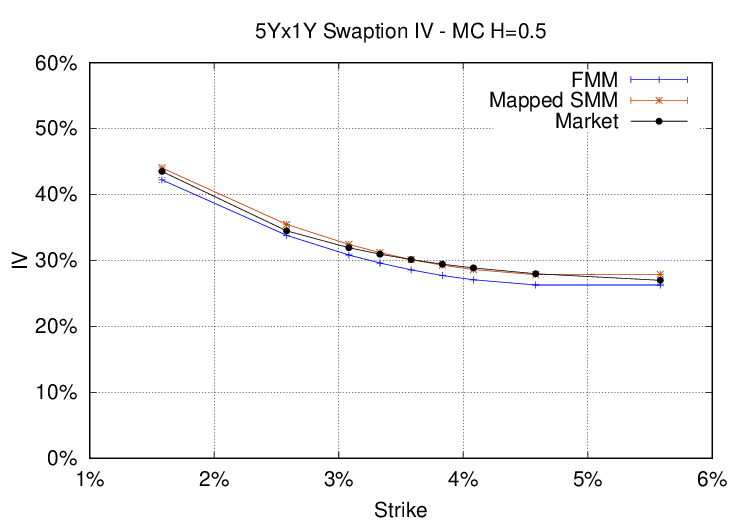} &
			\includegraphics[width=0.47\textwidth]{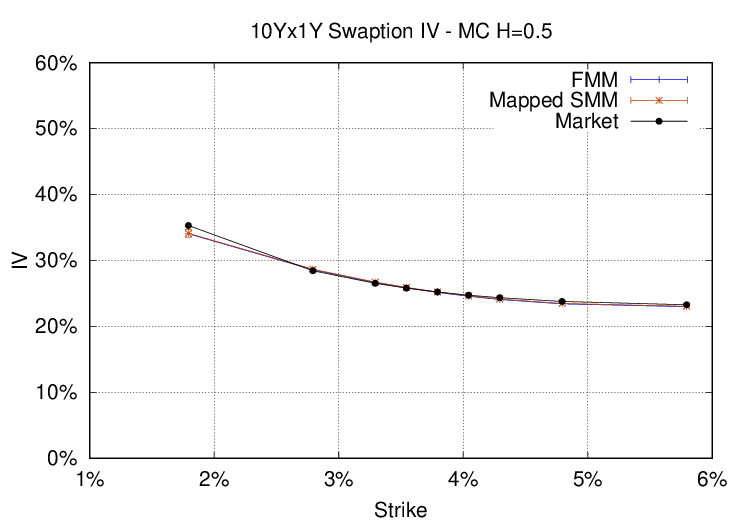}\\
		\end{tabular}
		\caption{Model and Market IVs of 1Y tenor at $H=0.5$ 2024/6/7}
		\label{figure:20240607MCH50A}
	\end{figure}

	\begin{figure}[!tbp]
		\centering
		\begin{tabular}{cc}
			\includegraphics[width=0.47\textwidth]{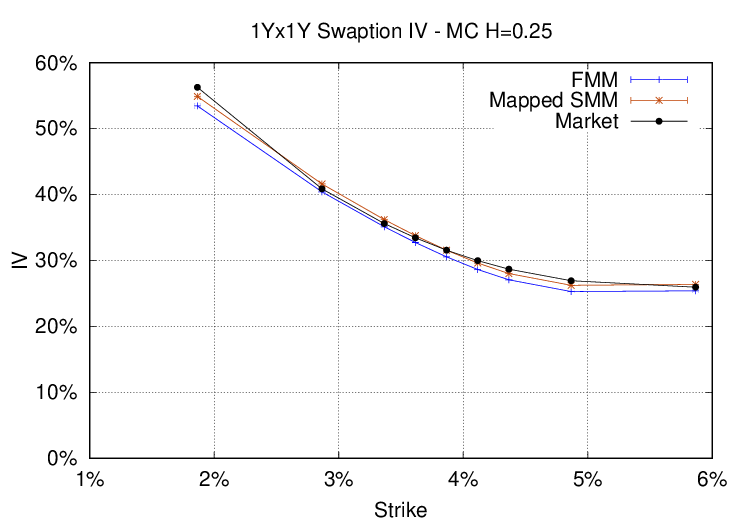} &
			\includegraphics[width=0.47\textwidth]{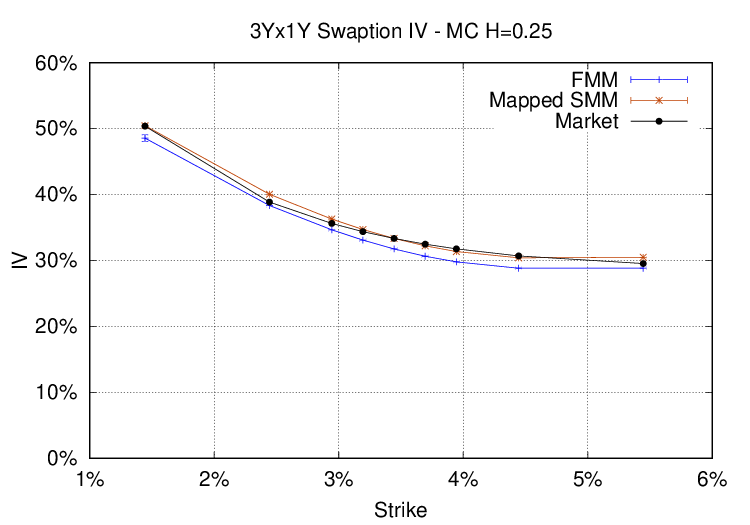} \\
			\includegraphics[width=0.47\textwidth]{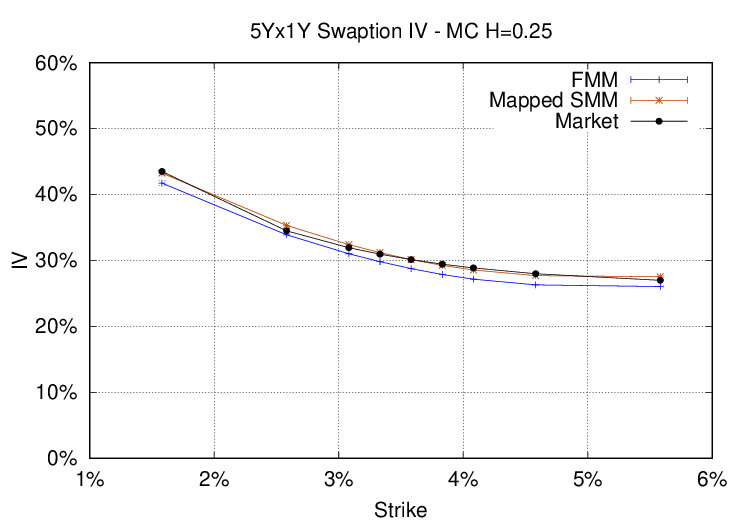} &
			\includegraphics[width=0.47\textwidth]{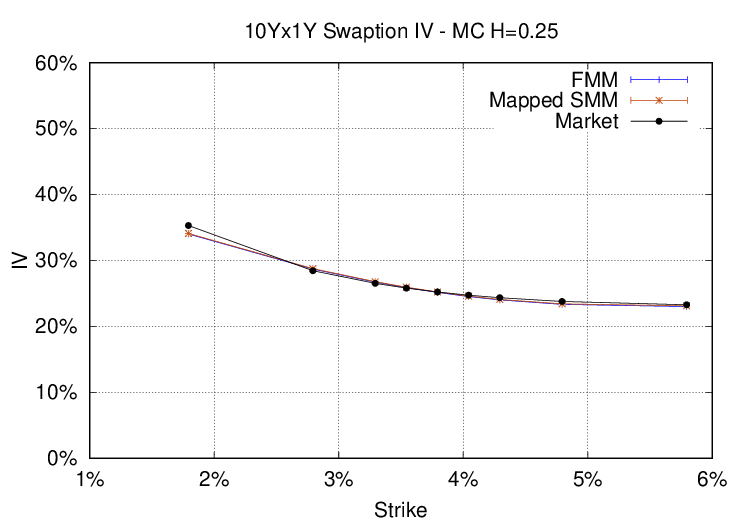}\\
		\end{tabular}
		\caption{Model and Market IVs of 1Y tenor at $H=0.25$ 2024/6/7}
		\label{figure:20240607MCH20A}
	\end{figure}
	
	\begin{figure}[!tbp]
		\centering
		\begin{tabular}{cc}
			\includegraphics[width=0.47\textwidth]{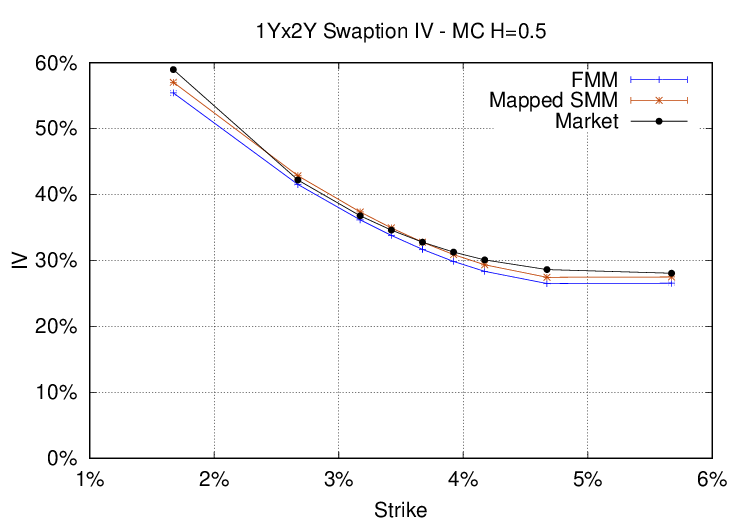} &
			\includegraphics[width=0.47\textwidth]{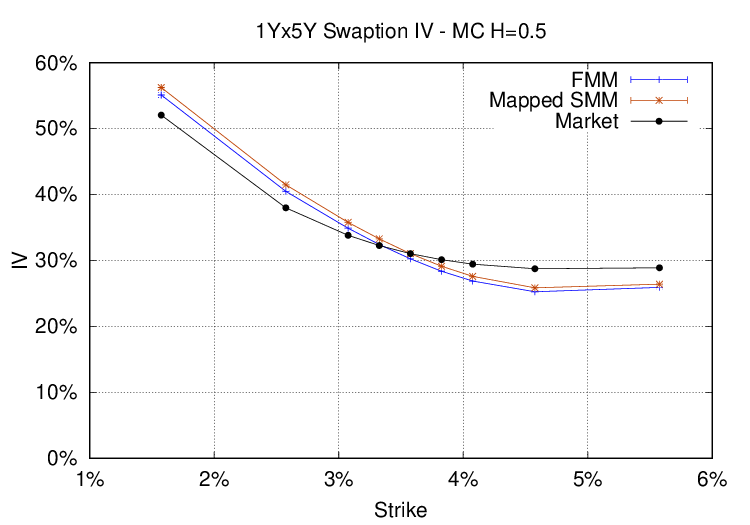} \\
			\includegraphics[width=0.47\textwidth]{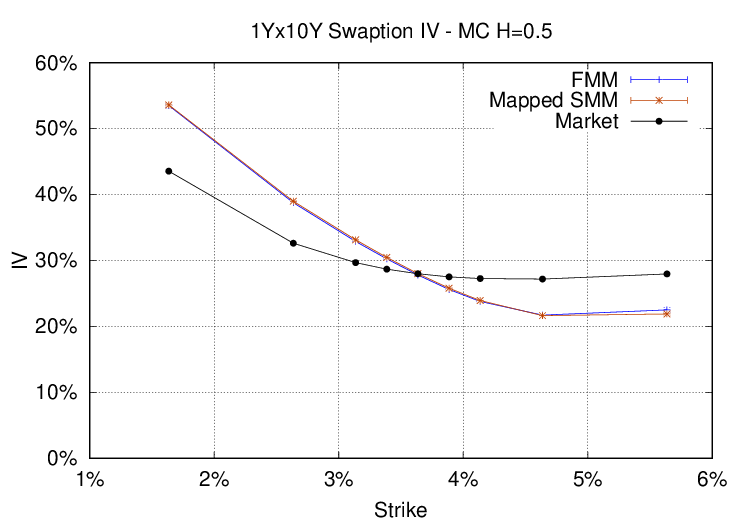} &
			\includegraphics[width=0.47\textwidth]{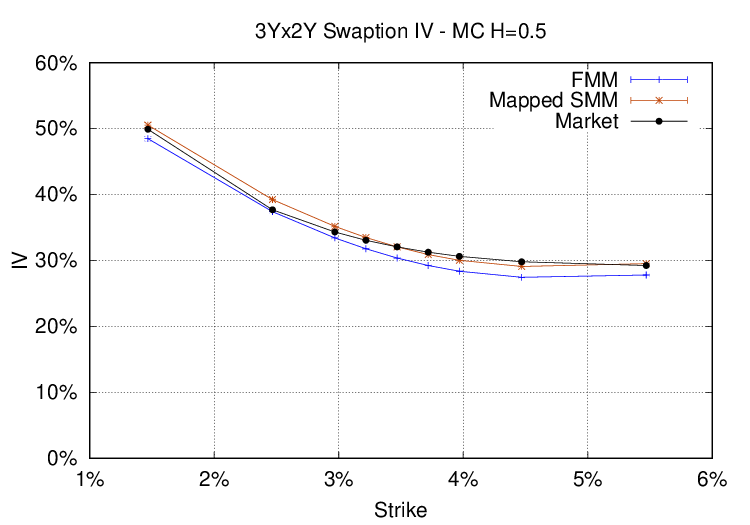} \\
			\includegraphics[width=0.47\textwidth]{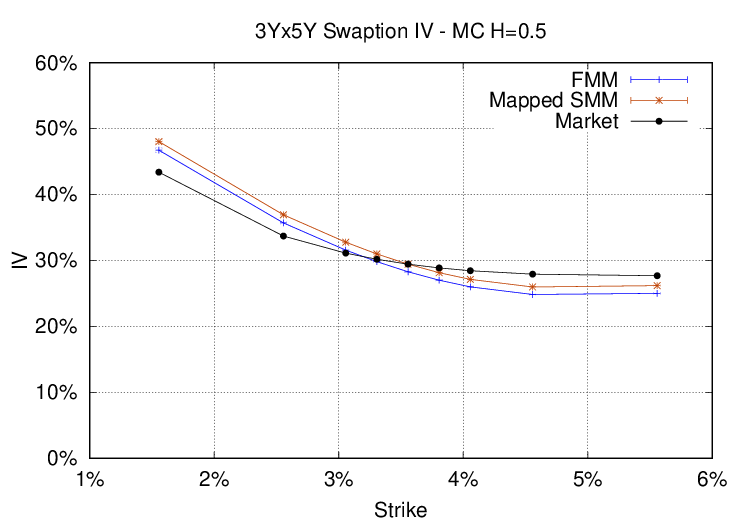} &
			\includegraphics[width=0.47\textwidth]{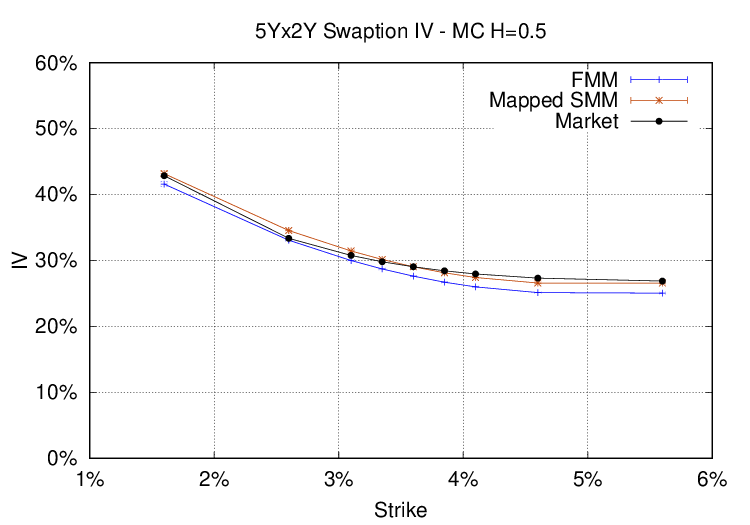} \\
			\includegraphics[width=0.47\textwidth]{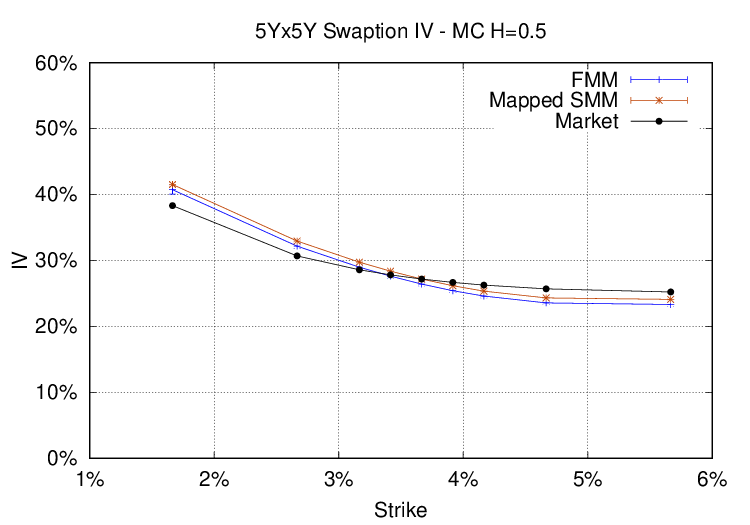} &
			\\
		\end{tabular}
		\caption{Model and Market IVs for tenors of 2Y or longer at $H=0.5$ 2024/6/7}
		\label{figure:20240607MCH50B}
	\end{figure}

	\begin{figure}[!tbp]
		\centering
		\begin{tabular}{cc}
			\includegraphics[width=0.47\textwidth]{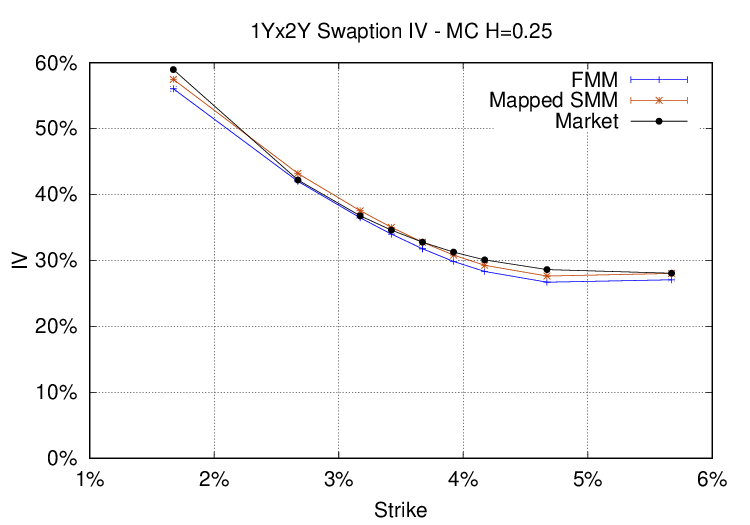} &
			\includegraphics[width=0.47\textwidth]{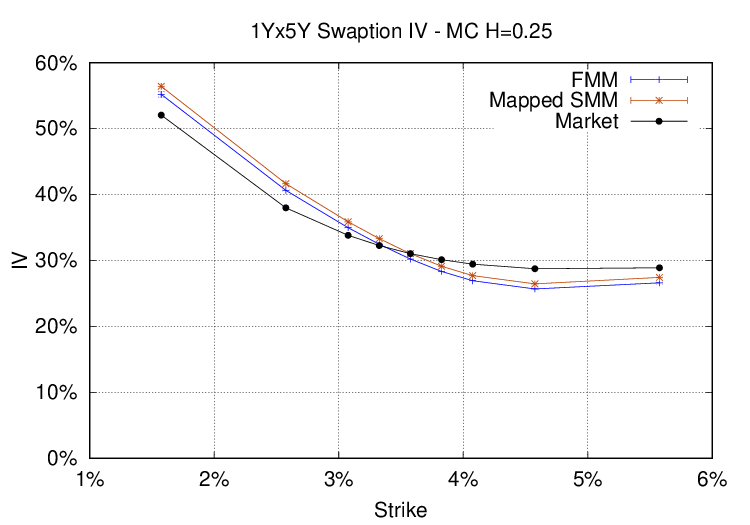} \\
			\includegraphics[width=0.47\textwidth]{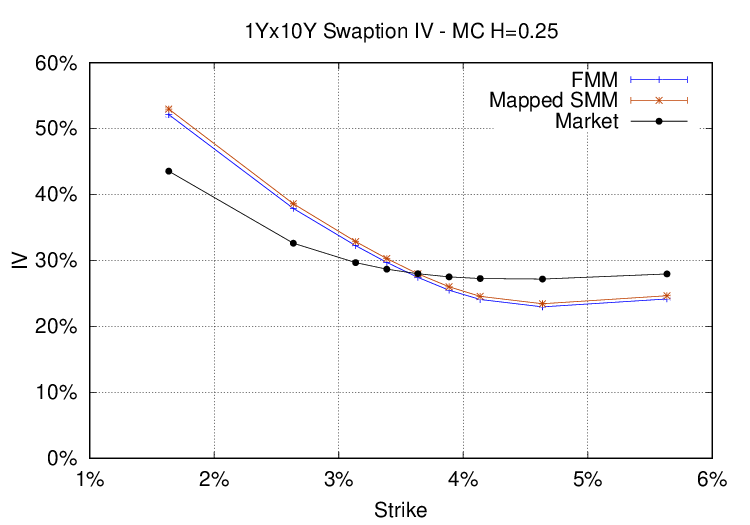} &
			\includegraphics[width=0.47\textwidth]{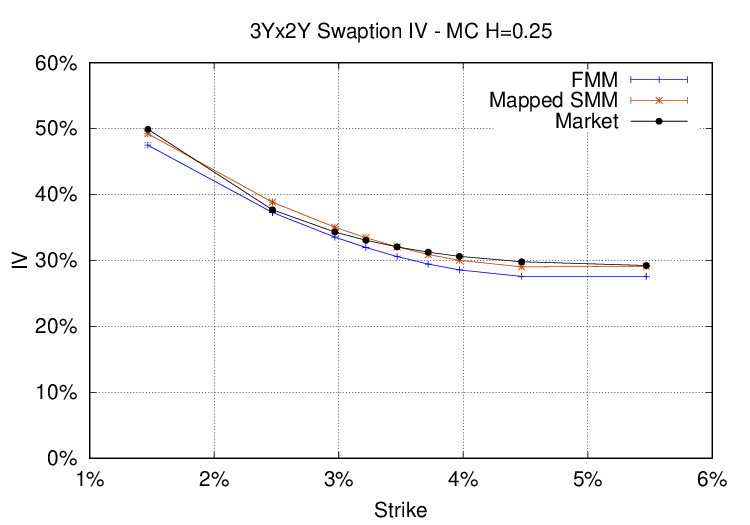} \\
			\includegraphics[width=0.47\textwidth]{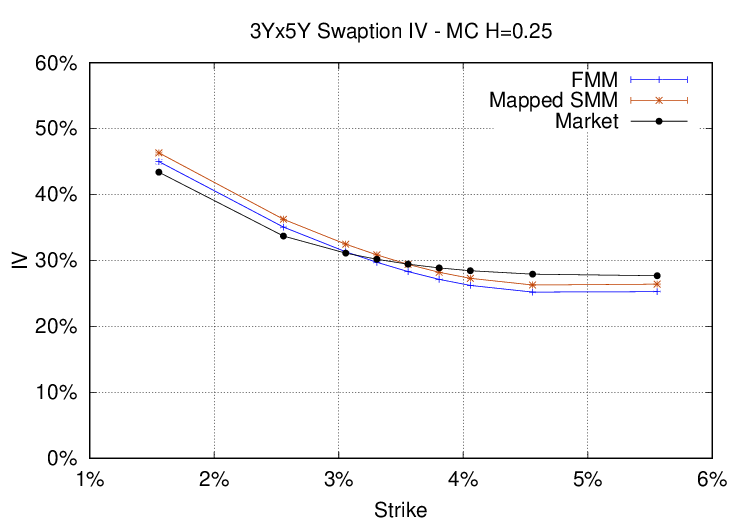} &
			\includegraphics[width=0.47\textwidth]{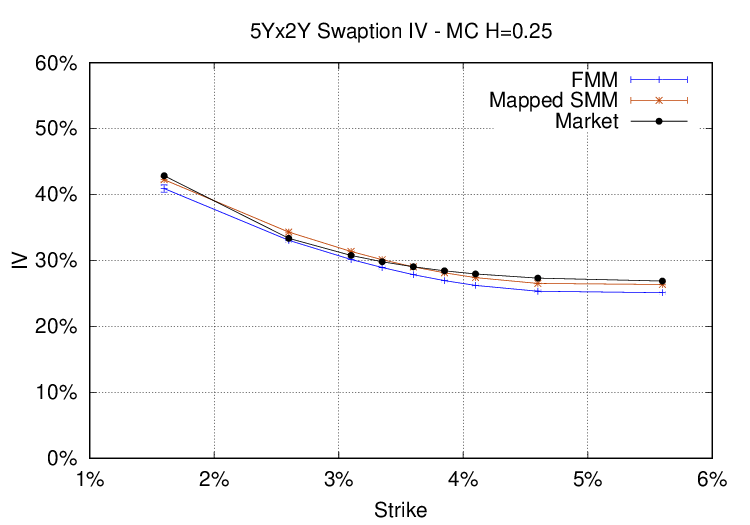} \\
			\includegraphics[width=0.47\textwidth]{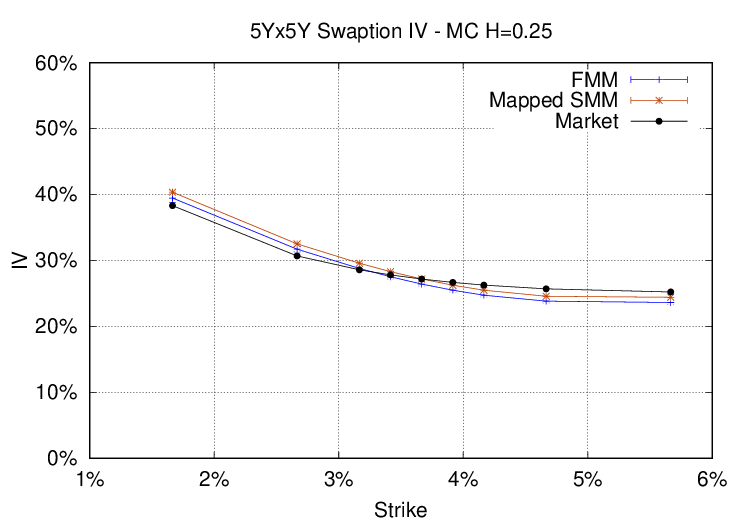} &
			\\
		\end{tabular}
		\caption{Model and Market IVs for tenors of 2Y or longer at $H=0.25$ 2024/6/7}
		\label{figure:20240607MCH20B}
	\end{figure}
	
		\begin{table}[]
		\tablefont
		\centering
		\caption{RMSE Comparison between FMM and Mapped SMM under Different Hurst Exponents (2024/6/7)}
		\label{table:RMSEComparison2024067}
		\begin{tabular}{|c|cc|cc|}
			\hline
			& \multicolumn{2}{c|}{FMM} & \multicolumn{2}{c|}{Mapped SMM} \\
			& H = 0.5 & H = 0.25 & H = 0.5 & H = 0.25 \\
			\hline
			1Yx1Y  & 1.64\% & 1.39\% & 0.73\% & 0.68\% \\
			3Yx1Y  & 1.55\% & 1.49\% & 0.88\% & 0.59\% \\
			5Yx1Y  & 1.38\% & 1.36\% & 0.53\% & 0.41\% \\
			10Yx1Y & 0.46\% & 0.49\% & 0.43\% & 0.46\% \\\hline
			1Yx2Y  & 1.73\% & 1.48\% & 0.88\% & 0.80\% \\
			1Yx5Y  & 2.31\% & 2.18\% & 2.44\% & 2.34\% \\
			1Yx10Y & 5.00\% & 4.16\% & 5.15\% & 4.37\% \\
			3Yx2Y  & 1.64\% & 1.67\% & 0.73\% & 0.63\% \\
			3Yx5Y  & 2.18\% & 1.74\% & 2.20\% & 1.62\% \\
			5Yx2Y  & 1.51\% & 1.44\% & 0.59\% & 0.58\% \\
			5Yx5Y  & 1.54\% & 1.19\% & 1.54\% & 1.13\% \\
			\hline
		\end{tabular}
	\end{table}

	\begin{table}[]
		\tablefont
		\centering
		\caption{First-Step Calibration Results 2025/6/10}
		\label{tb:firstStepCalibration20250610}
		\begin{tabular}{lc|cccc|cccc|c}
			\hline
			Hurst Exponent & $\kappa$ & $\alpha_{2}$ & $\alpha_{4}$ & $\alpha_{6}$ & $\alpha_{11}$ & $\rho_{0,2}$ & $\rho_{0,4}$ & $\rho_{0,6}$ & $\rho_{0,11}$ & RMSE \\
			\hline
			H=0.5 & 2.18032 & 0.329 & 0.322 & 0.325 & 0.393 & -67.3\% & -40.8\% & -37.1\% & -36.0\% & 1.308\% \\
			H=0.45 & 1.99582 & 0.327 & 0.311 & 0.303 & 0.348 & -67.1\% & -40.7\% & -37.1\% & -35.7\% & 1.286\% \\
			H=0.4 & 1.83877 & 0.324 & 0.301 & 0.286 & 0.311 & -67.0\% & -40.8\% & -37.4\% & -35.7\% & 1.245\% \\
			H=0.35 & 1.69842 & 0.321 & 0.292 & 0.271 & 0.279 & -66.8\% & -41.2\% & -37.9\% & -36.7\% & 1.187\% \\
			H=0.3 & 1.57150 & 0.315 & 0.283 & 0.257 & 0.252 & -66.8\% & -42.0\% & -39.1\% & -38.3\% & 1.117\% \\
			H=0.25 & 1.45492 & 0.307 & 0.271 & 0.243 & 0.230 & -67.1\% & -43.1\% & -40.7\% & -40.2\% & 1.038\% \\
			H=0.2 & 1.34230 & 0.294 & 0.256 & 0.227 & 0.208 & -68.2\% & -44.7\% & -42.8\% & -42.5\% & 0.951\% \\
			H=0.15 & 1.22944 & 0.274 & 0.235 & 0.206 & 0.186 & -71.3\% & -47.7\% & -46.3\% & -46.8\% & 0.865\% \\
			H=0.1 & 1.10295 & 0.238 & 0.202 & 0.177 & 0.158 & -81.2\% & -55.0\% & -53.4\% & -55.2\% & 0.786\% \\
			H=0.05 & 0.86551 & 0.181 & 0.159 & 0.140 & 0.126 & -100.0\% & -78.4\% & -76.8\% & -81.8\% & 0.999\% \\
			\hline
		\end{tabular}
	\end{table}
	
	\begin{table}[]
		\tablefont
		\centering
		\caption{First-Step Calibration Results 2025/6/10 time step 1000}
		\label{tb:firstStepCalibration20250610-1000}
		\begin{tabular}{lc|cccc|cccc|c}
			\hline
			Hurst Exponent & $\kappa$ & $\alpha_{2}$ & $\alpha_{4}$ & $\alpha_{6}$ & $\alpha_{11}$ & $\rho_{0,2}$ & $\rho_{0,4}$ & $\rho_{0,6}$ & $\rho_{0,11}$ & RMSE \\
			\hline
			H=0.5 & 2.170540 & 0.330 & 0.322 & 0.326 & 0.396 & -0.671 & -0.393 & -0.343 & -0.341 & 1.309\% \\
			H=0.45 & 1.992640 & 0.328 & 0.311 & 0.305 & 0.352 & -0.669 & -0.396 & -0.354 & -0.330 & 1.289\% \\
			H=0.4 & 1.834450 & 0.325 & 0.301 & 0.287 & 0.313 & -0.667 & -0.399 & -0.365 & -0.329 & 1.249\% \\
			H=0.35 & 1.695700 & 0.321 & 0.292 & 0.271 & 0.281 & -0.665 & -0.402 & -0.375 & -0.340 & 1.191\% \\
			H=0.3 & 1.572020 & 0.315 & 0.283 & 0.257 & 0.254 & -0.664 & -0.408 & -0.383 & -0.362 & 1.118\% \\
			H=0.25 & 1.456380 & 0.307 & 0.271 & 0.243 & 0.230 & -0.666 & -0.417 & -0.393 & -0.391 & 1.035\% \\
			H=0.2 & 1.345680 & 0.294 & 0.256 & 0.227 & 0.208 & -0.674 & -0.430 & -0.408 & -0.416 & 0.945\% \\
			H=0.15 & 1.236500 & 0.273 & 0.234 & 0.205 & 0.184 & -0.698 & -0.456 & -0.436 & -0.449 & 0.856\% \\
			H=0.1 & 1.121660 & 0.234 & 0.198 & 0.173 & 0.153 & -0.778 & -0.522 & -0.505 & -0.531 & 0.772\% \\
			H=0.05 & 0.917419 & 0.165 & 0.143 & 0.125 & 0.111 & -1.000 & -0.775 & -0.765 & -0.818 & 0.899\% \\
			\hline
		\end{tabular}
	\end{table}

	\begin{table}[htbp]
		\tablefont
		\centering
		\caption{Correlation Matrix Among Interest Rate Factors. (H=0.5)2025/6/10}
		\label{table:corrMCH520250610}
		\begin{tabular}{cccccccccc}
			\hline
			1.0000 & 0.9224 & 0.7097 & 0.6054 & 0.4645 & 0.3503 & 0.3993 & 0.3400 & 0.2534 & 0.1605\\
			0.9224 & 1.0000 & 0.8613 & 0.6768 & 0.5421 & 0.4134 & 0.3050 & 0.2315 & 0.1515 & 0.0865\\
			0.7097 & 0.8613 & 1.0000 & 0.8411 & 0.6622 & 0.4820 & 0.3546 & 0.2700 & 0.2292 & 0.1477\\
			0.6054 & 0.6768 & 0.8411 & 1.0000 & 0.8388 & 0.6814 & 0.5053 & 0.3959 & 0.2983 & 0.1722\\
			0.4645 & 0.5421 & 0.6622 & 0.8388 & 1.0000 & 0.8388 & 0.6913 & 0.5271 & 0.4209 & 0.3050\\
			0.3503 & 0.4134 & 0.4820 & 0.6814 & 0.8388 & 1.0000 & 0.8416 & 0.6968 & 0.5586 & 0.4746\\
			0.3993 & 0.3050 & 0.3546 & 0.5053 & 0.6913 & 0.8416 & 1.0000 & 0.8547 & 0.7344 & 0.6031\\
			0.3400 & 0.2315 & 0.2700 & 0.3959 & 0.5271 & 0.6968 & 0.8547 & 1.0000 & 0.8683 & 0.7453\\
			0.2534 & 0.1515 & 0.2292 & 0.2983 & 0.4209 & 0.5586 & 0.7344 & 0.8683 & 1.0000 & 0.8596\\
			0.1605 & 0.0865 & 0.1477 & 0.1722 & 0.3050 & 0.4746 & 0.6031 & 0.7453 & 0.8596 & 1.0000\\
			\hline
		\end{tabular}
	\end{table}
	
	\begin{table}[htbp]
		\tablefont
		\centering
		\caption{Correlation Matrix Among Interest Rate Factors. (H=0.1)2025/6/10}
		\label{table:corrMCH1020250610}
		\begin{tabular}{cccccccccc}
			\hline
			1.0000 & 0.9803 & 0.9339 & 0.9304 & 0.9222 & 0.9191 & 0.9104 & 0.9117 & 0.9269 & 0.9053\\
			0.9803 & 1.0000 & 0.9861 & 0.9845 & 0.9754 & 0.9728 & 0.9587 & 0.9612 & 0.9772 & 0.9497\\
			0.9339 & 0.9861 & 1.0000 & 1.0000 & 0.9915 & 0.9876 & 0.9707 & 0.9728 & 0.9902 & 0.9580\\
			0.9304 & 0.9845 & 1.0000 & 1.0000 & 0.9915 & 0.9876 & 0.9706 & 0.9726 & 0.9901 & 0.9577\\
			0.9222 & 0.9754 & 0.9915 & 0.9915 & 1.0000 & 0.9659 & 0.9732 & 0.9492 & 0.9903 & 0.9465\\
			0.9191 & 0.9728 & 0.9876 & 0.9876 & 0.9659 & 1.0000 & 0.9610 & 0.9612 & 0.9565 & 0.9409\\
			0.9104 & 0.9587 & 0.9707 & 0.9706 & 0.9732 & 0.9610 & 1.0000 & 0.9609 & 0.9550 & 0.9622\\
			0.9117 & 0.9612 & 0.9728 & 0.9726 & 0.9492 & 0.9612 & 0.9609 & 1.0000 & 0.9670 & 0.9692\\
			0.9269 & 0.9772 & 0.9902 & 0.9901 & 0.9903 & 0.9565 & 0.9550 & 0.9670 & 1.0000 & 0.9543\\
			0.9053 & 0.9497 & 0.9580 & 0.9577 & 0.9465 & 0.9409 & 0.9622 & 0.9692 & 0.9543 & 1.0000\\
			\hline
		\end{tabular}
	\end{table}

	\begin{figure}[!tbp]
		\centering
		\begin{tabular}{cc}
			\includegraphics[width=0.47\textwidth]{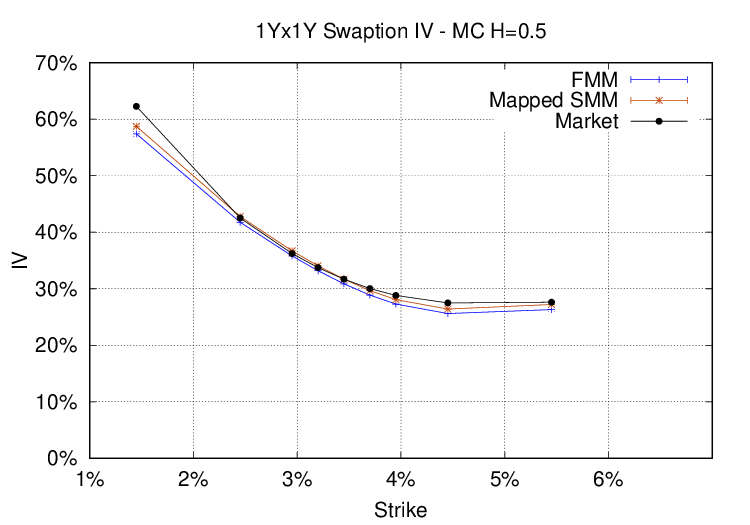} &
			\includegraphics[width=0.47\textwidth]{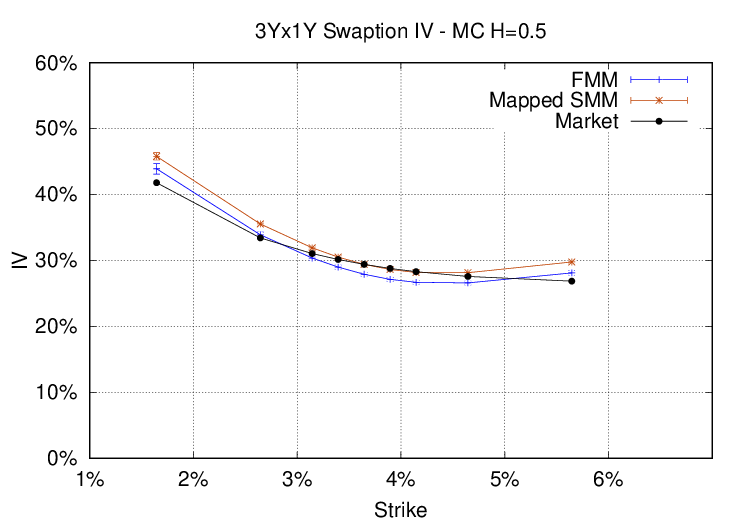} \\
			\includegraphics[width=0.47\textwidth]{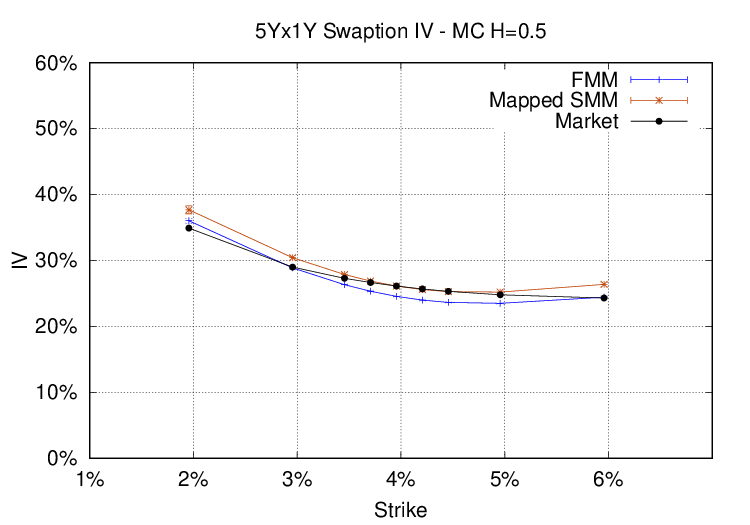} &
			\includegraphics[width=0.47\textwidth]{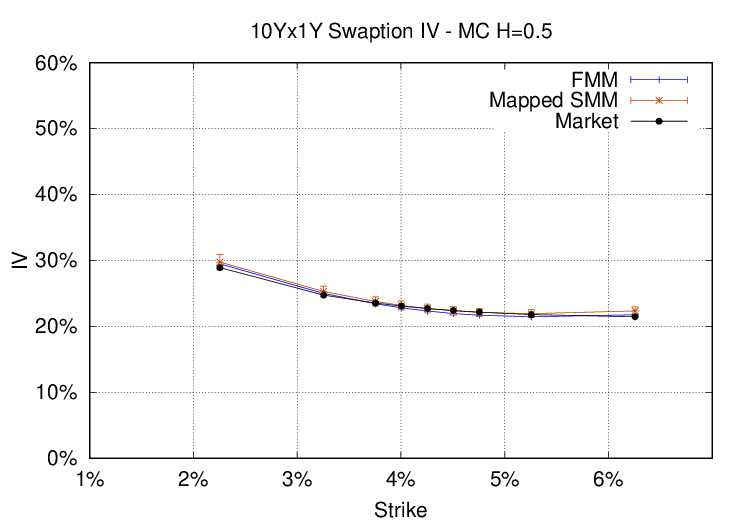}\\
		\end{tabular}
		\caption{Model and Market IVs of 1Y tenor at $H=0.5$ 2025/6/10}
		\label{figure:20250610MCH50A}
	\end{figure}

	\begin{figure}[!tbp]
		\centering
		\begin{tabular}{cc}
			\includegraphics[width=0.47\textwidth]{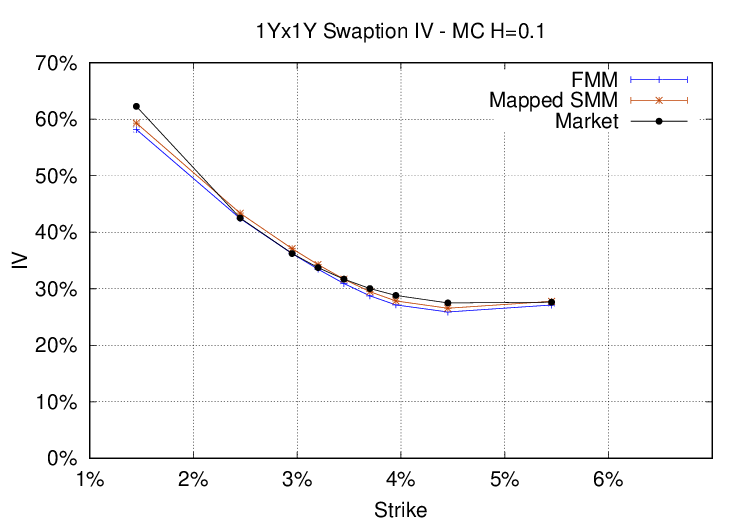} &
			\includegraphics[width=0.47\textwidth]{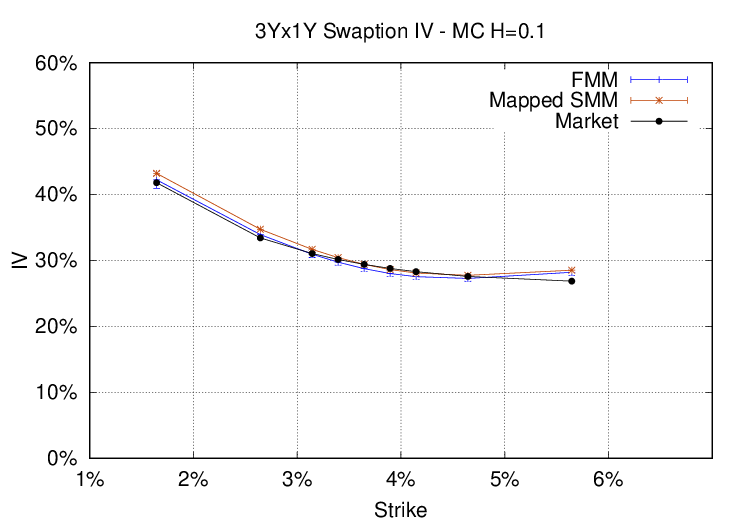} \\
			\includegraphics[width=0.47\textwidth]{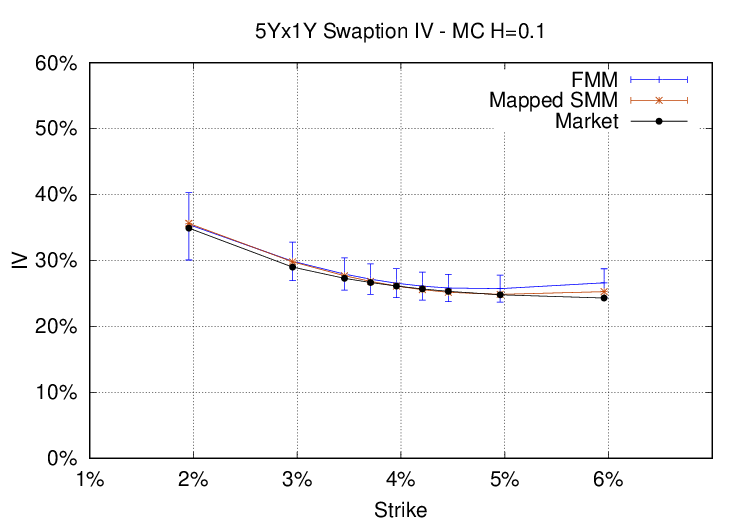} &
			\includegraphics[width=0.47\textwidth]{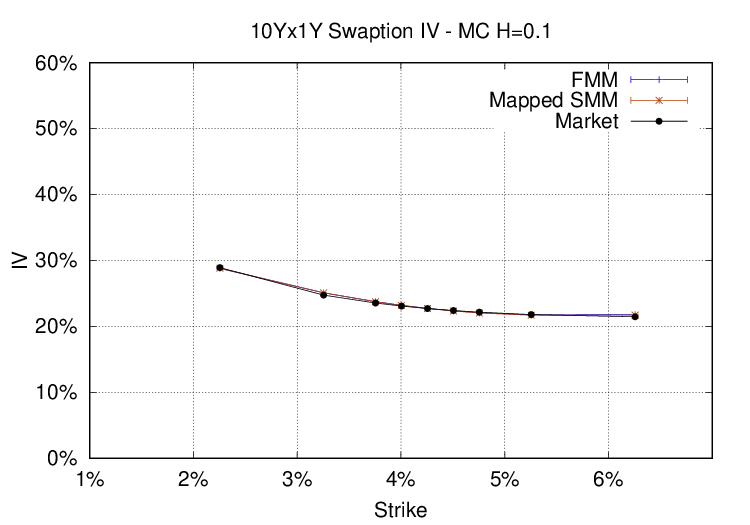}\\
		\end{tabular}
		\caption{Model and Market IVs of 1Y tenor at $H=0.1$ 2025/6/10}
		\label{figure:20250610MCH10A}
	\end{figure}
	
	\begin{figure}[!tbp]
		\centering
		\begin{tabular}{cc}
			\includegraphics[width=0.47\textwidth]{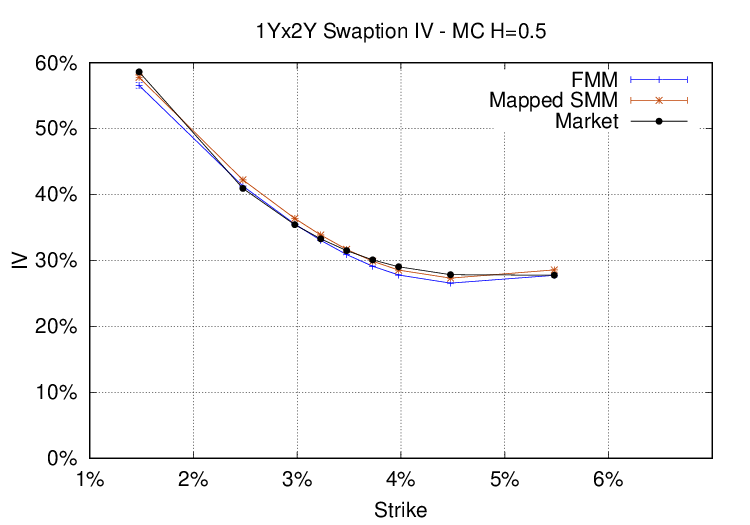} &
			\includegraphics[width=0.47\textwidth]{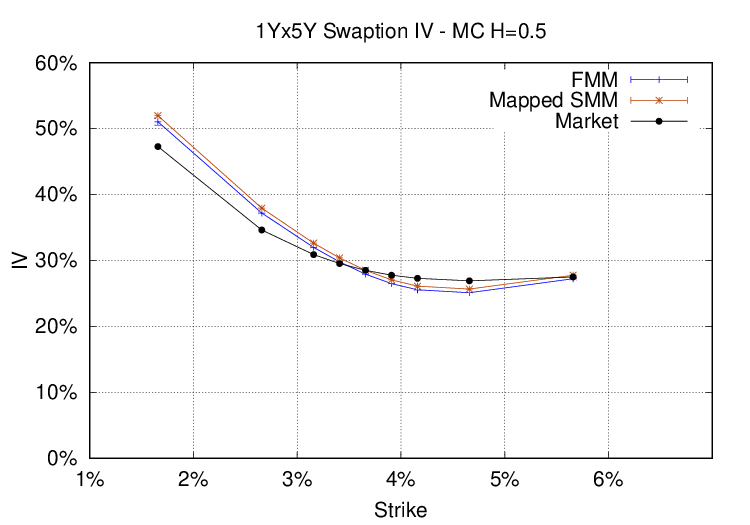} \\
			\includegraphics[width=0.47\textwidth]{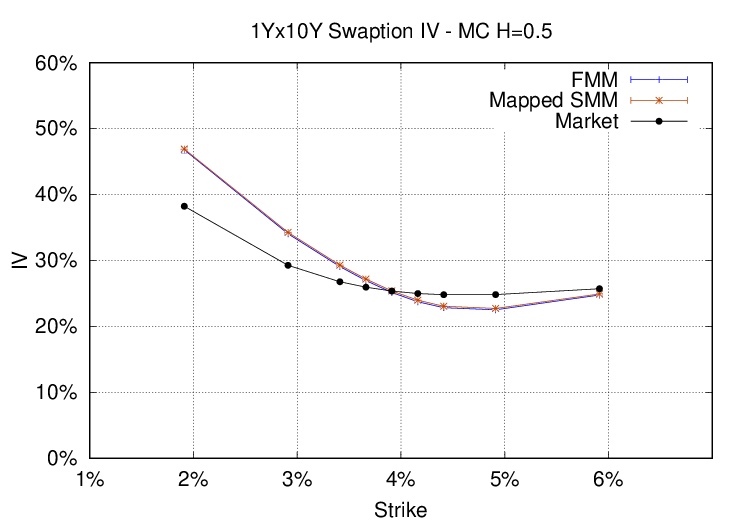} &
			\includegraphics[width=0.47\textwidth]{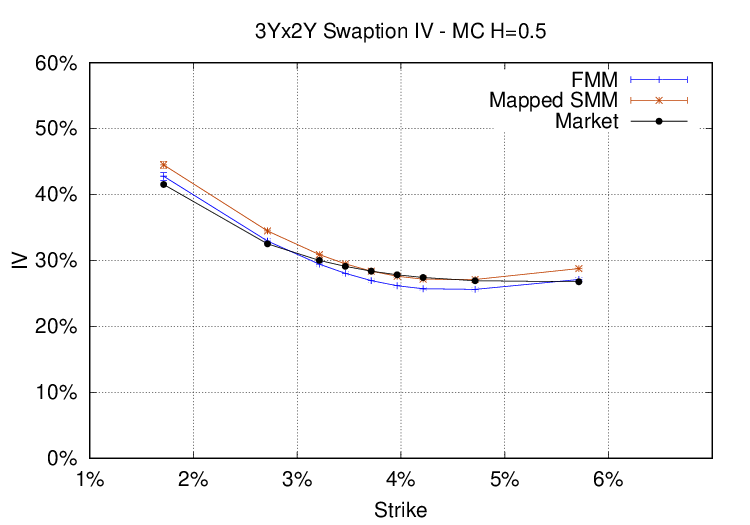} \\
			\includegraphics[width=0.47\textwidth]{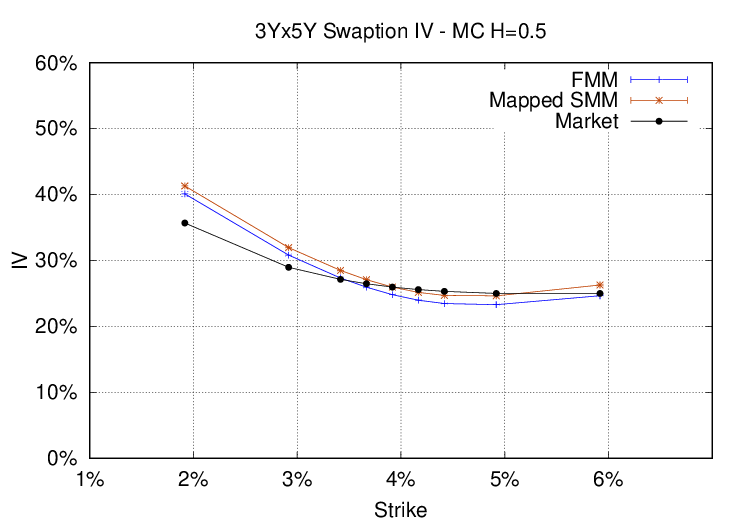} &
			\includegraphics[width=0.47\textwidth]{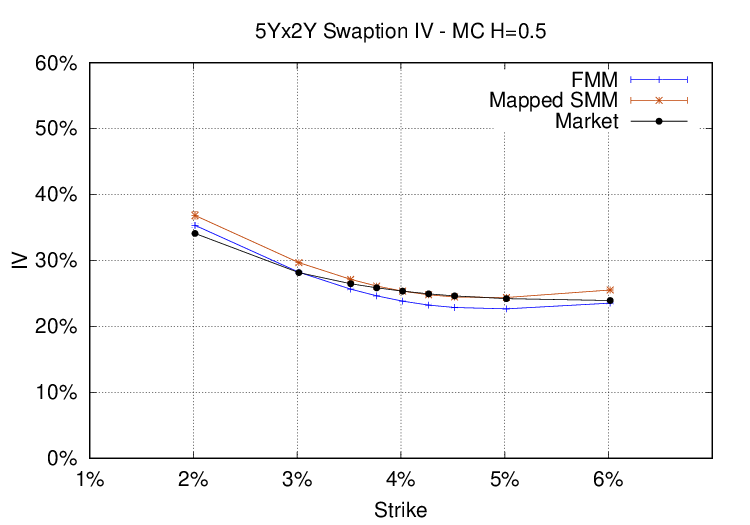} \\
			\includegraphics[width=0.47\textwidth]{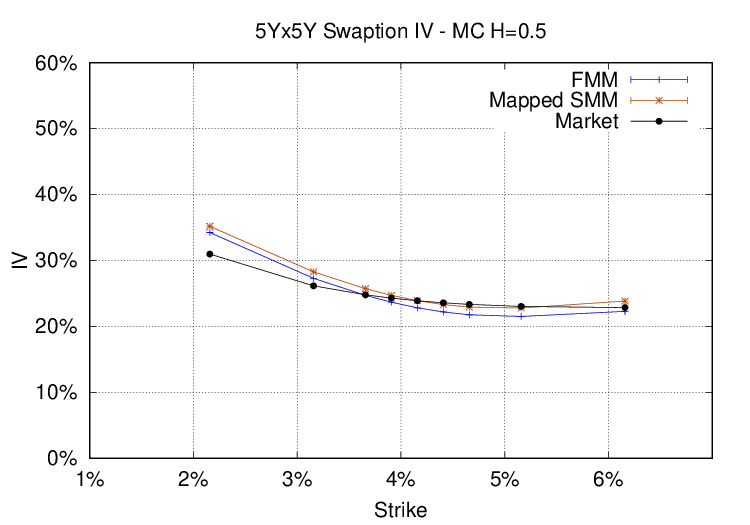} &
			\\
		\end{tabular}
		\caption{Model and Market IVs for tenors of 2Y or longer at $H=0.5$ 2025/6/10}
		\label{figure:20250610MCH50B}
	\end{figure}

	\begin{figure}[!tbp]
		\centering
		\begin{tabular}{cc}
			\includegraphics[width=0.47\textwidth]{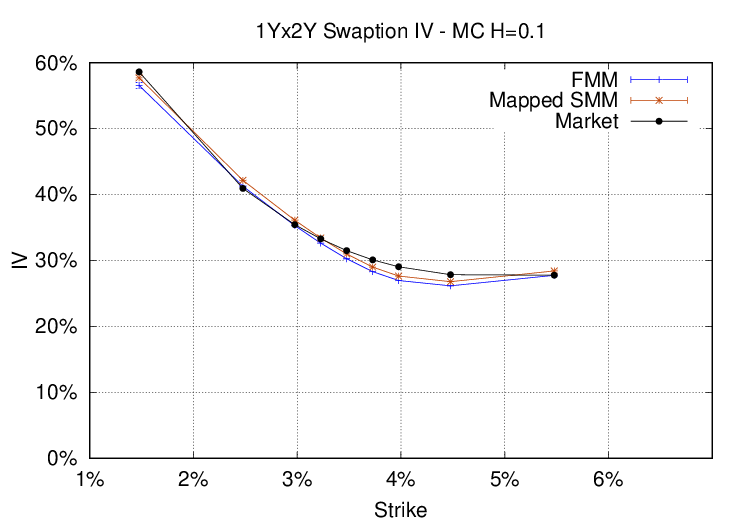} &
			\includegraphics[width=0.47\textwidth]{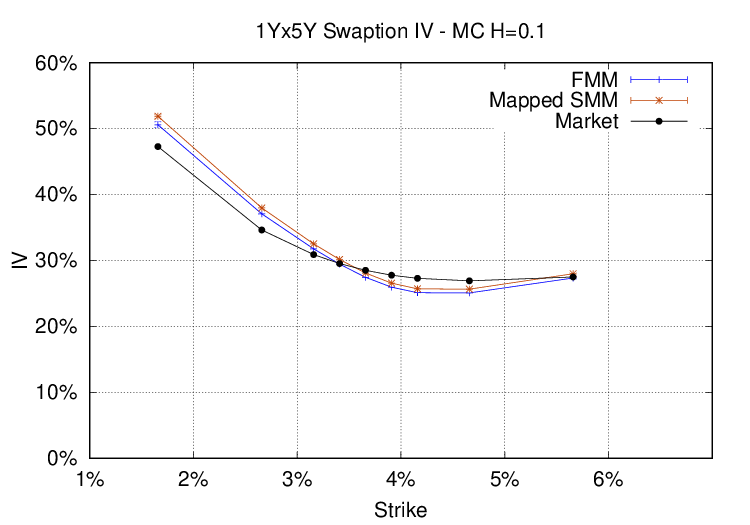} \\
			\includegraphics[width=0.47\textwidth]{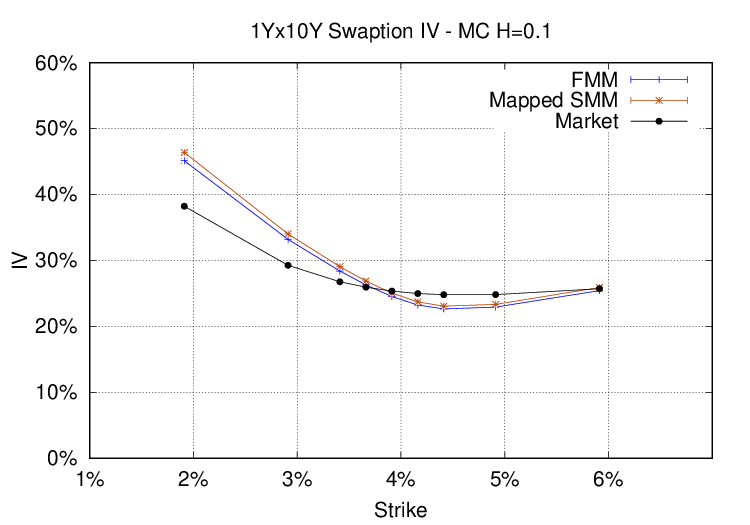} &
			\includegraphics[width=0.47\textwidth]{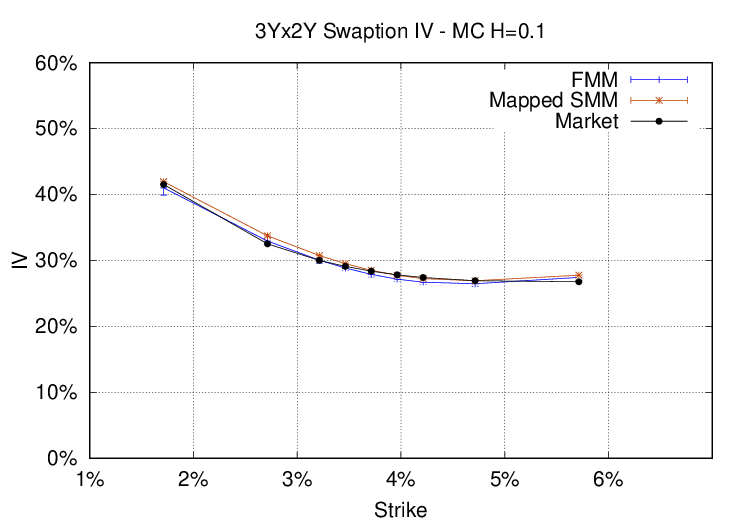} \\
			\includegraphics[width=0.47\textwidth]{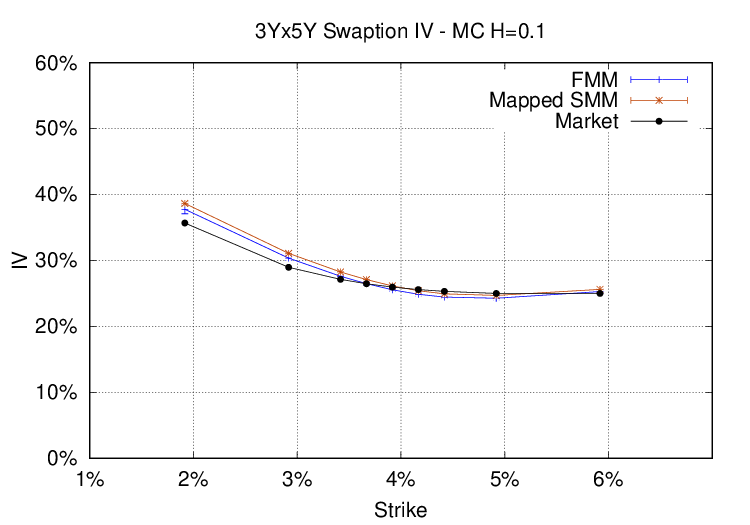} &
			\includegraphics[width=0.47\textwidth]{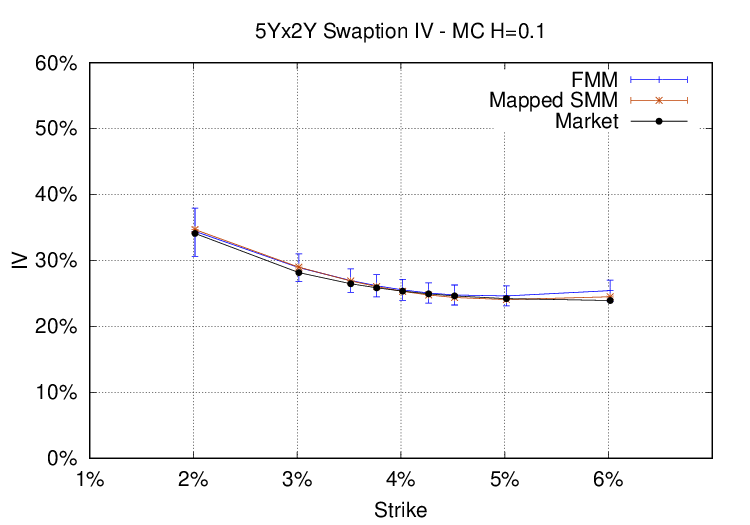} \\
			\includegraphics[width=0.47\textwidth]{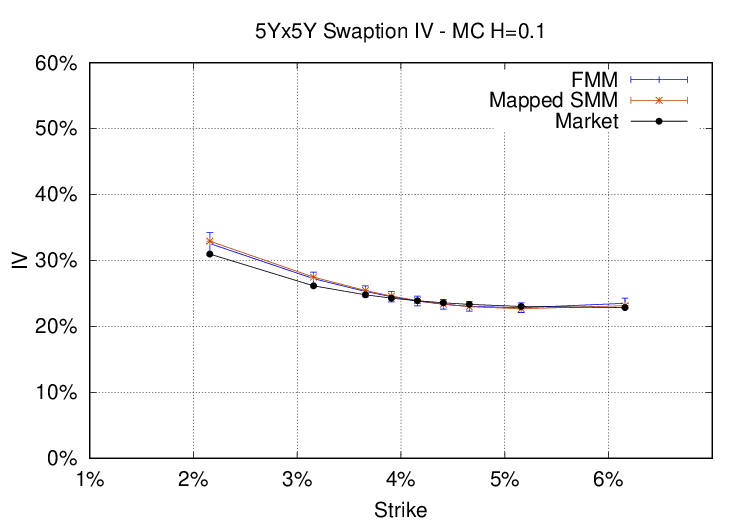} &
			\\
		\end{tabular}
		\caption{Model and Market IVs for tenors of 2Y or longer at $H=0.1$ 2025/6/10}
		\label{figure:20250610MCH10B}
	\end{figure}
	
	\begin{table}[]
		\tablefont
		\centering
		\caption{RMSE Comparison between FMM and Mapped SMM under Different Hurst Exponents (2025/6/10)}
		\label{table:RMSEComparison20250610}
		\begin{tabular}{|c|cc|cc|}
			\hline
			& \multicolumn{2}{c|}{FMM} & \multicolumn{2}{c|}{Mapped SMM} \\
			& H = 0.5 & H = 0.1 & H = 0.5 & H = 0.1 \\
			\hline
			1Yx1Y  & 1.95\% & 1.66\% & 1.29\% & 1.18\% \\
			3Yx1Y  & 1.36\% & 0.68\% & 1.83\% & 0.89\% \\
			5Yx1Y  & 1.24\% & 0.97\% & 1.27\% & 0.51\% \\
			10Yx1Y & 0.37\% & 0.18\% & 0.45\% & 0.18\% \\\hline
			1Yx2Y  & 1.00\% & 1.36\% & 0.74\% & 0.93\% \\
			1Yx5Y  & 1.84\% & 1.83\% & 2.11\% & 2.14\% \\
			1Yx10Y & 3.56\% & 2.94\% & 3.61\% & 3.37\% \\
			3Yx2Y  & 1.19\% & 0.51\% & 1.41\% & 0.62\% \\
			3Yx5Y  & 1.94\% & 0.97\% & 2.24\% & 1.32\% \\
			5Yx2Y  & 1.27\% & 0.62\% & 1.20\% & 0.43\% \\
			5Yx5Y  & 1.51\% & 0.72\% & 1.65\% & 0.85\% \\
			\hline
		\end{tabular}
	\end{table}

	Figures~\ref{figure:20240607MCH50A}--\ref{figure:20250610MCH10B} compare the model and market IVs on the two additional dates.
	On 7 June 2024, using $H=0.25$ rather than $H=0.5$ reduces the smile RMSE for nine of the eleven maturity--tenor pairs under the FMM and for ten under the mapped SMM (Table~\ref{table:RMSEComparison2024067}).
	On 10 June 2025, the corresponding counts for $H=0.1$ are ten and nine (Table~\ref{table:RMSEComparison20250610}).
	These results are consistent with the main experiment: allowing $H<0.5$ generally improves the fit to market smiles.